 \documentclass[prb,aps,superscriptaddress]{revtex4-1}
\usepackage{graphicx}
\usepackage{dcolumn}
\usepackage{bm}
\usepackage{lipsum}
\usepackage{ragged2e}
\usepackage{braket}
\usepackage{bbm}
\usepackage{color}
\usepackage{xcolor}
\usepackage{amsfonts}
\usepackage{amsmath}
\usepackage{amsmath,amssymb,latexsym}
\usepackage{wasysym,stmaryrd,marvosym}
\usepackage{mathrsfs}
\usepackage{natbib}
\providecommand{\keywords}[1]{\textbf{\textit{Index terms---}} #1}
\usepackage[mathlines]{lineno}
\usepackage[colorinlistoftodos]{todonotes}
\usepackage[colorlinks=true, allcolors=blue]{hyperref}

\begin{document}

\newcommand{\cnyn}{Centro de Nanociencias y Nanotecnolog\'ia,
Universidad Nacional Aut\'onoma de M\'exico, Apartado Postal 2681, 22800, Ensenada, Baja California, M\'exico.}
\newcommand{\uabc}{Facultad de Ciencias, Universidad Aut\'onoma de Baja California, Apartado Postal 1880, 22800 Ensenada, Baja California, M\'exico}

\title{Optical properties of massive anisotropic tilted Dirac systems}

\author{M. A. Mojarro}
\affiliation{\uabc}
\author{R. Carrillo-Bastos}
\affiliation{\uabc}
\author{Jes\'us A. Maytorena}
\affiliation{\cnyn}

\date{\today}

\begin{abstract}


We explore the effect of valley-contrasting gaps in the optical response of two-dimensional anisotropic tilted Dirac systems. We study the spectrum of intraband and interband transitions through the joint density of states (JDOS), the optical conductivity tensor and the Drude spectral weight. The energy bands present an indirect gap $2\tilde{\Delta}^{\xi}$ in each valley ($\xi$), with a reduced magnitude with respect to the nominal gap $2\Delta^{\xi}$ of the untilted system. Thus a new possibility opens for the position of the Fermi level, $\tilde{\Delta}^{\xi}\leqslant\varepsilon_F\leqslant\Delta^{\xi}$ (an ``indirect zone''), and for the momentum space for allowed transitions.  As a consequence, the JDOS near each gap displays a set of three van Hove singularities which are in contrast to the well known case of gapped graphene (an absorption edge at $\text{max}\{2\Delta,2\varepsilon_F\}$ only) or 8-$Pmmn$  borophene (two interband critical points made possible by the tilt). On the other hand, for the Fermi level lying in the gap the JDOS shows the usual linear dependence on  frequency, while when lying above $2\Delta^{\xi}$ it looks qualitatively similar to the borophene case. These spectral characteristics in each valley determine the prominent structure of the optical conductivity. The spectra of the longitudinal conductivity illustrate the strong anisotropy of the optical response. Similarly, the Drude weight is anisotropic and shows regions of nonlinear dependence on the Fermi level. The breaking of the valley symmetry leads to a finite Hall response and associated optical properties. The anomalous and valley Hall conductivities present graphene-like behavior with characteristic modifications due to the indirect zones. 
Almost perfect circular dichroism can be achieved by tuning the exciting frequency with an appropriate Fermi level position.
We also calculate the spectra of optical opacity and polarization rotation, which can reach magnitudes of tenths of radians in some cases. The spectral features of the calculated response properties are signatures of the simultaneous presence of tilt and mass, and suggest optical ways to determine the formation of different gaps in such class of Dirac systems.

\end{abstract}

\keywords{Suggested keywords}

\maketitle

\section{Introduction}

Relativistic effects are ubiquitous in two-dimensional materials, and light-matter interaction has become a powerful tool to test its intriguing consequences. From the spatial imaging of the spin Hall effect in two-dimensional electron gasses\cite{sih2005spatial} to the distinctive optical response in graphene\cite{nair2008fine} and the quantized Faraday and Kerr rotations in topological insulators\cite{wu2016quantized,universalFaraday}, optical techniques have not only served as a probe of non-conventional behaviour of these materials, but also as a way to extract parameter values for  effective models\cite{Determination01,Determination02}. Furthermore, optical properties can be very sensitive to broken symmetries present in the studied system \cite{nandkishore2011polar,AHE2DM,Dyrdal}. For example, the rotation of the polarization plane after passing through a medium, known as the Faraday effect, can serve as an indicator of the breaking of either time-reversal symmetry (TRS) or inversion symmetry; Even for the thinnest samples\cite{szechenyi2016transfer}, like graphene\cite{crassee2011giant} and the surface of topological insulators\cite{TsePhysRevLett.105.057401}, the Faraday angle can reach several degrees. Similarly, the polar Kerr effect has as a necessary condition the breaking of the TRS. Since both effects are directly related with the ac conductivity, they offer a contact-free manner to measure the electronic transport properties of materials\cite{nandkishore2011polar}. 

When materials display a relativistic-like linear spectrum they are called Dirac materials\cite{DiracMatter2014}. Most of these materials present an isotropic spectrum in momentum space\cite{DiracMatter2015}, a symmetric Dirac cone. Nevertheless, it has been recently found that some of them present anisotropic linear spectra, i.e. tilted anisotropic Dirac cones; such as the case for 8-{\it Pmmn} borophene\cite{Two-Dimensional-Boron,DFT-Borophene2016,xu2016hydrogenated,nakhaee2018tight}, quinod-type graphene\cite{goerbig2008tilted} and the organic conductor $\alpha$-(BEDT-TTF)$_{2}$I$_{3}$\cite{kajita2006massless,MesslessFermions}. In general, the presence of a cone tilt does lead to qualitatively different behaviour compared with the untilted system\cite{Effects-of-tilt}; In particular, when interacting with light it gives raise to different optical and electronic properties. Sadhukhan and Agarwal\cite{Plasmons2017} found anisotropic plasmon dispersion, and Sarí, et. al. a unique intervalley damping effect for magnetoplasmons\cite{MagnetoplasmonsPRB,MagnetoplasmonsPRB2}. Likewise, it has been found that the dc conductivity becomes strongly anisotropic between the parallel and perpendicular direction to the tilt \cite{rostamzadeh2019large}, while the frequency dependent optical conductivity acquires a non-monotonic behaviour with energy that allows to extract the tilting parameter from optical measurements\cite{verma2017effect,TiltingDiracCond,ConductivityOrganic}. Being semimetals, these materials have zero energy gap, but a gap can be generated artificially\cite{kibis,oka2009photovoltaic,syzranov2008effect,calvo2011tuning,champo2019metal, ibarra2019dynamical,yuan2017ideal,sandoval2020floquet}. In general, the emergence of a gap can be related with the breaking of a symmetry. For example, in graphene the otherwise semimetallic behaviour, can be changed by breaking inversion symmetry\cite{komatsu2018observation}, which results in the opening of a band gap and the consequent valley Hall effect. Perhaps, the most noticeable example of this, is the quantum Hall effect in graphene\cite{novoselov2006unconventional},  which is obtained by placing graphene in a perpendicular magnetic field.
Haldane\cite{Haldane1988} showed with an example that the presence of a magnetic field was not necessary condition, but the breaking of TRS. An alternative way, to generate a Hall conductivity by breaking TRS without a magnetic field, is the spin-texture proposed by Hill et al.\cite{hill2011valley}. In Hill's model, the localized spins of ad-atoms doping one of the sublattices of graphene arrange themselves creating a spin configuration with tilted spins, which can be described by an effective tight-binding Hamiltonian with valley dependent gaps. The latter effect drastically modifies the density of states (DOS) profile, as well as the optical longitudinal and anomalous Hall conductivities. Aside of the spin-textured graphene, there are other Dirac systems with valley asymmetric gap, like gated silicene\cite{Stille-silicene}, $\alpha$-(BEDT-TTF)$_{2}$I$_{3}$ with magnetic modulations\cite{osada2017chern}, and the modified Haldane model\cite{Saito-Haldane}. Although the optical properties of tilted anisotropic Dirac systems have been subject of intense research\cite{nishine2010tilted,Nonlinear-Agarwal,zabolotskiy2016strain,space-time-platform,space-time-platform,zhang2017two,zhang2018oblique,Polarization2019,xu2019insights,Vibrational2016,sengupta2018anomalous,superconducting,Minkowski2019}, up to our knowledge the effect of valley dependent\cite{xu2018electrically,jotzu2014NATURE} gaps\cite{gap2019,gap2020} in the optical properties of a tilted anisotropic Dirac system,  has not been reported yet; In this paper we present such a study.

The outline of the paper is the following. In Sec.\,II we present the Dirac-like Hamiltonian 
and its energy band structure, identifying the effect of tilting and anisotropy, the
nonuniform gapped valleys, and Fermi contours. In Sec.\,III we study the optical transitions
near the gaps. We first study  the joint density of states in order to identify critical points, which will
determine the prominent spectral features of the optical response (Sec.\,III\,A). The electrical conductivity
tensor, due to intra and interband transitions, is calculated in Sec.\,III\,B within the Kubo formalism.
The Drude weight is discussed in Sec.\,III\,C. In Sec.\,IV we present optical properties of our system.
The anisotropy of the response, circular dichroism spectrum 
and valley polarization are studied in Sec.\,IV\,A. The
anomalous and valley Hall conductivities are obtained in Sec.\,IV\,B, and compared to the model
of gapped graphene with broken valley symmetry developed by Hill et al.\,\cite{hill2011valley}.
Spectra of transmission as a function of angles of incidence and polarization for several positions
of the Fermi level is considered in Sec.\,IV\,C. In Sec.\,IV\,D we calculate spectra of Kerr and Faraday rotation. Finally,  we present our conclusions in Sec.\,V. There are two 
appendices with expressions of Fermi contours and related quantities, and of the Fresnel amplitudes
describing the problem of refraction of a 2D system between two dielectrics.

\section{The Hamiltonian: anisotropy, tilt and valley-contrasting gaps}

We consider a 2D anisotropic tilted Dirac system with momentum-space Hamiltonian
 \begin{equation} \label{Hborogap}
H^{\xi}({\bf k})=\xi(\hslash v_tk_y\mathbb{I}+\hslash v_xk_x\sigma_x+\xi\hslash v_yk_y\sigma_y) + \Delta^{\xi}\sigma_z \ ,
\end{equation}
where $\sigma_i$ are the Pauli matrices acting on the pseudospin space, $\mathbb{I}$ is the identity matrix and $\xi=\pm$ (or $K, K'$) is a valley index; the
electron wave vector ${\bf k}=(k_x,k_y)$ is measured from the nominal Dirac point in each valley.
In addition to the broken particle-hole symmetry (PHS), the model include a  valley-contrasting mass, 
$\Delta^+\neq\Delta^-$, which breaks the time-reversal symmetry (TRS). For $\Delta^{\xi}=0$ the Hamiltonian
describes the low lying excitations of two tilted Dirac cones  like in
some 2D graphene-type materials or some organic conductors subjected to pressure and uniaxial strain 
\cite{zabolotskiy2016strain,verma2017effect,rostamzadeh2019large,goerbig2008tilted,MagnetoplasmonsPRB}. 
The Hamiltonian of graphene is recovered by additionally taking $v_t = 0$ and $v_x = v_y = v_F$.
Following 8-$Pmmn$ borophene as a reference we shall take $v_x=0.86\,v_F, v_y=0.69\,v_F$ with
$v_F= 10^6\,$m/s, and for the tilting velocity $v_t=0.32\,v_F$ \cite{zabolotskiy2016strain}.

The energy-momentum dispersion relation corresponding to the Hamiltonian in Eq.(\ref{Hborogap}) is
\begin{equation} \label{Bands1}
\varepsilon^{\xi}_{\lambda}(k_x,k_y)=\xi\alpha_tk_y+\lambda\sqrt{\alpha_x^2k_x^2+\alpha_y^2k_y^2
+(\Delta^{\xi})^2} \ ,
\end{equation}
where $\alpha_i=\hslash v_i\, (i=x,y,t,F)$, and the index $\lambda=\pm$ defines the energy branch and the helicity of the states in the conduction ($\lambda=+$) and the valence  ($\lambda=-$) bands in each valley.
 
 \begin{figure}[h]
    \centering
    \includegraphics[scale=0.4]{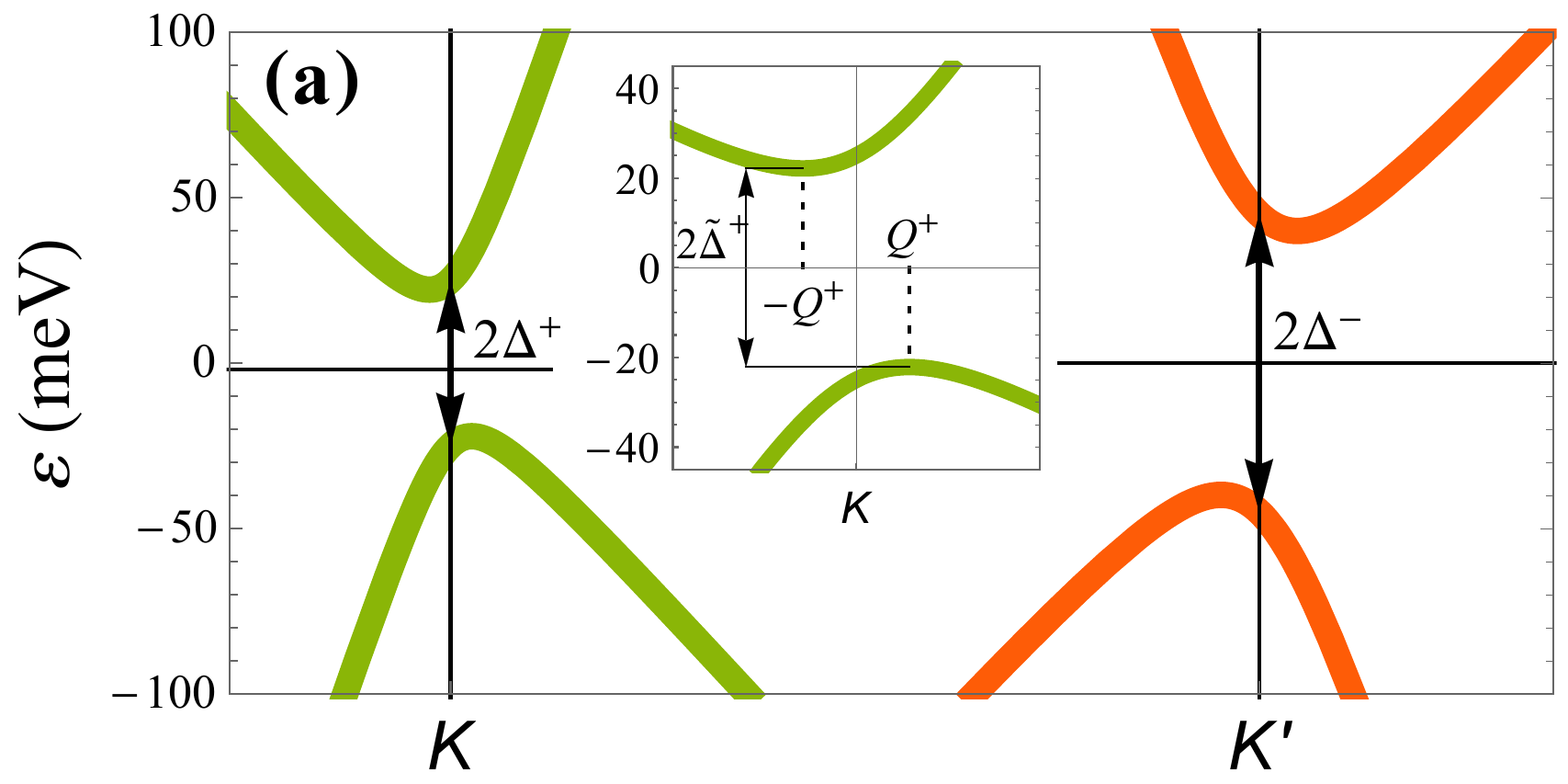}
     \includegraphics[scale=0.4]{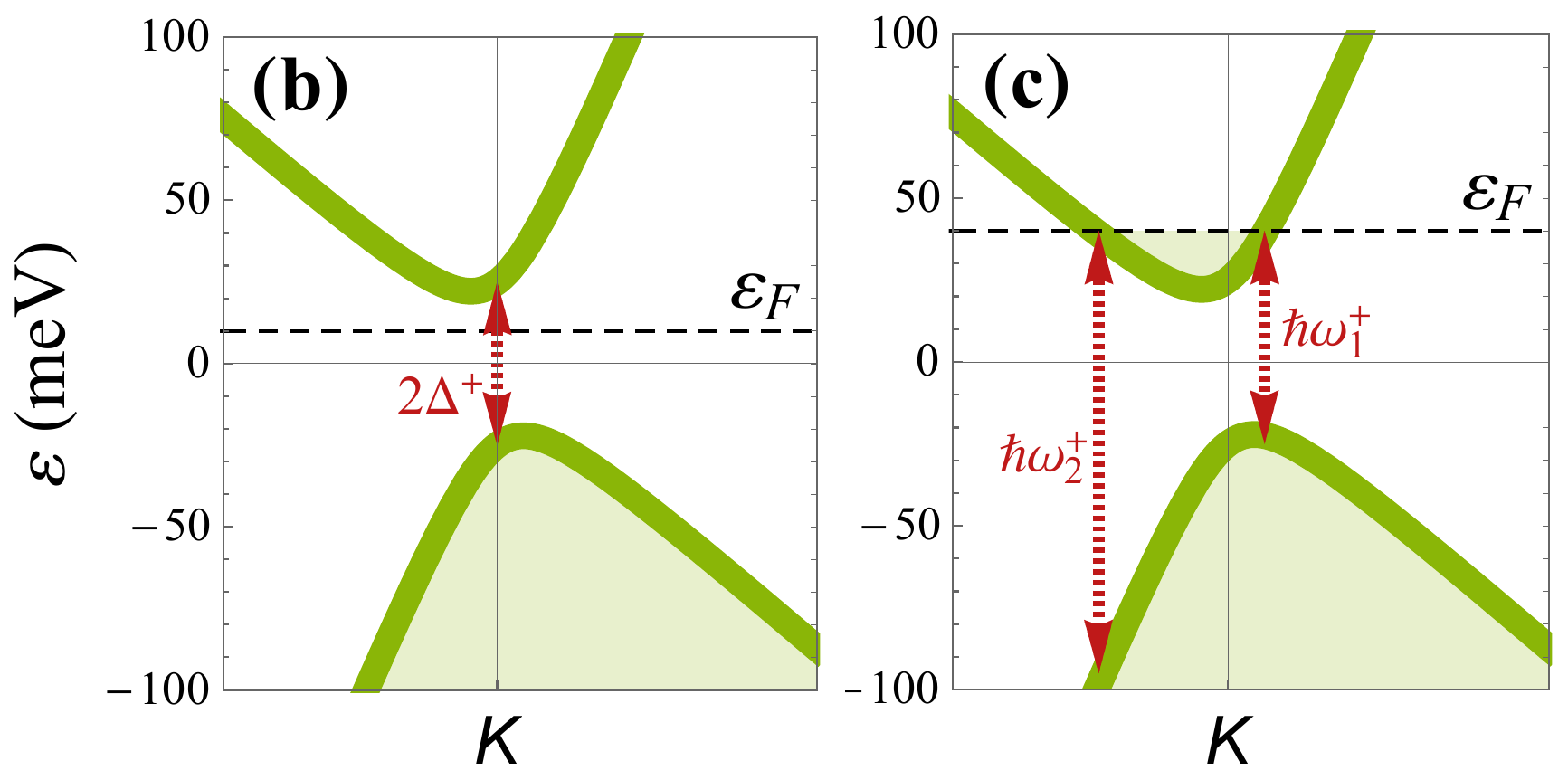}
    \caption{(a) 
    Energy spectrum of an anisotropic tilted Dirac system with valley symmetry breaking (Eq.(\ref{Bands1})). The smallest vertical energy differences $2\Delta^{\xi}$ for each valley are shown ($\Delta^+=25$ meV, $\Delta^-=45$ meV). The indirect nature of the gaps is illustrated in the inset ($K$ valley). The points ${\bf k}^*=(0,\mp Q^{+})$ correspond to the minimum and maximum of the branches. (b) Fermi level lying within a gap, $\varepsilon_F<\tilde{\Delta}^{+}$. (c) Fermi level lying above a gap, $\varepsilon_F>\Delta^{+}$ with allowed interband transitions. The energies $\hbar\omega_1^+$, $\hbar\omega_2^+$ give the minimum and maximum energies needed for interband transition.}
    \label{fig:spectrum}
\end{figure}
 
 The bands (\ref{Bands1}) have critical points at ${\bf k}^*=(0,-\xi\lambda Q^{\xi})$, where 
 $Q^{\xi}=(|\Delta^{\xi}|/\alpha_y)(\gamma/\sqrt{1-\gamma^2})$,
 yielding a minimum of $\varepsilon^{\xi}_+$ at $(0,-\xi Q^{\xi})$ and a
maximum of $\varepsilon^{\xi}_-$ at $(0,\xi Q^{\xi})$ (see the inset in Fig. \ref{fig:spectrum}(a)); we have introduced the tilting parameter
$\gamma=v_t/v_y\,$ ($0\leqslant\gamma<1$, non-overtilted cones).
Note then that $\text{min}\{\varepsilon^{\xi}_+({\bf k})\}-\text{max}\{\varepsilon^{\xi}_-({\bf k})\}=
\varepsilon^{\xi}_+(0,-\xi Q^{\xi})-\varepsilon^{\xi}_-(0,\xi Q^{\xi})=
2|\Delta^{\xi}|\sqrt{1-\gamma^2}$ is smaller than the smallest vertical energy difference
$\text{min}\{\varepsilon^{\xi}_+({\bf k})-\varepsilon^{\xi}_-({\bf k})\}=\varepsilon^{\xi}_+({\bf 0})-\varepsilon^{\xi}_-({\bf 0})
=2|\Delta^{\xi}|$. Thus, as a consequence of the simultaneous presence of tilting ($\gamma$) and mass ($\Delta^{\xi}$) there is an indirect band gap at each valley
around the nominal Dirac point,  see Fig. \ref{fig:spectrum}.
This implies for example that a Fermi level in the gap means now $|\varepsilon_F|<\text{min}\{|\tilde{\Delta}^+|, |\tilde{\Delta}^-|\}$, 
where $\tilde{\Delta}^{\xi}=\Delta^{\xi}\sqrt{1-\gamma^2}<\Delta^{\xi}$. 
Indeed, for each valley the following scenarios are now possible according to the position of the Fermi level: 
$(i) \,|\varepsilon_F|>|\Delta^{\xi}|$ (Fermi level above the nominal
direct gap),  $(ii)\, |\tilde{\Delta}^{\xi}|<|\varepsilon_F|<|\Delta^{\xi}|$ (Fermi level in an ``indirect gap'' region), 
$(iii)\, |\varepsilon_F|<|\tilde{\Delta}^{\xi}|$ (Fermi level in the absolute gap); see Fig. \ref{fig:spectrum}. This will cause additional structure in the optical response in contrast to the untilted case ($\gamma=0$). Now the Fermi contours, defined by the curves $C^{\xi}_{\lambda}=\{{\bf k}|\,\varepsilon^{\xi}_{\lambda}({\bf k})=\varepsilon_F\}$, are the displaced ellipses 
$(1-\gamma^2)\alpha_x^2(k_{\lambda,x}^{\xi})^2+(1-\gamma^2)^2\alpha_y^2\left(k_{\lambda,y}^{\xi}+\xi\lambda Q^{\xi}
|\varepsilon_F|/|\tilde{\Delta}^{\xi}|\right)^2=\varepsilon_F^2-(\tilde{\Delta}^{\xi})^2$ centered at 
$(0,-\xi\lambda Q^{\xi}|\varepsilon_F|/|\tilde{\Delta}^{\xi}|)$ with the major semiaxis along the
$k_y$-direction. Note that when the Fermi level lies in an indirect zone $|\tilde{\Delta}^{\xi}|<|\varepsilon_F|<|\Delta^{\xi}|$
the ellipse resides completely in the region $k_y<0 \,(k_y>0)$ for $\lambda=+ \,(\lambda=-)$ (see Fig.\,\ref{fig:contours}(d)) which modifies
significantly the momentum space available for interband transitions, with respect to that of the case 
$|\varepsilon_F|>|\Delta^{\xi}|$.
The roots  of equation 
$\varepsilon^{\xi}_{\lambda}({\bf k})=\varepsilon_F$ are displayed in Appendix \ref{FermiLines}.

\section{Optical transitions near an indirect gap}
\begin{figure}[h]
    \centering
    \includegraphics[scale=0.35]{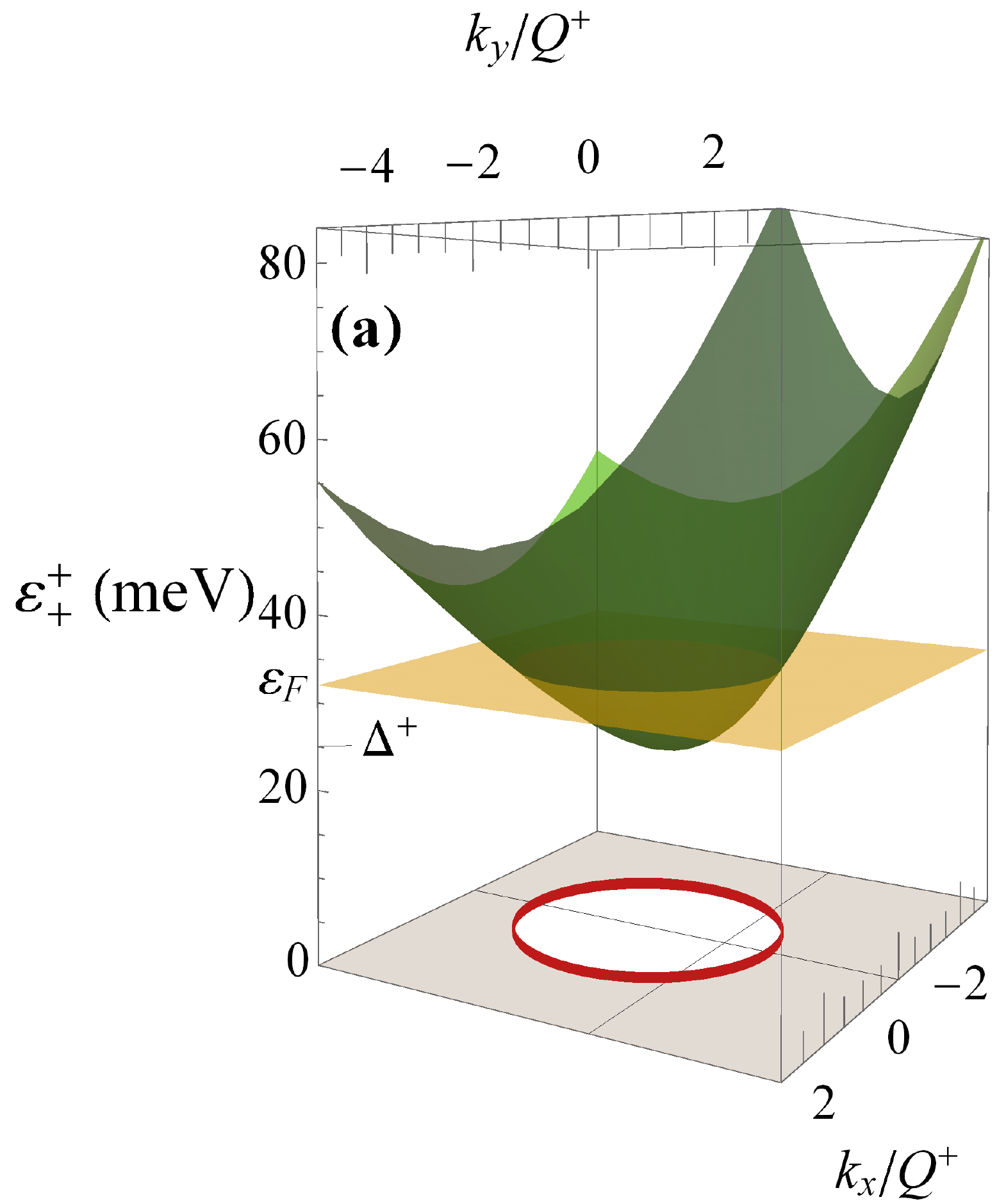}\hspace*{.7cm}
    \includegraphics[scale=0.28]{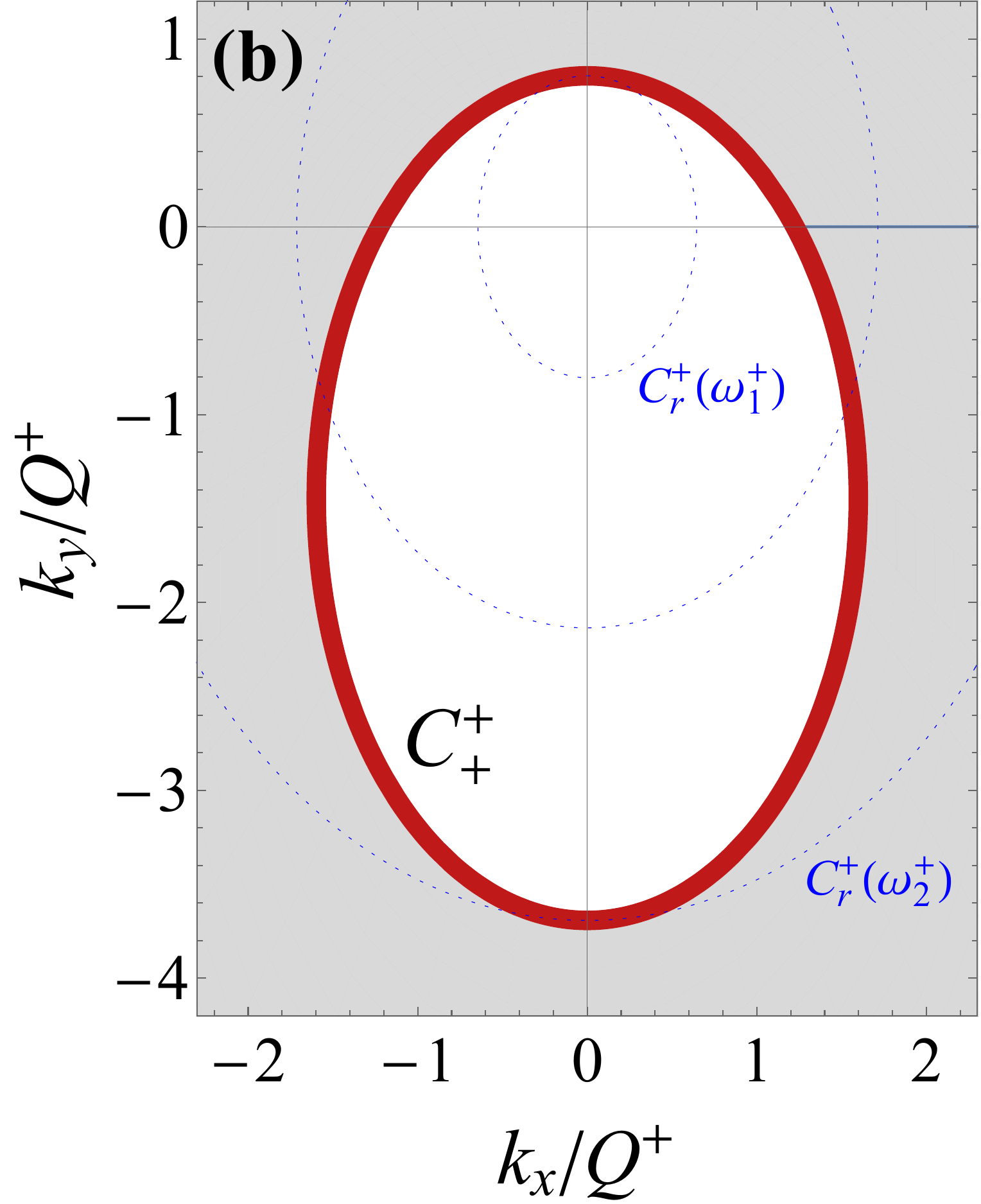}\\
    \includegraphics[scale=0.35]{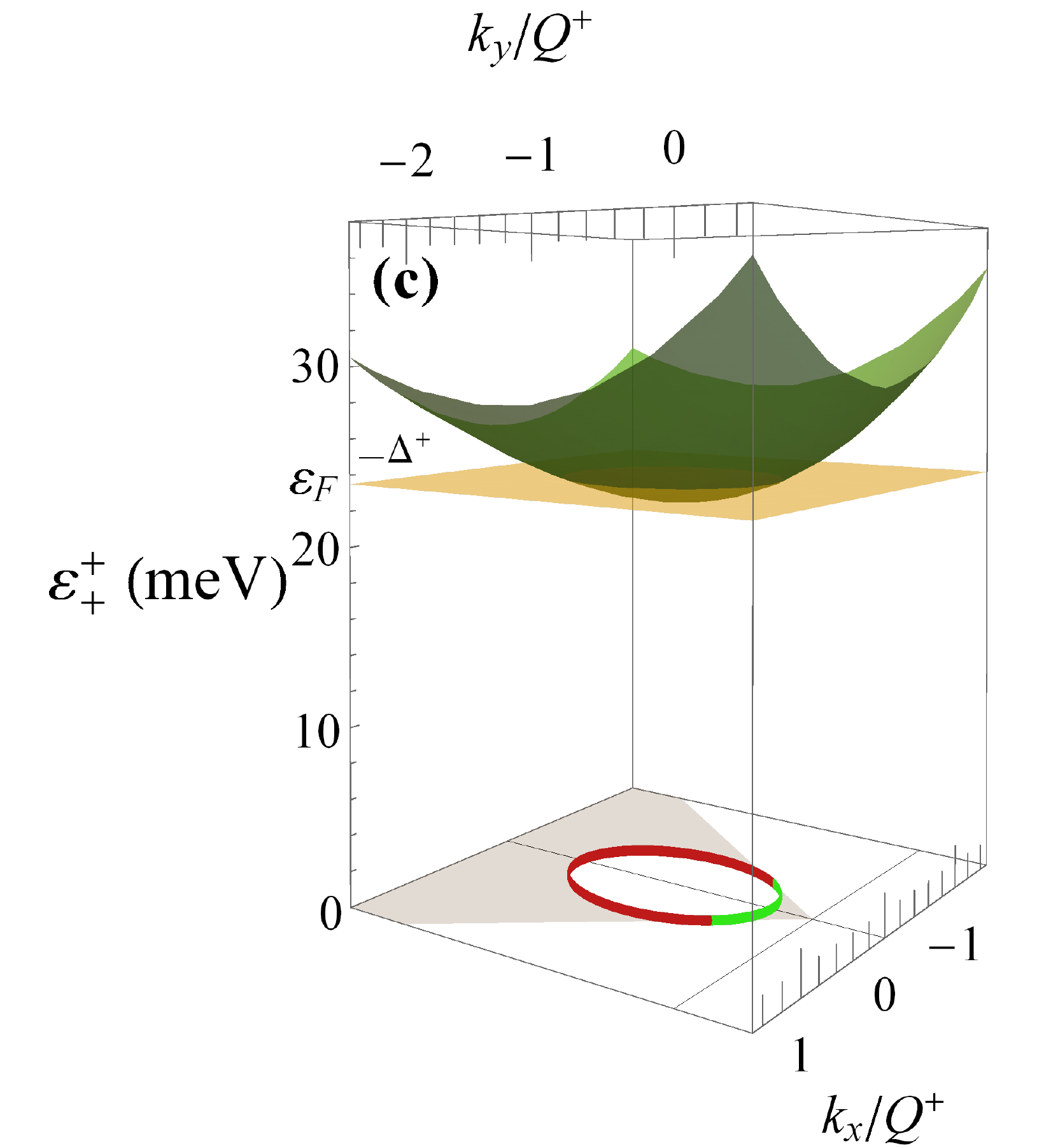}\hspace*{.7cm}
    \includegraphics[scale=0.28]{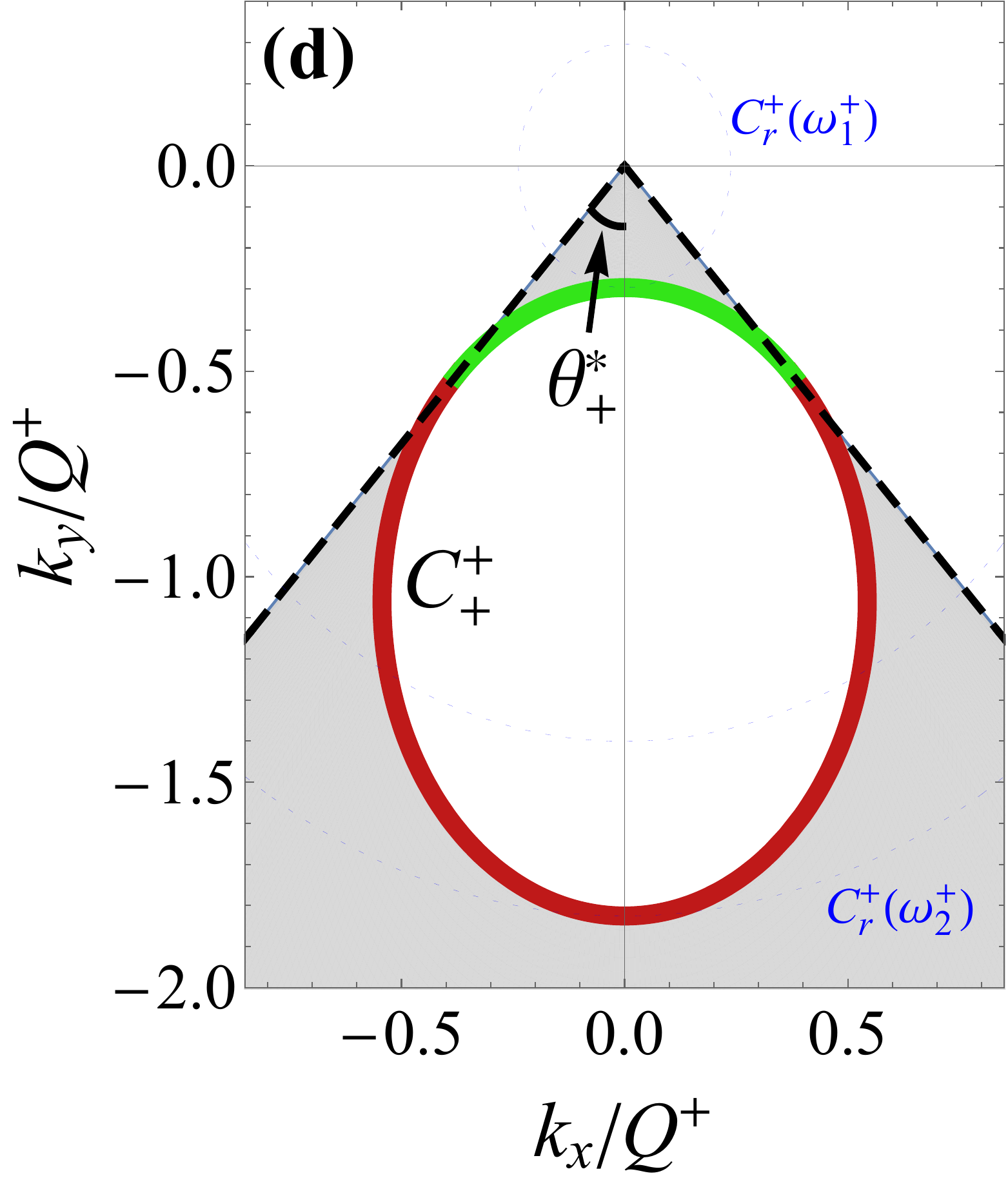}
    \caption{Conduction energy band $\varepsilon^+_+({\bf k})$ at the $K$ valley. The gap parameter is $\Delta^+=25\,$meV.
    At the Fermi level, the contour line $C_+^+=\{{\bf k}|\varepsilon_+^+({\bf k})=\varepsilon_F\}$ is an ellipse
    displaced along the $k_y$-axis by $Q^+(\varepsilon_F/\tilde{\Delta}^+)$, with $\tilde{\Delta}^+=\Delta^+\sqrt{1-\gamma^2}$. (a) $\varepsilon_F>\Delta^+$, with $\varepsilon_F=32\,$meV.
    (b) Fermi contour $k_{+,F}^{+}(\theta)$ displayed in (a).
    The gray shaded region indicates the momentum space 
    available for optical transitions, such that $\varepsilon_-^+({\bf k})<\varepsilon_F<\varepsilon_+^+({\bf k})$. (c) Fermi level in the indirect zone $\tilde{\Delta}^+<\varepsilon_F<\Delta^+$,
    with $\varepsilon_F=23.5\,$meV. (d) Fermi contour showed in (c). It is defined by the two arcs
    $q_{+,F}^{+,-}(\theta)$ (green) and $q_{+,F}^{+,+}(\theta)$ (red) in the half-space $k_y<0$, within the sector $|\theta-3\pi/2|<\theta^*_+$. As a consequence, the corresponding ${\bf k}$-region for interband transitions is strongly reduced. In (b) and (d) the resonance curve 
    ${\cal C}_r^+(\omega)=\{{\bf k}|\varepsilon_+^+({\bf k})-\varepsilon_-^+({\bf k})=\hslash\omega\}$ is presented for several frequencies (dotted lines). It is an ellipse centered at the origin, defined for $\hslash\omega>2\Delta^+$. At the critical frequencies $\omega_1^+,\omega_2^+$ (see Fig.\,\ref{fig:spectrum}(b)) ${\cal C}_r^+$ is tangential to the Fermi lines. At a given frequency, only the arcs of ${\cal C}_r^+$ lying in the shaded region contribute to the optical response.}
    \label{fig:contours}
\end{figure}

As a previous step to the calculation of the optical conductivity tensor and to understand the spectral features of the optical response of the system, we first consider the joint density of states (JDOS).
In the following we adopt the generic notation of a two-band model
and write the Hamiltonian and its spectrum as $H^{\xi}({\bf k})=\varepsilon^{\xi}_0({\bf k})\mathbb{I}+\boldsymbol{\sigma}\cdot {\bf d}^{\xi}({\bf k})$ and
$\varepsilon^{\xi}_{\lambda}({\bf k})=\varepsilon^{\xi}_0({\bf k}) +\lambda d^{\xi}({\bf k})$, where
$\varepsilon_0^{\xi}({\bf k})=\xi\alpha_tk_y$, $\,{\bf d}^{\xi}({\bf k})=\xi\alpha_xk_x{\bf\hat{x}}+
\alpha_yk_y{\bf\hat{y}}+\Delta^{\xi}{\bf\hat{z}}$, and $d^{\xi}({\bf k})=|{\bf d}^{\xi}({\bf k})|$. In polar coordinates we write
${\bf k}=k\cos\theta{\bf\hat{x}}+k\sin\theta{\bf\hat{y}}$ and
\begin{equation}
\varepsilon^{\xi}_{\lambda}(k,\theta)=\xi\alpha_Fkh(\theta)+\lambda\sqrt{\alpha_F^2k^2g^2(\theta)+(\Delta^{\xi})^2} \ ,
\end{equation}
in terms of the adimensional quantities $h(\theta)=(\alpha_t/\alpha_F)\sin\theta$, which characterize the tilting dependence,
and $g^2(\theta)=(\alpha_x^2/\alpha_F^2)\cos^2\theta+(\alpha_y^2/\alpha_F^2)\sin^2\theta$ which accounts for the anisotropic dispersion.
The roots of equation $\varepsilon^{\xi}_{\lambda}(k(\theta),\theta)=\varepsilon_F$ are denoted by $k_{\lambda,F}^{\xi}(\theta)$ when
$|\varepsilon_F|>|\Delta^{\xi}|$ (Fig.\,\ref{fig:contours}(b)), and by $q_{\lambda,F}^{\xi,\pm}(\theta)$ if $|\tilde{\Delta}^{\xi}|<|\varepsilon_F|<|\Delta^{\xi}|$
(Fig.\,\ref{fig:contours}(d)); 
see Appendix \ref{FermiLines} for expressions of these roots.
In the later case, the $\pm$ sign refers to the two arcs [in green (red) for the sign $-(+)$ in Fig.\,\ref{fig:contours}(d)] forming the ellipse 
which lies completely in the $k_y<0$ ($C_+^{+}$, $C_-^{-}$) or $k_y>0$ ($C_-^{+}$, $C_+^{-}$) half space;
the roots $q_{\lambda,F}^{\xi,\pm}(\theta)$ are defined only in the sector $|\theta-3\pi/2|<\theta^*_{\xi}$ if $\xi\lambda=+$, or
for $|\theta-\pi/2|<\theta^*_{\xi}$ if $\xi\lambda=-$, where 
$\tan\theta^*_{\xi}=(v_y/v_x)\sqrt{[\varepsilon^2_F-(\tilde{\Delta}^{\xi})^2]/[(\Delta^{\xi})^2-\varepsilon_F^2]}$.

\subsection{Joint density of states}

The number of pair of states in conduction (unoccupied) and valence (occupied) bands separated by a given energy $\hbar\omega$ is given by
$\mathcal{J}(\omega)=\sum_{\xi=\pm}\mathcal{J}^{(\xi)}(\omega)$ with
\begin{equation} \label{JDOS1}
\mathcal{J}^{(\xi)}(\omega)=g_s\int'\!\!\frac{d^2k}{(2\pi)^2}\,\delta(\varepsilon^{\xi}_+({\bf k})-\varepsilon^{\xi}_-({\bf k})-\hslash\omega),
\end{equation}
where $g_s=2$ is the spin degeneracy and the prime on the integral indicates a range of integration 
restricted to that region of ${\bf k}$-space for which
$\varepsilon^{\xi}_-({\bf k})<\varepsilon_F<\varepsilon^{\xi}_+({\bf k})$.
On the other hand, the $\delta$ function restricts the integration to points lying on the resonance curve 
${\cal C}^{\xi}_r(\omega)=\{(k_x,k_y)|\,2d^{\xi}(k_x,k_y)=\hslash\omega\}$. Thus, the integral (\ref{JDOS1}) 
has to be carried out over those portions of the curve ${\cal C}^{\xi}_r(\omega)$ lying within the ${\bf k}$-region for
which the previous inequality (Pauli blocking) is satisfied.
The curve ${\cal C}^{\xi}_r(\omega)$ is the ellipse centered at the origin 
$\alpha_x^2k_{r,x}^2+\alpha_y^2k_{r,y}^2=(\hslash\omega/2)^2-(\Delta^{\xi})^2$, defined only for 
$\hslash\omega\geqslant 2|\Delta^{\xi}|$,
or in polar coordinates $2\alpha_Fk_r(\theta)g(\theta)=\sqrt{(\hslash\omega)^2-(2\Delta^{\xi})^2}$ (Fig.\,\ref{fig:contours}(b),(d)).
Indeed, for a given frequency the expression (\ref{JDOS1}) can be written as a line integral of $(\hslash v_{\bf k})^{-1}=|\nabla_{\bf k}[2d^{\xi}({\bf k})]|^{-1}$
over those portions of the curve of constant interband energy ${\cal C}^{\xi}_r(\omega)$ lying in the regions imposed by Pauli blocking.
Peaks in the JDOS will appear due to electronic excitations involving states with allowed wave vectors on  ${\cal C}^{\xi}_r(\omega)$  
such that $v_{\bf k}$ takes extreme values. 
The energy difference $2d^{\xi}(k,\theta)$ between the conduction and valence bands at the Fermi lines $k_{\lambda,F}^{\xi}(\theta)$
and $q_{\lambda,F}^{\xi,\pm}(\theta)$ will be denoted by $\hslash\omega^{\xi}_{\lambda}(\theta)$ and $\hslash\nu^{\xi,\pm}_{\lambda}(\theta)$, respectively
(see Appendix\,\ref{FermiLines}).
For $\gamma=0$ these energies reduce to the same value $2|\varepsilon_F|$, which is the threshold (above the gap) for interband transitions in gapped graphene.
Given the tilt of the bands around each valley, it is verified that $\hslash\omega^{\xi}_{-}(\theta)=\hslash\omega^{\xi}_{+}(\theta\pm\pi)$ and
$\hslash\nu^{\xi,\pm}_{-}(\theta)=\hslash\nu^{\xi,\pm}_{+}(\theta\pm\pi)$. The minimum and the maximum of these energy differences 
take place at $\theta=\pi/2$ or $3\pi/2$, and they are all given by the same functions of the Fermi level,
\begin{eqnarray}
\hslash\omega_1^{\xi} (\varepsilon_F) &=&  \frac{2}{1-\gamma^2}\,\left(|\varepsilon_F|-
\gamma\sqrt{\varepsilon_F^2-(\tilde{\Delta}^{\xi})^2}\right), \label{omega1} \\
\hslash\omega_2^{\xi}(\varepsilon_F) &=&  \frac{2}{1-\gamma^2}\,\left(|\varepsilon_F|+
\gamma\sqrt{\varepsilon_F^2-(\tilde{\Delta}^{\xi})^2}\right) \ ,\label{omega2}
\end{eqnarray}
such that $\text{min}_{\theta}\{\hslash\omega^{\xi}_{\pm}(\theta)\}=\text{min}_{\theta}\{\hslash\nu_{\pm}^{\xi,-}(\theta)\}=\hbar\omega_1^{\xi}<2|\varepsilon_F|$
and $\text{max}_{\theta}\{\hslash\omega^{\xi}_{\pm}(\theta)\}=\text{max}_{\theta}\{\hslash\nu_{\pm}^{\xi,+}(\theta)\}=
\hbar\omega_2^{\xi}>2|\varepsilon_F|$.
As was mentioned above, $\hslash\omega_1^{\xi}(\gamma=0)=\hslash\omega_2^{\xi}(\gamma=0)=2|\varepsilon_F|$. 
Equations \eqref{omega1} and \eqref{omega2} suggests optical measurements of $\omega_{1/2}^{\xi}$ to determine the tilting and gap parameters. Indeed, we can recover them through the
expressions $\gamma=\sqrt{1-(4|\varepsilon_F|/\varepsilon_s)}$
and $[\Delta^{\xi}(\omega_1^{\xi},\omega_2^{\xi})]^2=
|\varepsilon_F|[\hslash\omega_1^{\xi}\hslash\omega_2^{\xi}
-\varepsilon_s|\varepsilon_F|]/(\varepsilon_s-4|
\varepsilon_F|)$, where $\varepsilon_s=\hslash\omega_1^++\hslash\omega_2^+=
\hslash\omega_1^-+\hslash\omega_2^-$.

\begin{figure}
    \centering
    \includegraphics[scale=0.31]{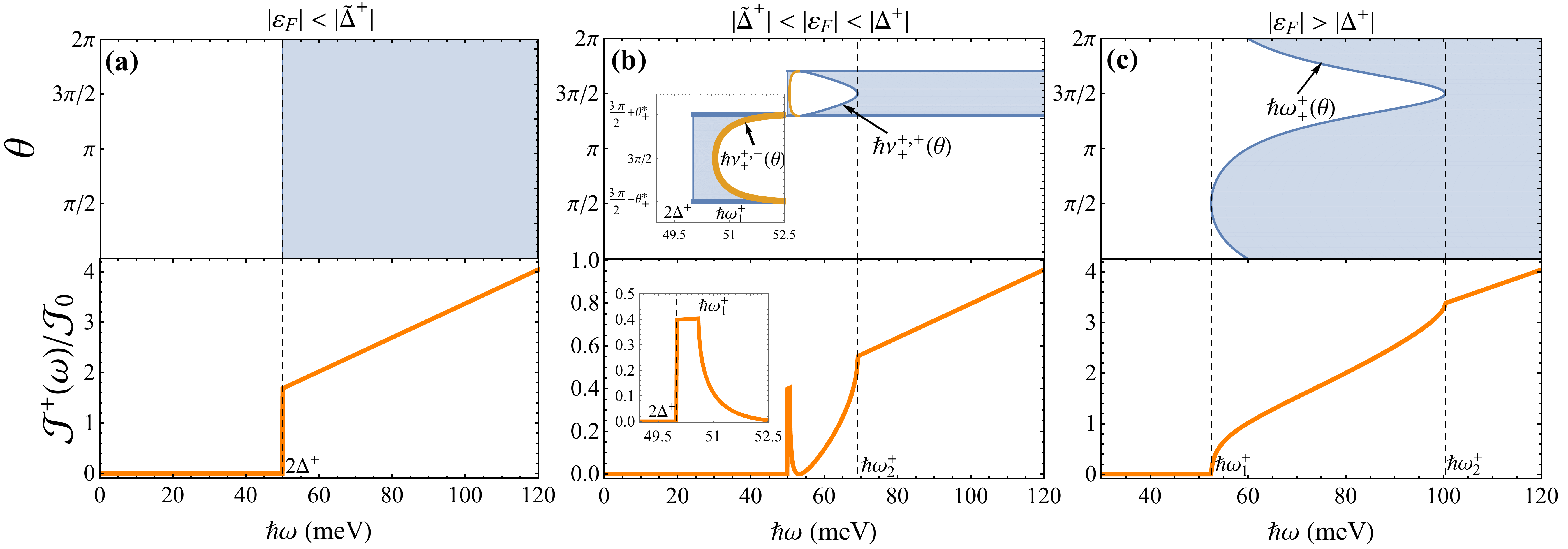}
    \caption{Joint density of states for transitions
    at the $K$ valley: (a) $|\varepsilon_F|<|\tilde{\Delta}^+|$, (b) $|\tilde{\Delta}^+|<|\varepsilon_F|<|\Delta^+|$, and (c)  $|\varepsilon_F|>|\Delta^+|$. We take $\Delta^+=25$ meV and $\mathcal{J}_0=(g_s/8\pi)(2|\Delta^+|/\alpha_F^2)$.}
    \label{fig:my_labelJ}
\end{figure}

The integral (\ref{JDOS1}) looks different according to the position of the Fermi level:

$(i)\ |\varepsilon_F|>|\Delta^{\xi}|$:

In this case Pauli blocking and energy conservation restricts to $k_r(\theta,\omega)\geqslant k_{\lambda,F}^{\xi}(\theta)$, which leads to the result
\begin{equation}\label{JDOS2}
\mathcal{J}^{(\xi)}(\omega)=g_s\,\frac{\hslash\omega}{8\pi\alpha_F^2}\,\frac{1}{2\pi}\int_0^{2\pi}\!\!\frac{d\theta}{g^2(\theta)}\,
\Theta[\omega-\omega^{\xi}_{\lambda}(\theta)] \ ,
\end{equation}
where the sign $\lambda=+$ ($-$) is used when $\varepsilon_F>0 \,(<0)$; $\Theta(x)$ is the Heaviside unit step function. 
In Fig.\,\ref{fig:my_labelJ}(c) the JDOS $\mathcal{J}^{(+)}(\omega)$ as
calculated from Eq. (\ref{JDOS2}) is shown. For photon energies $\hslash\omega_1^{\xi}<\hslash\omega<\hslash\omega^{\xi}_2$, the angular region in momentum space
available for vertical transitions is no longer $0\leqslant\theta\leqslant 2\pi$ as in the untilted case, but a reduced region with a boundary
determined by $\hslash\omega^{+}_{+}(\theta)$ (top panel). This is
in contrast to the well known onset $\Theta(\hslash\omega-2\varepsilon_F)$ for interband transitions
between bands with electron-hole symmetry. The characteristic and unique threshold $2\varepsilon_F$ observed in 
the optical response of graphene\cite{CarbottePRL96} becomes
 a region bounded by the critical energies $\hslash\omega^+_1$ and $\hslash\omega^+_2$, where the frequency dependence is no longer lineal (bottom panel), because of the tilt of the bands.
The JDOS vanishes for $\hslash\omega<\hslash\omega_1^{\xi}$.
When $\hslash\omega>\hslash\omega_2^{\xi}$ the whole angular region
$0\leqslant\theta\leqslant 2\pi$ contributes, giving the result $\mathcal{J}^{(\xi)}(\omega>\omega_2^{\xi})=(g_s/8\pi)(\hslash\omega/\alpha_x\alpha_y)$, which is
independent of the tilting parameter $\alpha_t$ and shows the
usual linear $\omega$-dependence of Dirac systems.
Globally, Fig.\,\ref{fig:my_labelJ}(c) displays qualitatively a similar
behavior as that reported by Verma et al. for borophene.
\cite{verma2017effect}

$(ii)\  |\tilde{\Delta}^{\xi}|<|\varepsilon_F|<|\Delta^{\xi}|$:

Now the momentum space available for direct transitions is restricted to 
$k_r(\theta,\omega)\geqslant q_{\lambda,F}^{\xi,+}(\theta)$ or $0\leqslant k_r(\theta,\omega)\leqslant q_{\lambda,F}^{\xi,-}(\theta)$ (Fig.\,\ref{fig:contours}(d)) and the JDOS for the valley
$\xi$ reads as
\begin{equation} \label{JDOS3}
\mathcal{J}^{(\xi)}(\omega) = g_s\,\frac{\hslash\omega}{8\pi\alpha_F^2}\frac{1}{2\pi}\Theta(\hslash\omega-2|\Delta^{\xi}|)
 \int_{\theta_0-\theta_{\xi}^*}^{\theta_0+\theta_{\xi}^*}\!\!\frac{d\theta}{g^2(\theta)}
\left\{\Theta[\nu_{\lambda}^{\xi,-}(\theta)-\omega]+\Theta[\omega-\nu_{\lambda}^{\xi,+}(\theta)] \right\}, 
\end{equation}
where $\theta_0=3\pi/2 \, (\pi/2)$ when $\xi\lambda=+\,(-)$. We also find that
$\mathcal{J}^{(\xi)}(2|\Delta^{\xi}|\leqslant\hslash\omega\leqslant\hslash\omega_1^{\xi})=\mathcal{J}^{(\xi)}(\hslash\omega>\hslash\omega_2^{\xi})=(g_s/8\pi)(\hslash\omega/\alpha_x\alpha_y)[1-(2\beta_{\xi}/\pi)]/2$, where $\tan\beta_{\xi}=\sqrt{[(\Delta^{\xi})^2-\varepsilon_F^2]/[\varepsilon_F^2-(\tilde{\Delta}^{\xi})^2]}$.
The JDOS (\ref{JDOS3}) for the valley $\xi=+$ is shown in Fig.\,\ref{fig:my_labelJ}(b). The spectrum displays  van Hove singularities at 
$2\Delta^+, \hslash\omega_1^+$, and $\hslash\omega_2^+$, and  a reduced overall size in comparison to the cases  
$|\varepsilon_F|<|\tilde{\Delta}^{\xi}|$ or $|\varepsilon_F|>|\tilde{\Delta}^{\xi}|$. 
Now a linear behavior as a function of photon energy $\hslash\omega$ appears, with a 
lower slope, in two separated domains only. Moreover, the number of interband transitions is strongly diminished
between $\hslash\omega_1^+$ and $\hslash\omega_2^+$ because the angular space available for transitions is considerably smaller, as
is illustrated in the top panel of Fig.\,\ref{fig:my_labelJ}(b). 
The insets show how the contributing angular region narrows for $\omega_1^+<\omega<\nu^{+,-}_+(\theta)$
or $\nu^{+,+}_+(\theta)<\omega<\omega_2^+$, while the whole sector $|\theta-3\pi/2|<\theta^*_+$ contributes when
$2\Delta^+\leqslant\hslash\omega\leqslant\hslash\omega_1^+$ or $\hslash\omega>\hslash\omega_2^+$.
The appearance of three critical energies instead of one (for $|\varepsilon_F|<|\tilde{\Delta}^{\xi}|$)
or two (for $|\varepsilon_F|>|\Delta^{\xi}|$) constitutes an optical signature of the indirect gap.

$(iii)\, |\varepsilon_F|<|\tilde{\Delta}^{\xi}|$:

For the Fermi level within the gap the JDOS becomes \begin{equation} \label{JDOS4}
    \mathcal{J}^{(\xi)}(\omega)=(g_s/8\pi)(\hslash\omega/\alpha_x\alpha_y)
\Theta(\hslash\omega-2|\Delta^{\xi}|) \, .
\end{equation}
Besides the reduction of the absolute gap ($|\tilde{\Delta}^{\xi}|<|\Delta^{\xi}|$),
we note that this result is independent of the tilting parameter.
The JDOS looks very similar to that corresponding to gapped graphene but now it
involves the geometric mean $\sqrt{v_xv_y}$, instead of velocity $v_F$, due to the anisotropy of the energy dispersion;
see Fig.\,\ref{fig:my_labelJ}(a).

\begin{figure}[h]
    \centering
    \includegraphics[scale=0.298]{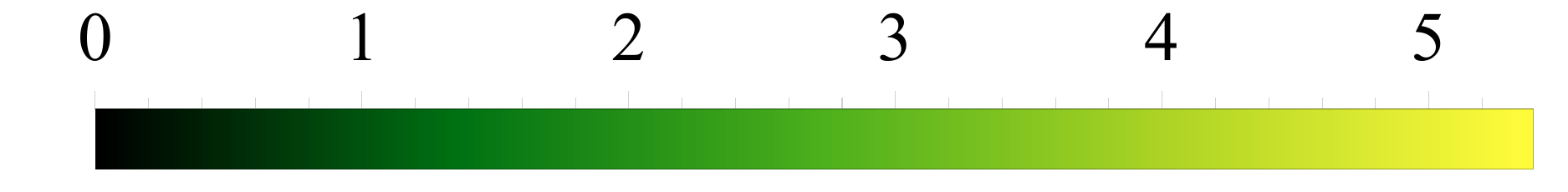}
    \includegraphics[scale=0.298]{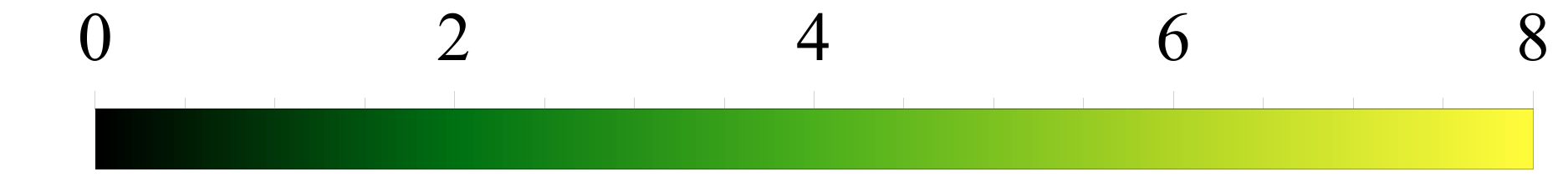}
    \includegraphics[scale=0.298]{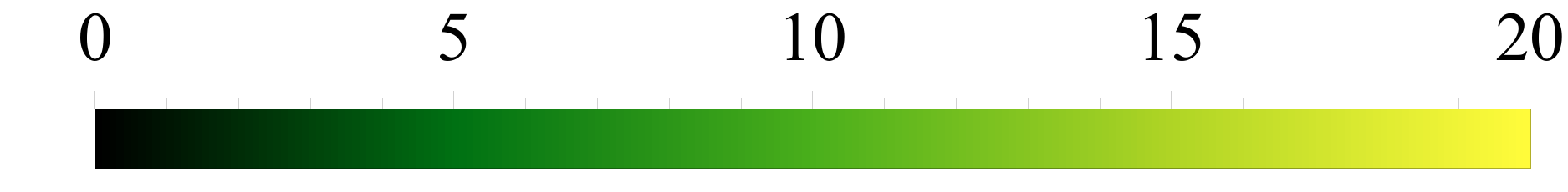}
    \includegraphics[scale=0.315]{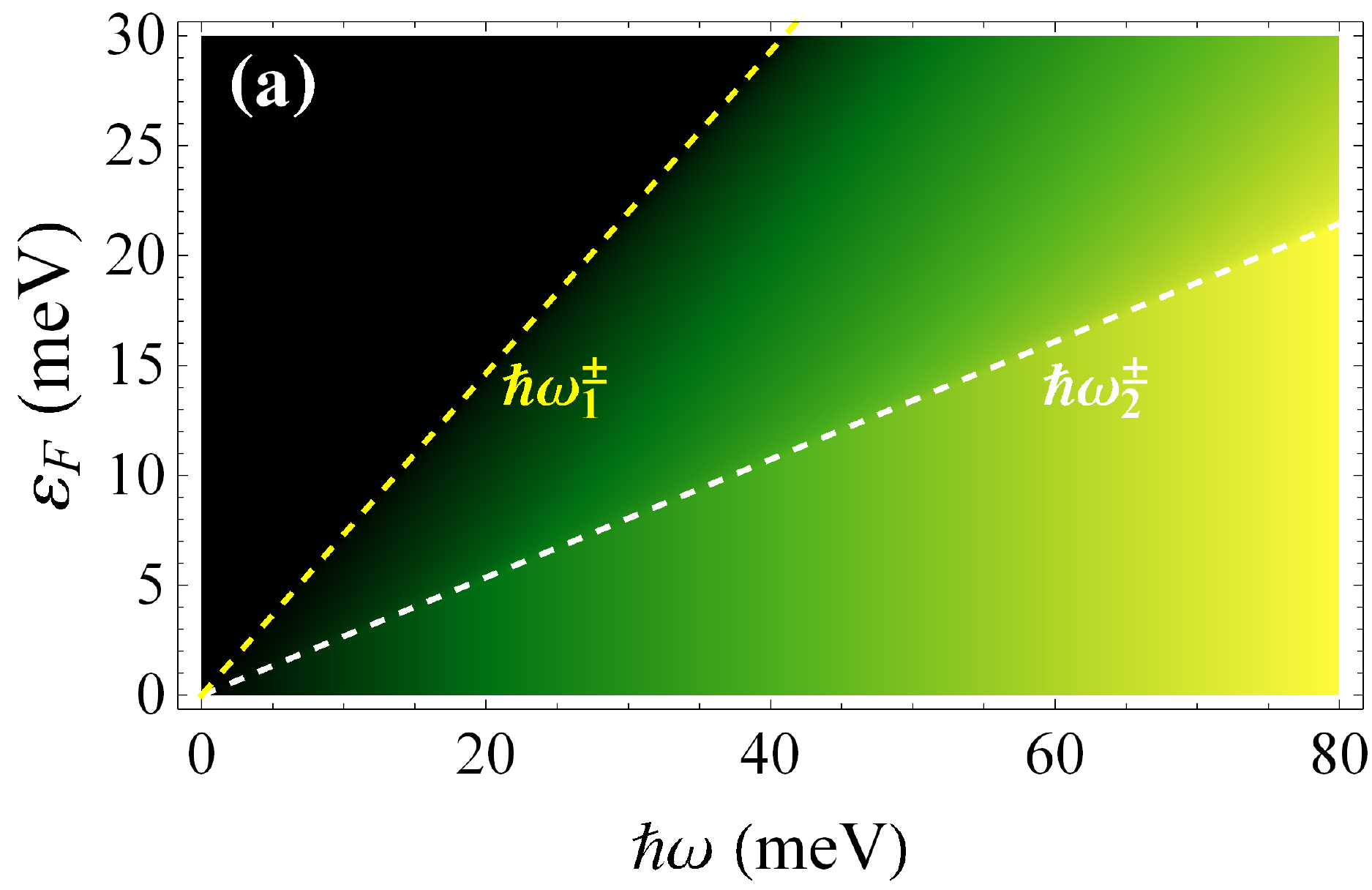}
    \includegraphics[scale=0.30]{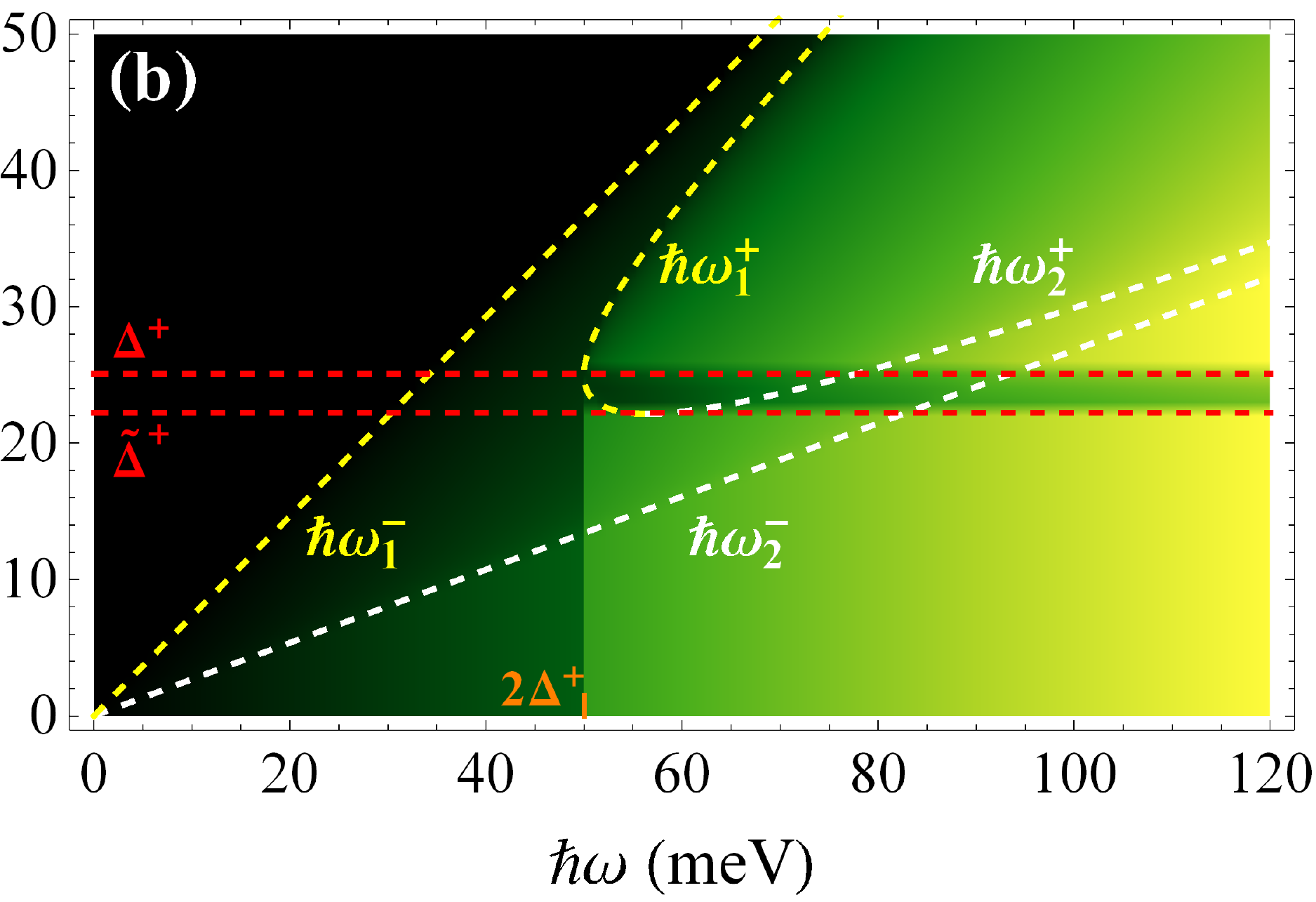}
    \includegraphics[scale=0.30]{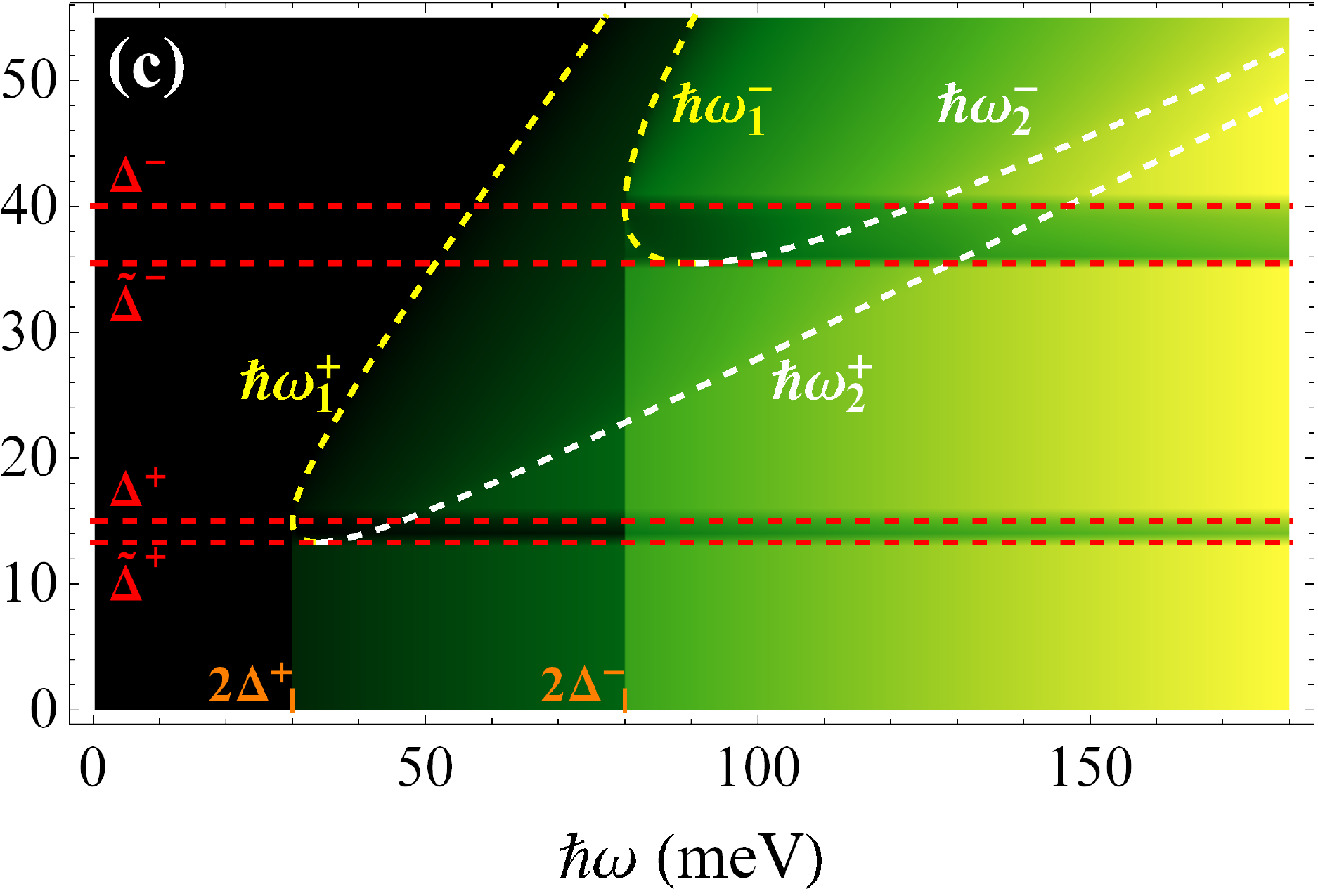}
    \caption{Total joint density of states $\mathcal{J}(\omega;\varepsilon_F)/\mathcal{J}_0$ considering; (a) $\Delta^+=\Delta^-=0$, (b) $\Delta^+=25$ meV and $\Delta^-=0$, and (c) $\Delta^+=15$ meV and $\Delta^-=40$ meV. In (b) and (c) we take $\mathcal{J}_0$ as in Fig.\,\ref{fig:my_labelJ}. In (a) we use $\Delta^+=25$ meV in $\mathcal{J}_0$}.
    \label{fig:my_labelJ2}
\end{figure}

All these results are used to evaluate
the frequency dependence of the total JDOS $\sum_{\xi}\mathcal{J}^{(\xi)}(\omega;\varepsilon_F)$ for
a continuous variation of the Fermi level.  
Figure\,\ref{fig:my_labelJ2} illustrates well the appearance of onsets and how the lines defined by the critical frequencies evolve and shape
the spectrum of JDOS when a gap is open.
In Fig.\,\ref{fig:my_labelJ2}(a) we show the result for borophene ($\Delta^+=\Delta^-=0$),
where the corresponding critical energies $\hslash\omega_{1,2}(\varepsilon_F)$ are also indicated.
The case with only one gap, $\Delta^+\neq 0, \Delta^-=0$, is shown in Fig.\,\ref{fig:my_labelJ2}(b).
The contribution $\mathcal{J}^{(-)}$ behaves as for borophene, with critical frequencies $\hslash\omega^-_{1,2}= 2\varepsilon_F/(1\pm\gamma)$.
On the other hand, for transitions near the gapped valley, the spectrum of the contribution $\mathcal{J}^{(+)}$ displays the threshold at
$2\Delta^+$ (Eq.\,(\ref{JDOS4})), the borders defined by critical frequencies $\hslash\omega^+_{1,2}$ with a nonlinear dependence on $\varepsilon_F$
(Eqs.(\ref{omega1}) and (\ref{omega2})), and an indirect zone where the joint density is strongly suppressed. 
Figure\,\ref{fig:my_labelJ2}(c) displays the case with valley-contrasting gaps $|\Delta^-|>|\Delta^+|$.

\subsection{Optical conductivity tensor}

Within the framework of the linear response theory, we find that the conductivity tensor
$\sigma_{ij}(\omega)=\sum_{\xi}\sigma_{ij}^{(\xi)}(\omega)$, which determines the 
electrical current induced in  the system  by an external homogeneous electric field of frequency
$\omega$, has the form
\begin{eqnarray}
\text{Re}\,\sigma_{ii}^{(\xi)}(\omega) &= & D^{(\xi)}_{ii}\delta(\omega)+\text{Re}\,\sigma^{(\xi),inter}_{ii}(\omega) \label{sii} \\
\text{Im}\,\sigma_{ii}^{(\xi)}(\omega) &=& \sigma^{(\xi),intra}_{ii}(\omega)+\text{Im}\,\sigma^{(\xi),inter}_{ii}(\omega) \\
\sigma_{xy}^{(\xi)}(\omega)  &=& -\sigma_{yx}^{(\xi)}(\omega) = \sigma^{(\xi),inter}_{xy}(\omega) \ ,
\end{eqnarray}
where the label intra (inter) refers to contributions due to intraband (interband) transitions. 
 According to the Kubo formula, these are  obtained from (at zero temperature)
\begin{eqnarray}
\sigma^{(\xi),intra}_{ii}(\omega) &=&ig_s\frac{\sigma_0}{4\pi\hslash\omega}\sum_{\lambda}\int\!\!d^2k\left[V_{\lambda,i}^{\xi}({\bf k})\right]^2
\delta(\varepsilon^{\xi}_{\lambda}({\bf k})-\varepsilon_F) \\
\mbox{Re}\,\sigma_{ij}^{(\xi),inter}(\omega)  &=&  g_s\frac{\sigma_0}{4\hslash\omega} \int'\!\!d^{2}k\,
\frac{S^{\xi}_{ij}({\bf k})}{(d^{\xi}({\bf k}))^2}\,
\delta(\hslash\omega-2d^{\xi}({\bf k})) \label{Resig_sp} \\
\mbox{Im}\,\sigma_{ij}^{(\xi),inter}(\omega) &=&
g_s \frac{\sigma_0}{4\pi\hslash\omega}\,\mathscr{P}\!\!\int'\!\!d^{2}k\,\frac{S^{\xi}_{ij}({\bf k})}{(d^{\xi}({\bf k}))^3}\,
\frac{(\hslash\omega)^2}{(\hslash\omega)^2-(2d^{\xi}({\bf k}))^2} \label{dispersive1}\\
\mbox{Re}\,\sigma_{xy}^{(\xi),inter}(\omega)  &=& g_s \frac{\sigma_0}{2\pi} \,\mathscr{P}\!\!\int'\!\!d^{2}k\,
\frac{T^{\xi}_{xy}({\bf k})}{d^{\xi}({\bf k})} \,
\frac{1}{(\hslash\omega)^2-(2d^{\xi}({\bf k}))^2} 
\label{dispersive2} \\
\nonumber \\
\mbox{Im}\,\sigma_{xy}^{(\xi),inter}(\omega) &=&  -g_s\frac{\sigma_0}{8}\int'\!\!d^{2}k\,
\frac{T^{\xi}_{xy}({\bf k})}{(d^{\xi}({\bf k}))^2} \,
\delta(\hslash\omega-2d^{\xi}({\bf k})) 
\label{Imsig_T} \ ,
\end{eqnarray}
where $\sigma_0=2e^2/h$, $\mathscr{P}$ means Principal Value integral, 
$V_{\lambda,i}^{\xi}({\bf k})=\frac{\partial\varepsilon^{\xi}_0({\bf k})}{\partial k_i}+\frac{\lambda}{d^{\xi}}\,
{\bf d}^{\xi}\cdot\frac{\partial{\bf d}^{\xi}}{\partial k_i}$, 
$\,S^{\xi}_{ij}({\bf k})=\left({\bf d}^{\xi}\times\frac{\partial {\bf d}^{\xi}}{\partial k_i}\right)\cdot\left({\bf d}^{\xi}\times
\frac{\partial {\bf d}^{\xi}}{\partial k_j}\right)$, and $T^{\xi}_{xy}({\bf k})={\bf d}^{\xi}\cdot
\left(\frac{\partial {\bf d}^{\xi}}{\partial k_x}\times\frac{\partial {\bf d}^{\xi}}{\partial k_y}\right)$.
The prime indicates integration over domains which depend on
the position of the Fermi energy according to the condition $\varepsilon^{\xi}_-({\bf k})<\varepsilon_F<\varepsilon^{\xi}_+({\bf k})$. 
We have included in (\ref{sii}) the Drude weight\cite{Stauber_2013}
$D^{(\xi)}_{ii}=\pi\lim_{\omega\to 0}[\omega\,\mbox{Im}\,\sigma^{(\xi)}_{ii}(\omega)]
=\pi\lim_{\omega\to 0}[\omega\,\mbox{Im}\,\sigma_{ii}^{(\xi),intra}(\omega)]$.
When the Fermi level lies in the gap, the intraband conductivity is null, only transitions from
the valence into the conduction band contribute.
We will work within the infinite band limit.

\begin{figure}
    \centering
    \includegraphics[scale=0.35]{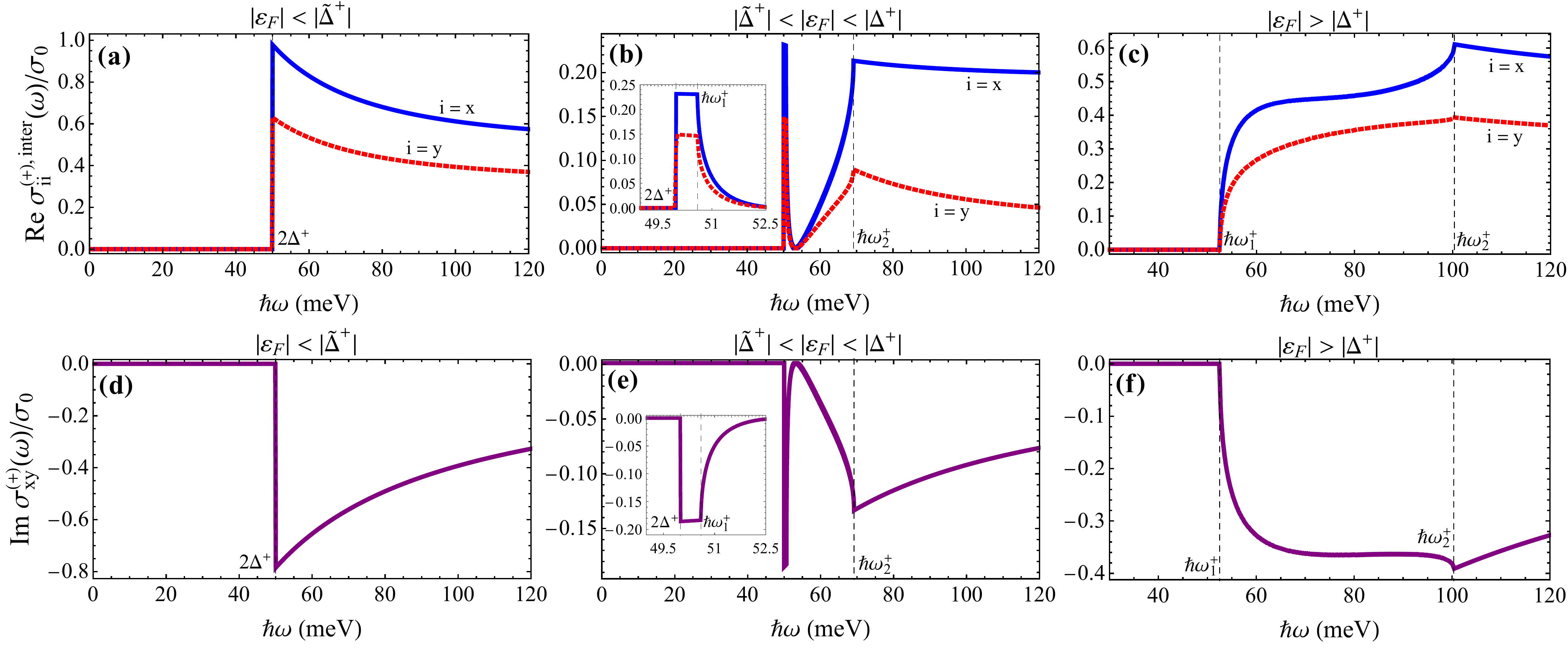}
    \caption{Dissipative components of the interband conductivity at the $K$ valley, 
    for $|\varepsilon_F|<|\tilde{\Delta}^+|$ (a) and (d),  $|\tilde{\Delta}^+|<|\varepsilon_F|<|\Delta^+|$ (b) and (e),   $|\varepsilon_F|>|\Delta^+|$ (c) and (f). We take $\Delta^+=25$ meV.}
    \label{fig:my_label_opt1} 
\end{figure}

Fig.\,\ref{fig:my_label_opt1} shows the dissipative components of the optical conductivity tensor for several positions of the Fermi level in the valley $\xi=+$.
In accordance to the spectral behavior of JDOS, these spectra show interband critical points and a characteristica
reduction of the response when the Fermi level lies in the indirect zone. In contrast to graphene and borophene there
is a finite transverse response $\propto\Delta^{\xi}$. On the other hand, the result $\sigma^{\xi}_{xx}(\omega)\neq\sigma^{\xi}_{yy}(\omega)$
reveals the anisotropic character of the optical response of the system. 

For $|\varepsilon_F|<|\tilde{\Delta}^{\xi}|$ (Fig.\,\ref{fig:my_label_opt1}(a) and (d)) we obtain
\begin{eqnarray}
\text{Re}\,\sigma_{ii}^{(\xi),inter}(\omega) &=& g_s\frac{\sigma_0\pi}{16}\left[1+\left(\frac{2\Delta^{\xi}}{\hslash\omega}\right)^2\right]
\frac{v_i}{v_x}\frac{v_i}{v_y}\,\Theta(\hslash\omega-2|\Delta^{\xi}|) \label{ii1}\\
\text{Im}\,\sigma_{xy}^{(\xi),inter}(\omega)  &=&  -\xi g_s\sigma_0\pi\,\frac{\Delta^{\xi}}{4\hslash\omega}\,
\Theta(\hslash\omega-2|\Delta^{\xi}|)\label{xy1} \ .
\end{eqnarray}
For the diagonal components it is verified that $v_y^2\,\text{Re}\,\sigma_{xx}^{(\xi),inter}(\omega)=v_x^2\,\text{Re}\,\sigma_{yy}^{(\xi),inter}(\omega)$,
while the Hall component is independent of $v_x, v_y$, and $v_t$. These results look very similar to those of gapped graphene, but with the additional 
 anisotropy factor $v_i^2/v_xv_y\neq 1$ in the diagonal elements.

When $|\varepsilon_F|>|\Delta^{\xi}|$ (Fig.\,\ref{fig:my_label_opt1}(c) and (f)),
\begin{eqnarray}
\mbox{Re}\,\sigma_{ii}^{(\xi),inter}(\omega)  &=& g_s\frac{\sigma_0}{16}\left\{\frac{\alpha_x^2}{\alpha_F^2}\frac{\alpha_y^2}{\alpha_F^2}
\left[1-\left(\frac{2\Delta^{\xi}}{\hslash\omega}\right)^2\right]
\int_0^{2\pi}\!\!d\theta\,\frac{(\delta_{ix}\sin^2\theta+\delta_{iy}\cos^2\theta)}{g^4(\theta)}\,\Theta[\omega-\omega^{\xi}_{\lambda}(\theta)] \right. \label{SpR1} \\
&& \hspace*{6.5cm} \left.\ +\ \frac{\alpha_i^2}{\alpha_F^2}\left(\frac{2\Delta^{\xi}}{\hslash\omega}\right)^2
\int_0^{2\pi}\!\!\frac{d\theta}{g^2(\theta)}\,\Theta[\omega-\omega^{\xi}_{\lambda}(\theta)]\right\}   \nonumber \\
\text{Im}\,\sigma_{xy}^{(\xi),inter}(\omega)   &=&  - \xi g_s\sigma_0\pi\,\frac{\Delta^{\xi}}{4\hslash\omega}\,\frac{\alpha_x}{\alpha_F}
\frac{\alpha_y}{\alpha_F}\,\frac{1}{2\pi}\int_0^{2\pi}\!\!\frac{d\theta}{g^2(\theta)}\,\Theta[\omega-\omega^{\xi}_{\lambda}(\theta)] \ ,\label{ApIm1}
\end{eqnarray}
where $\lambda=+$ ($-$) if $\varepsilon_F>0 \,(<0)$. We find that $\text{Re}\,\sigma_{ii}^{(\xi),inter}(\omega<\omega_1^+)=0$ and 
\begin{equation} \label{Spcte}
\text{Re}\,\sigma_{ii}^{(\xi),inter}(\omega>\omega_2^+) =
 g_s\frac{\sigma_0\pi}{16}\left[1+\left(\frac{2\Delta^{\xi}}{\hslash\omega}\right)^2\right]\frac{v_i}{v_x}\frac{v_i}{v_y} \ . 
\end{equation}
In contrast to the result for borophene ($\Delta^{\xi}=0$), expression (\ref{Spcte}) shows a dependence on frequency;
only for high enough frequency the universal result $\text{Re}[\sigma^{inter}_{xx}(\omega>\omega_2^+)]
\times \text{Re}[\sigma^{inter}_{yy}(\omega>\omega_2^+)]=(e^2/4\hslash)^2$ reported by Verma et al. \cite{verma2017effect} can be recovered.
On the other hand,
between $\omega_1^+$ and $\omega_2^+$ Eq. (\ref{SpR1}) gives a frequency dependence very similar to that of borophene, although
with slightly different critical frequencies. Similarly, $\text{Im}\,\sigma_{xy}^{(\xi),inter}(\omega<\omega_1^+)=0$ and
\begin{equation}
\text{Im}\,\sigma_{xy}^{(\xi),inter}(\omega>\omega_2^+)=-\xi g_s\sigma_0\pi\frac{\Delta^{\xi}}{4\hslash\omega} \ .
\end{equation}
Again, this result is very close to that of gapped graphene for the $\xi$ valley $\text{Im}\,\sigma_{xy}^{gr}(\omega)=-\xi g_s\sigma_0\pi(\Delta^\xi/4\hslash\omega)
\Theta(\hslash\omega-2\text{max}\{|\varepsilon_F|,|\Delta^\xi|\})$.

Distinctly different behavior occurs when the Fermi level is located within the indirect zone $|\tilde{\Delta}^{\xi}|<|\varepsilon_F|<|\Delta^{\xi}|$
(Fig.\,\ref{fig:my_label_opt1}(b) and (e)) because of the strong reduction of the momentum space available for optical transitions, as was discussed about
the JDOS. In this narrow window for the Fermi energy we have
\begin{eqnarray}
\mbox{Re}\,\sigma_{ii}^{(\xi),inter}(\omega) &=& g_s\frac{\sigma_0}{16}\,\Theta(\hslash\omega-2|\Delta^{\xi}|)
\left\{\frac{\alpha_x^2}{\alpha_F^2}\frac{\alpha_y^2}{\alpha_F^2}
\left[1-\left(\frac{2\Delta^{\xi}}{\hslash\omega}\right)^2\right] \times \right. \\
&& \hspace*{4cm} \left. \int_{\theta_0-\theta_{\xi}^*}^{\theta_0+\theta_{\xi}^*}
\!d\theta\,\frac{\delta_{ix}\sin^2\theta+\delta_{iy}\cos^2\theta}{g^4(\theta)}\left(\Theta[\nu^{\xi,-}_\lambda(\theta)-\omega]+
\Theta[\omega-\nu^{\xi,+}_\lambda(\theta)]\right) \right. \nonumber \\
&& \hspace*{4cm} \left.\ +\  \frac{\alpha_i^2}{\alpha_F^2}\left(\frac{2\Delta^{\xi}}{\hslash\omega}\right)^2
\int_{\theta_0-\theta_{\xi}^*}^{\theta_0+\theta_{\xi}^*}\!\frac{d\theta}{g^2(\theta)}\,\left(\Theta[\nu^{\xi,-}_\lambda(\theta)-\omega]
+\Theta[\omega-\nu^{\xi,+}_\lambda(\theta)]\right)\right\} \ , \nonumber \\
\text{Im}\,\sigma_{xy}^{(\xi),inter}(\omega) &=& -\xi g_s\sigma_0\pi\,\Theta(\hslash\omega-2|\Delta^{\xi}|)\,
\frac{\Delta^{\xi}}{4\hslash\omega}\,\frac{\alpha_x}{\alpha_F}
\frac{\alpha_y}{\alpha_F}\,\frac{1}{2\pi}\int_{\theta_0-\theta_{\xi}^*}^{\theta_0+\theta_{\xi}^*}\!\!\frac{d\theta}{g^2(\theta)}\,
\left(\Theta[\nu^{\xi,-}_\lambda(\theta)-\omega]+\Theta[\omega-\nu^{\xi,+}_\lambda(\theta)]\right) \ . \label{xy3}
\end{eqnarray}

As was mentioned above, the spectral features associated to three critical frequencies in the optical response serve as a fingerprint of the simultaneous
presence of tilting and mass in the band structure of a 2D Dirac system at low energies. From Eqs.(\ref{ii1})-(\ref{xy3}), it is verified that
for $|\Delta^+|=|\Delta^-|$, $\,\text{Re}\,\sigma_{ii}^{(+)}=
\text{Re}\,\sigma_{ii}^{(-)}$, while for $\Delta^+=\pm\Delta^-$,
$\text{Im}\,\sigma_{xy}^{(+)}=\mp\,\text{Im}\,\sigma_{xy}^{(-)}$.

Now, we comment on the total response function $\sigma_{ij}(\omega)=\sum_{\xi}\sigma_{ij}^{(\xi)}(\omega)$.
In contrast to pristine or (uniformly) gapped graphene, where the conductivity tensor is diagonal and isotropic with an absorption edge defined by $2|\Delta^{\xi}|$ or
$2|\varepsilon_F|$, in our system we have two pairs of tilted cones with different gaps in each valley. It is verified that $\sigma_{xy}(\omega)=0$ when $\Delta^+=\Delta^-$, because of
the recovery of the time-reversal symmetry. Correspondingly,
the inversion symmetry is broken, which anticipates a
valley sensitive response.

The three distinct possibilities
for the position of the chemical potential, and the corresponding spectral characteristics of the optical conductivity of each valley,
open a number of scenarios for the total response. In the following we list the distinctive cases:
\begin{enumerate}
\item A closed gap in one valley and an open gap in the other. For instance $\Delta^+\neq 0$, $\Delta^-=0$:

$(i) \ |\varepsilon_F|<|\tilde{\Delta}^+|$

$(ii) \ |\Tilde{\Delta}^+|<\varepsilon_F<|\Delta^+|$

$(iii)\ |\varepsilon_F|>|\Delta^+|$
\item $\varepsilon_F$ within the absolute gap, $|\varepsilon_F|<\text{min}\{|\tilde{\Delta}^+|,|\tilde{\Delta}^-|\}$.
\item Non-overlapping indirect zones,  $|\tilde{\Delta}^+|<|\Delta^+|<|\tilde{\Delta}^-|<|\Delta^-|$.

$(i)$ $\varepsilon_F$ at the indirect zone at $K$ but within the gap at $K'$, 
$|\tilde{\Delta}^+|<|\varepsilon_F|<|\Delta^+|<|\tilde{\Delta}^-|$:

$(ii)$  $\varepsilon_F$ above the direct zone at $K$, but in the gap at $K'$,
 $|\tilde{\Delta}^+|<|\Delta^+|<|\varepsilon_F|<|\tilde{\Delta}^-|$.

$(iii)$ $\varepsilon_F$ above the direct zone at $K$, but in the indirect zone at $K'$,
$|\Delta^+|<|\tilde{\Delta}^-|<|\varepsilon_F|<|\Delta^-|$.

$(iv)$ $\varepsilon_F$ above the direct zones at $K$ and $K'$, $|\varepsilon_F|>\text{max}\{|\Delta^+|,|\Delta^-|\}$.

\item $\varepsilon_F$ lying at overlapping indirect zones at $K$ {\it and} $K'$, 
$|\tilde{\Delta}^+|<|\tilde{\Delta}^-|<|\varepsilon_F|<|\Delta^+|<|\Delta^-|$.
\end{enumerate}

\begin{figure}
    \centering
    \includegraphics[scale=0.36]{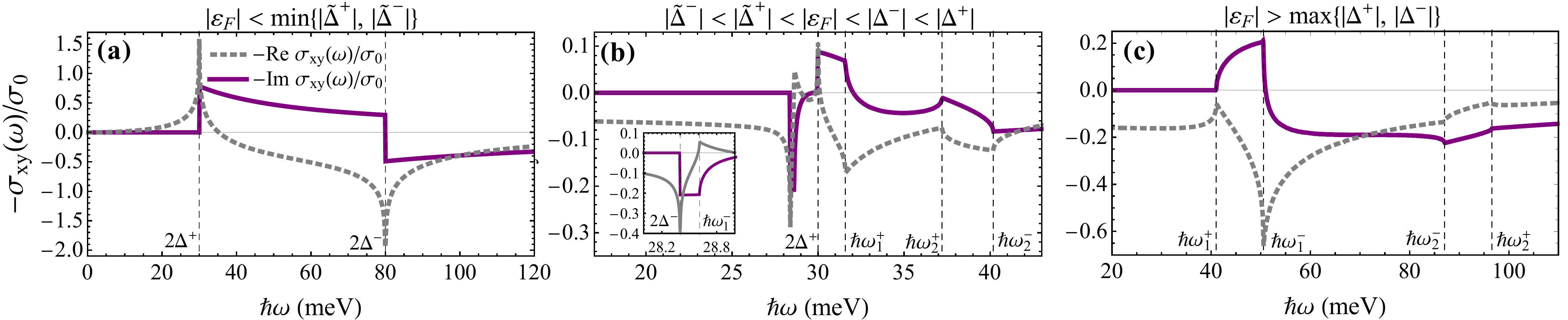}
    \caption{Real and imaginary part of the total Hall response
    $\sigma_{xy}(\omega)=\sum_{\xi}\sigma_{xy}^{(\xi)}(\omega)$
    for (a) Fermi level within the absolute gap, with $\Delta^+=15$ meV, $\Delta^-=40$ meV, $\varepsilon_F=0$ meV, (b) Fermi energy lying at overlapped indirect zones, with $\Delta^+=15$ meV, $\Delta^-=14.2$ meV,  $\varepsilon_F=13.5$ meV, and (c) Fermi level above the direct zones $\Delta^+=15$ meV, for $\Delta^-=25$ meV,  $\varepsilon_F=27$ meV.}
    \label{fig:my_labelH}
\end{figure}

Thus, from the spectral characteristics of the response of an individual valley it is possible to anticipate the spectral features of
the total response. For example, in the case 1 the Hall conductivity will arise from the transverse response of the valley at $K$ exclusively (Fig.\,\ref{fig:my_label_opt1}(d)-(f)),
in the case 2 the spectrum will display features at the onsets $2\Delta^+$ and $2\Delta^-$, four critical frequencies will be present
in the case $3(iv)$, while six will shape the spectrum in the case 4. To illustrate this variability of the optical response,
in Fig.\,\ref{fig:my_labelH}
we show $\sigma_{xy}(\omega)$ for the scenarios 2, 3$(iv)$, and 4 only, in the name of brevity.

\subsection{Drude weight}

\begin{figure}[h]
    \centering
    \includegraphics[scale=0.455]{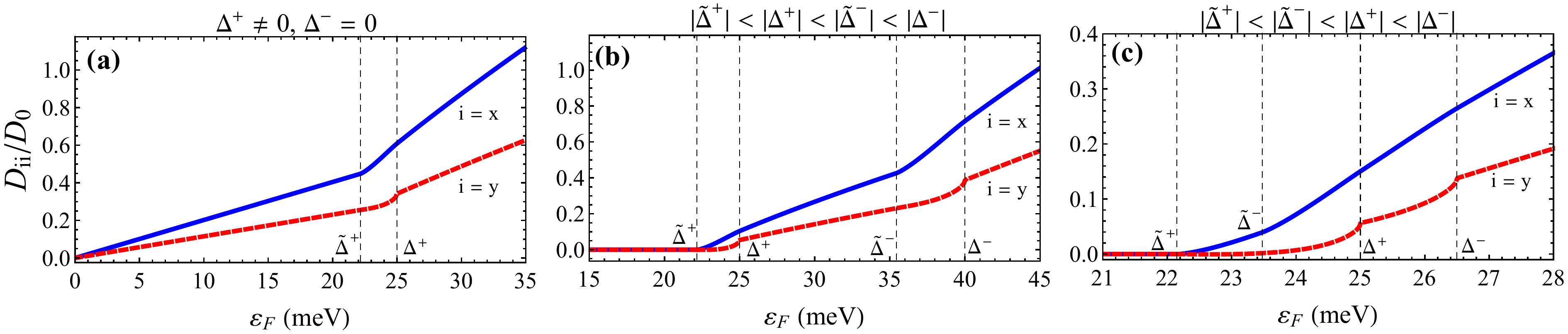}
    \caption{Drude weight as a function of Fermi energy when
    $\Delta^+=25$ meV for
    (a) scenario 1, with $\Delta^-=0$,
    (b) scenario 3, with $\Delta^-=40$ meV,
    and (c) scenario 4, with $\Delta^-=26.5$ meV. The scale is normalized to the Drude weight of pristine graphene  $D_0=(e^2/\hslash)(\varepsilon_F^0/\hslash)$, where $\varepsilon_F^0=\hslash v_F\sqrt{\pi n_e}$ is the Fermi energy in pristine graphene, with electron density $n_e=1\times10^{15}$ m$^{-2}$.}
    \label{fig:my_labelDrude}
\end{figure}

The intraband optical conductivity spectral weight is anisotropic, as expected, and shows a specific behavior due to the
presence of unequal gaps.  
For gappless, tilted or untilted pair of cones, the Drude weight $D_{ii}^{(\xi)} \propto |\varepsilon_F|$, while for gapped graphene
is $D_{\Delta^\xi}=\Theta(|\varepsilon_F|-|\Delta^\xi|)(e^2/\hslash^2)(\varepsilon_F^2-(\Delta^\xi)^2)/|\varepsilon_F|$. In our case this behavior is modified in each valley
because of the indirect nature of the gap. In addition, the breaking of the valley symmetry yields $D_{ii}^{(+)}\neq D_{ii}^{(-)}$.
The total Drude weight
$D_{ii}=\sum_{\xi}D^{(\xi)}_{ii}$ as a function of (positive) Fermi energy is shown in
Figure\,\ref{fig:my_labelDrude} for scenarios 1,3, and 4. When the valley $K'$ is gapples, its contribution to the total weight
displays the characteristic linear dependence up to $\tilde{\Delta}^+$, after which the contribution of the gapped valley $K$ starts,
presenting a specific nonlinear behavior in the indirect zone $\tilde{\Delta}^+<\varepsilon_F<\Delta^+$, 
and a gapped-graphene-like $D_{\Delta^\xi}$ function
above the nominal direct gap $\Delta^+$ (Fig.\,\ref{fig:my_labelDrude}(a)). The situation looks rather different when both valleys are gapped. For the scenario
with non-overlapping indirect zones $|\tilde{\Delta}^+|<|\Delta^+|<|\tilde{\Delta}^-|<|\Delta^-|$ (Fig.\,\ref{fig:my_labelDrude}(b)), $D_{ii}(\varepsilon_F)$ is zero
for $\varepsilon_F$ below $\tilde{\Delta}^+$, and then follows a behavior similar to that of Fig.\,\ref{fig:my_labelDrude}(a) for
 $\tilde{\Delta}^+<\varepsilon_F< \tilde{\Delta}^-$, determined only by transitions in the branch $\varepsilon_+^+$. Above $\tilde{\Delta}^-$, the
 transitions in the band $\varepsilon^-_+$ are added,  leading to a variation similar to that in Fig.\,\ref{fig:my_labelDrude}(a).
 The function
$D_{ii}(\varepsilon_F)$ notably changes  when the indirect zones overlap, $\tilde{\Delta}^+<\tilde{\Delta}^-<\Delta^+<\Delta^-$ (Fig.\,\ref{fig:my_labelDrude}(c)).
Between $\tilde{\Delta}^+$ and $\tilde{\Delta}^-$ only intraband transitions in the indirect zones of the branch $\varepsilon^+_+$ contribute, 
while for $\tilde{\Delta}^-<\varepsilon_F<\Delta^+$  transitions in the indirect zone of each valley start to count. In the range between
$\Delta^+$ and $\Delta^-$, $D_{ii}^{(-)}$ is due to transitions in the band $\varepsilon_+^-$ which take place only in the corresponding indirect zone. 
Above $\Delta^+$ and $\Delta^-$, the Drude weight behaves as can be identified in Fig.\,\ref{fig:my_labelDrude}(a) or (b). 
Globally, a nonlinear dependence on $\varepsilon_F$, 
with an overall reduction of its magnitude, is observed
for the total weight.

\section{Optical properties}
\subsection{Anisotropic response, circular dichroism, and valley polarization}

\begin{figure}
    \centering
    \includegraphics[scale=0.31]{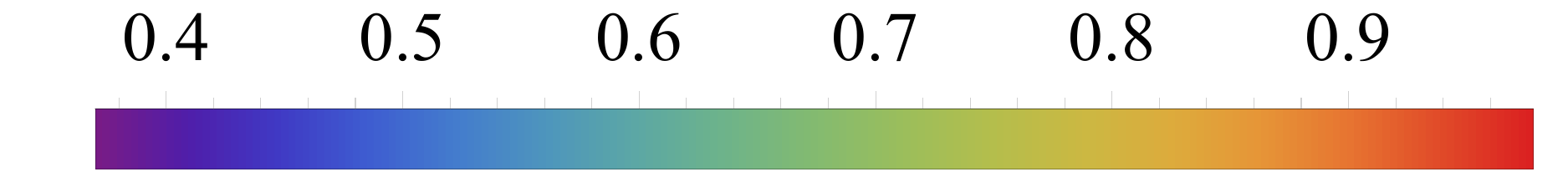}
    \includegraphics[scale=0.31]{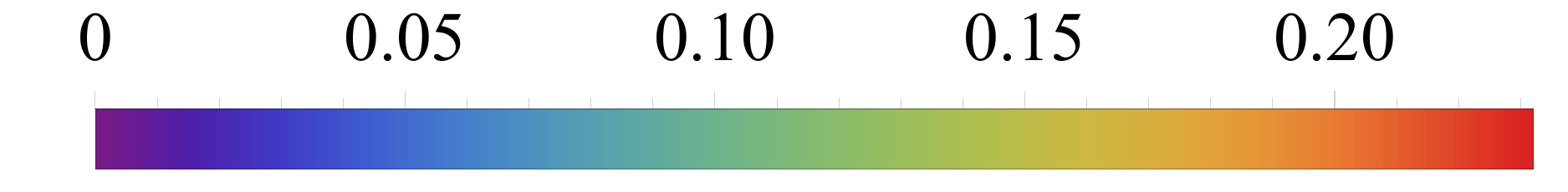}
    \includegraphics[scale=0.31]{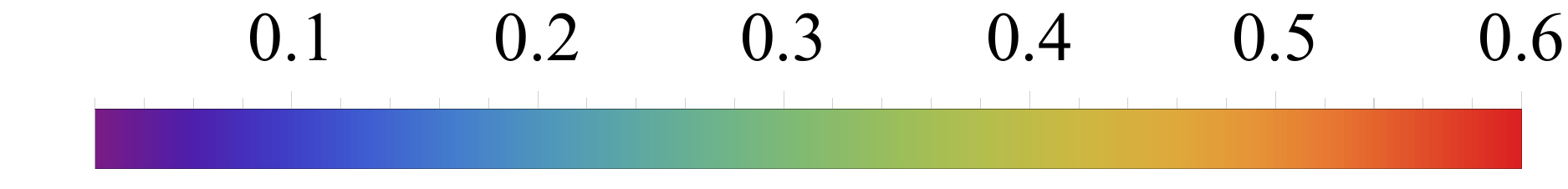}
    \includegraphics[scale=0.31]{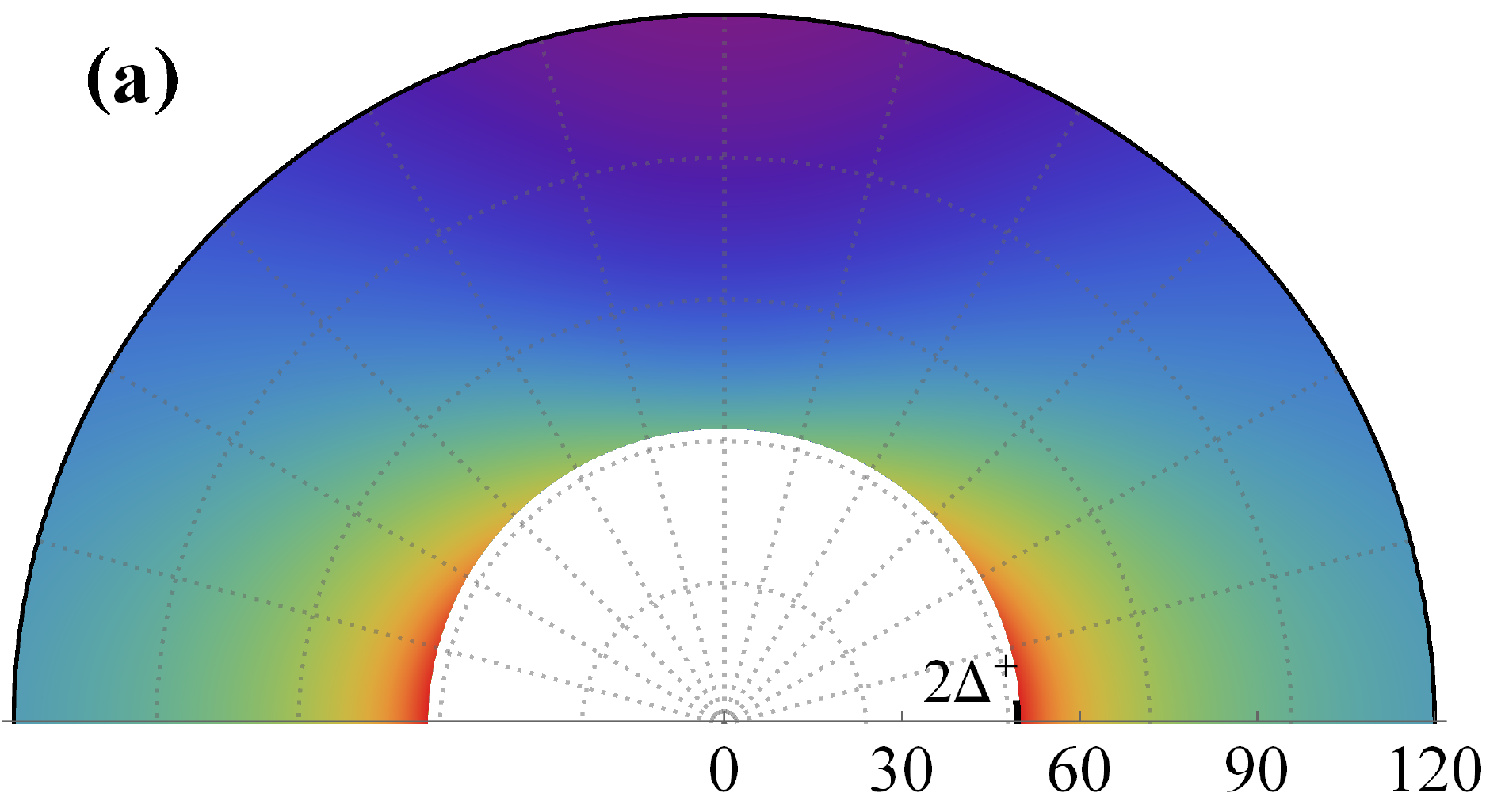}
    \includegraphics[scale=0.31]{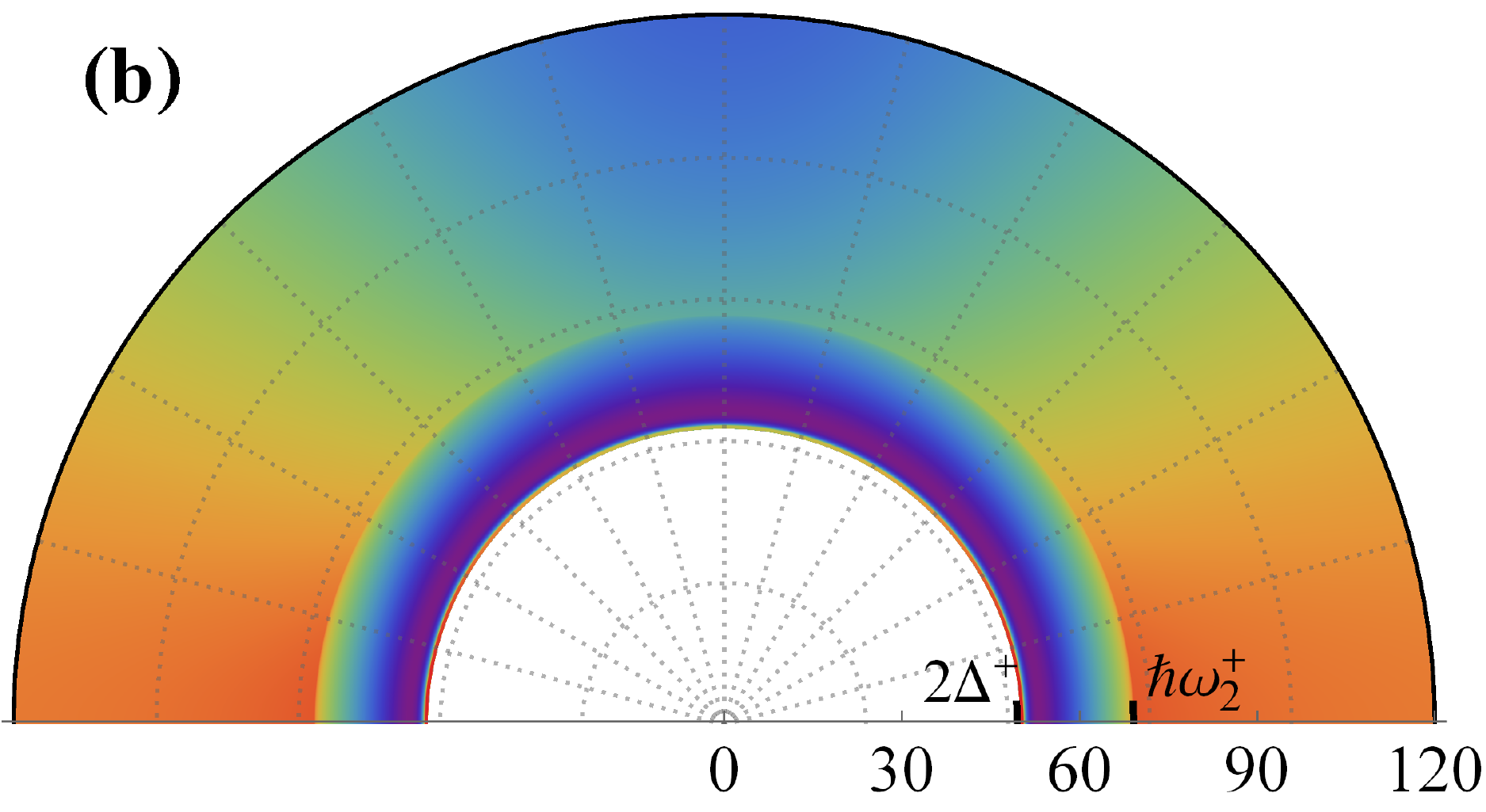}
    \includegraphics[scale=0.31]{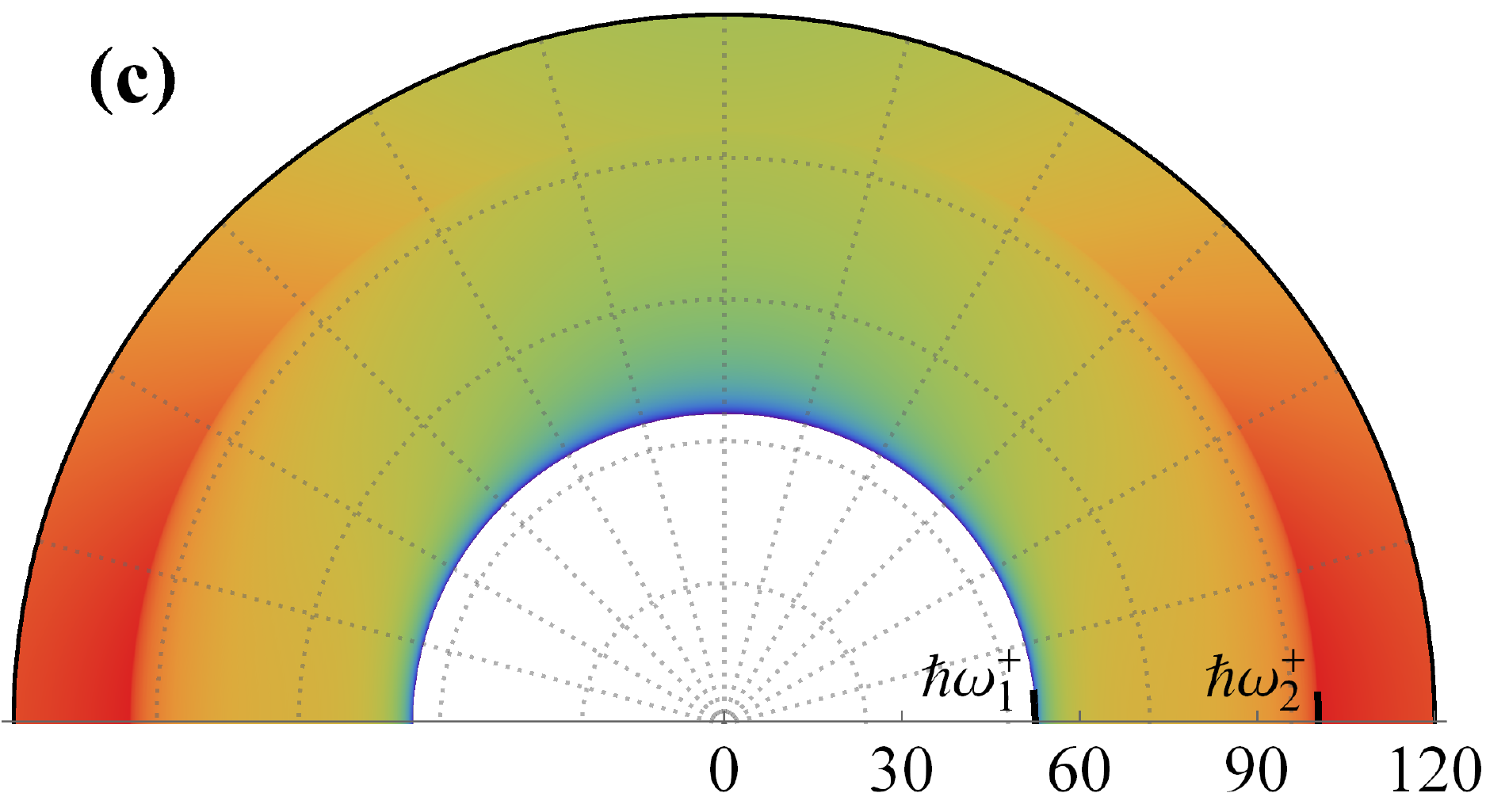}
    \includegraphics[scale=0.33]{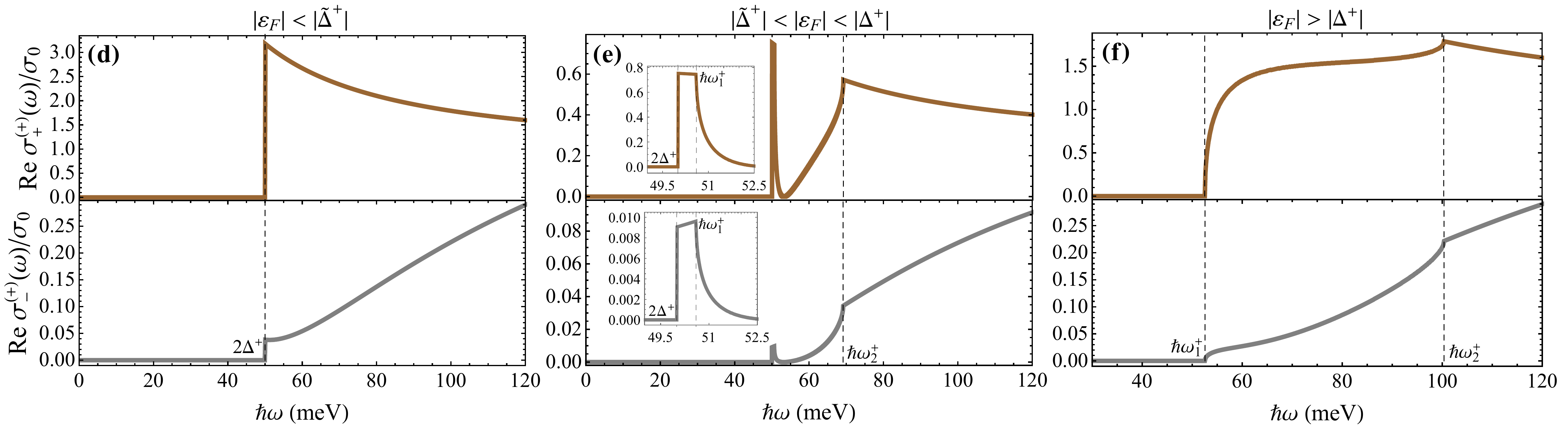}
    \caption{Longitudinal conductivity $\text{Re}[\sigma_{\parallel}^{(+)}(\omega;\varphi)]$ ((a)-(c)) and circular dichroism response $\text{Re}[\sigma^{(+)}_{\pm}(\omega)]$ ((d)-(f)), for
    three distinct positions of the Fermi level
    in the vicinity of the $K$ point with $\Delta^+=25$ meV.}
    \label{fig:circDich}
\end{figure}

The anisotropy expressed by the result $\sigma_{xx}\neq\sigma_{yy}$  can also be presented through the longitudinal
conductivity $\sigma_{\parallel}(\omega; \varphi)=\sum_{\xi}\sigma^{(\xi)}_{\parallel}(\omega; \varphi)$. This scalar response function
determines the density current induced along the direction of the external field, ${\bf J}_{\parallel}=\sigma_{\parallel}{\bf E}$,  and it is given by
$\sigma_{\parallel}(\omega; \varphi)=\hat{q}_i\sigma_{ij}(\omega)\hat{q}_j$ (sum over repeated indices is implied), where ${\bf \hat{q}}=\cos\varphi\,{\bf\hat{x}}+\sin\varphi\,{\bf\hat{y}}$ 
gives the direction of the external field ${\bf E}\parallel {\bf \hat{q}}$.  The quantity $\mbox{Re}\,\sigma_{\parallel}(\omega;\varphi)$ determines
the dissipation (power absorption/area) for linearly polarized fields. For the valley $\xi$,
\begin{equation} \label{sigmalong}
\sigma^{(\xi)}_{\parallel}(\omega;\varphi)=\sigma^{(\xi)}_{xx}(\omega)\cos^2\varphi+\sigma^{(\xi)}_{yy}(\omega)\sin^2\varphi \ .
\end{equation}
The off diagonal components of the conductivity tensor does not appear in this quantity because of its antisymmetry,
$\sigma^{(\xi)}_{yx}=-\sigma^{(\xi)}_{xy}$. 
Polar plots of the longitudinal conductivity (\ref{sigmalong}), made of transitions in the vicinity of the valley $\xi=+$, are shown in
Fig.\,\ref{fig:circDich}(a)-(c) as color maps for three positions of the level $\varepsilon_F$. 
We observe that, as a function of the direction $\varphi$, the response follows the same functional angular dependence as the
function $g(\theta)$ when $|\varepsilon_F|<|\tilde{\Delta}^+|$ for frequencies above the gap, and when $|\varepsilon_F|>|\Delta^+|$
for $\omega>\omega_2^+$. Indeed, given that $\text{Re}\,\sigma^{(\xi)}_{ii}(\omega)\propto \alpha_i^2$ we find that
 $\text{Re}\,\sigma^{\xi}_{\parallel}(\omega;\varphi)=\bar{\sigma}(\omega)\Theta(\hslash\omega-2|\Delta^{\xi}|)g^2(\varphi)$ for the former case
 (Fig.\,\ref{fig:circDich}(a)) and $\text{Re}\,\sigma^{\xi}_{\parallel}(\omega>\omega_2^{\xi};\varphi)=\bar{\sigma}(\omega)g^2(\varphi)$ in the latter
 (Fig.\,\ref{fig:circDich}(c)),  where $\bar{\sigma}(\omega)=g_s(\sigma_0\pi/16)[1+(2\Delta^{\xi}/\hslash\omega)^2]\alpha_F^2/\alpha_x\alpha_y$. This
 dependency on $\varphi$ is otherwise modified for $\omega_1^{\xi}\leqslant\omega\leqslant\omega_2^{\xi}$ (Fig.\,\ref{fig:circDich}(b)-(c)) due to the
 spectral characteristics of the allowed transitions in this frequency range. In Fig.\,\ref{fig:circDich}(b), the anisotropy
 is hardly noticeable below $\hslash\omega_2^+$ because of the strong suppression of the spectrum there.

On the other hand, owing to the nonvanishing Hall component $\sigma_{xy}(\omega)$
the medium absorb right- and left-circularly polarized light differently, revealing the circular
dichroism of the medium. The appropriate conductivity for circularly polarized external field ${\bf E}_{\pm}$ is the
quantity
$\sigma_{\pm}(\omega)=\sigma_{xx}(\omega)+\sigma_{yy}(\omega)\pm i[\sigma_{xy}(\omega)-\sigma_{yx}(\omega)]$,
which gives the induced current ${\bf J}=\sigma_{\pm}{\bf E}_{\pm}$. In this case, the power absorption/area is
determined by 
\begin{equation}
\text{Re}\,\sigma_{\pm}(\omega)=\text{Re}[\sigma_{xx}(\omega)+\sigma_{yy}(\omega)]\mp \text{Im}[\sigma_{xy}(\omega)-\sigma_{yx}(\omega)].
\end{equation}


Figure \ref{fig:circDich}(d)-(f) displays $\text{Re}\,\sigma^{(+)}_+(\omega)$ and $\text{Re}\,\sigma^{(+)}_-(\omega)$ for several values of the level $\varepsilon_F$.
As expected, for $|\varepsilon_F|<|\tilde{\Delta}^{\xi}|$ (Fig.\,\ref{fig:circDich}(d)) and $|\varepsilon_F|>|\Delta^{\xi}|$ (Fig.\,\ref{fig:circDich}(f)) the spectra show a graphene-like
and borophene-like response, respectively (see Fig.\,\ref{fig:my_label_opt1}(a),(c),(d),(f)).
On the other hand, the new scenario (Fig.\,\ref{fig:circDich}(e)) opened by the indirect nature of the gap (Fig.\,\ref{fig:circDich}(e)) presents
a spectrum which, in the interval $\omega_1^+\leqslant\omega\leqslant\omega_2^+$,  breaks that graphene-like behavior (observed in the narrow 
 range $2\Delta^+\leqslant\hslash\omega\leqslant\hslash\omega_1^+$) and the borophene-like behavior (started above $\omega_2^+$).
 
Circular dichroism can also be illustrated through the Hall angle
\begin{equation}
\tan\Theta_{\text{H}}(\omega;\varepsilon_F)=\frac{\text{Re}\,\sigma_+-\text{Re}\,\sigma_-}{\text{Re}\,\sigma_++\text{Re}\,\sigma_-}=
\frac{-\text{Im}(\sigma_{xy}-\sigma_{yx})}{\text{Re}(\sigma_{xx}+\sigma_{yy})} \ .
\end{equation}

\begin{figure}[h]
    \centering
    \includegraphics[scale=0.65]{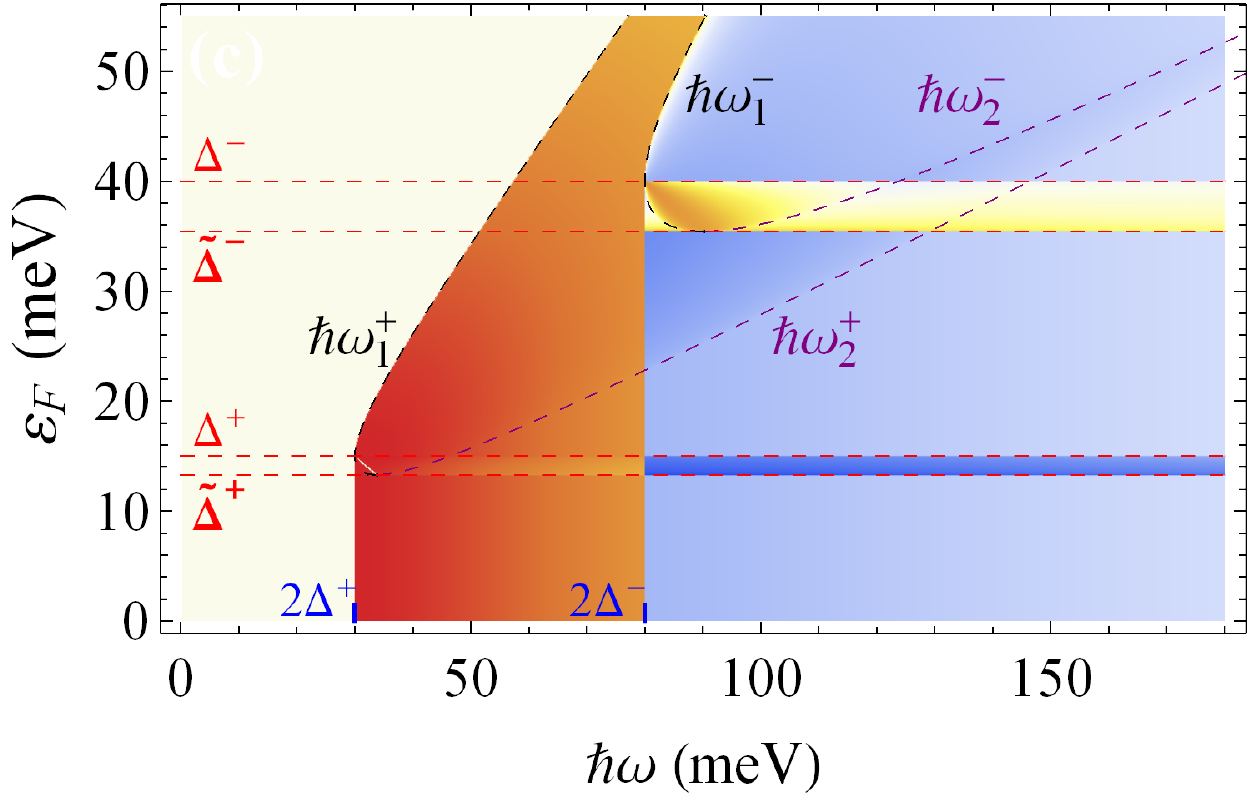}\includegraphics[scale=0.67]{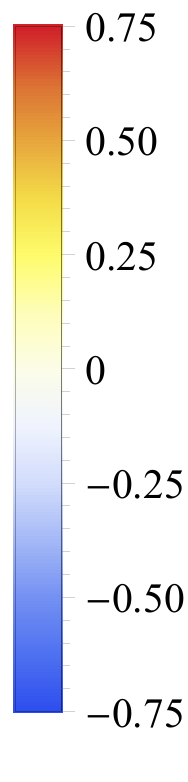}
    \caption{Hall angle $\Theta_\text{H}(\omega;\varepsilon_F)$ (radians), for $\Delta^+=15$ meV, $\Delta^-=40$ meV.}
    \label{fig:my_labelTh}
\end{figure}

In Fig.\,\ref{fig:my_labelTh} we show the Hall angle spectrum as a function of the (positive) Fermi energy, encompassing the scenario 2 and the cases of scenario 3. When 
$\varepsilon_F<\tilde{\Delta}^+$ (scenario 2) the spectrum starts at the energy gap
$2\Delta^+$, with decreasing magnitude until the onset $2\Delta^-$ where is an abrupt change of color due to a change of sign of $-\text{Im}[\sigma_{xy}(\omega)]$ (Fig.\,\ref{fig:my_labelH}(a)). For $\varepsilon_F>\Delta^-$
(case $3(iv)$),
the spectrum present four critical points, with a change of
sign at $\omega^-_1$ (Fig.\,\ref{fig:my_labelH}(c)).
When $\tilde{\Delta}^+<\varepsilon_F<\Delta^+$ (case $3(i)$), the angle $\Theta_\text{H}(\omega)$ show three
critical points associated to the van Hove singularities of the JDOS in the indirect zone of the valley at $K$ point, and a change of sign (from positive to negative) of $-\text{Im}[\sigma_{xy}(\omega)]$ at $2\Delta^-$. In contrast, at higher values of $\varepsilon_F$, lying within the other indirect zone (case $3(iii)$), we see that the spectrum of dichroism will display five critical energies, without any change of sign. With respect to the case $3(ii)$, where the Fermi level is outside the indirect zones but inside the $K'$ valley gap, the corresponding spectrum is molded by three critical points and a change of sign of $-\text{Im}[\sigma_{xy}(\omega)]$.

We note that there are regions in the $\varepsilon_F$-$\omega$ diagram with the Hall angle close to $\pm\pi/4$ indicating an almost perfect
circular dichroism, where the system absorbs mostly left circularly polarized light and very little the opposite
handed polarization, or vice versa. Indeed, we note that for
$\varepsilon_F<\tilde{\Delta}^+$ and $\hslash\omega\gtrsim 2\Delta^+$, the dichroism arises from
the absorption at the $K$ valley only. Thus 
$\text{Re}\,\sigma_{\pm}=\text{Re}\,\sigma_{\pm}^{(+)}$,
where $\text{Re}\,\sigma_{+}^{(+)}\gg \text{Re}\,\sigma_-^{(+)}$,
as can be seen in Fig.\,\ref{fig:circDich}(d) for example, leading to $\tan\Theta_{\text{H}}\approx +1$. On the other hand, when $\tilde{\Delta}^+<\varepsilon_F<\Delta^+<\tilde{\Delta}^-$, 
$\text{Re}\,\sigma_{\pm}\approx\text{Re}\,\sigma_{\pm}^{(-)}$
for $\hslash\omega\gtrsim 2\Delta^-$, because of the decreasing of transitions at $K$ valley (see Fig.\,\ref{fig:circDich}(e)).
This leads to $\tan\Theta_{\text{H}}\approx -2\,\text{Im}\,\sigma_{xy}^{(-)}/\text{Re}[\sigma_{xx}^{(-)}+\sigma_{yy}^{(-)}]$. From Eqs. (\ref{ii1}) and (\ref{xy1}), we obtain
$\tan\Theta_{\text{H}}\approx -2/[(v_x/v_y)+(v_y/v_x)]\approx -1$. This is remarkable because such possibility
occurs due to the existence of an indirect zone 
and the associated significant reduction of the
dynamical response. The
result is in sharp contrast to the untilted case, where in a map like that of Fig.\,\ref{fig:my_labelTh} only
the value $\tan\Theta_{\text{H}} \approx +1$ can be achieved, for $\hslash\omega$ slightly above $2\Delta^+$.

It is interesting to consider the valley polarization
expressed by the angle $\Theta_{\text{V},\pm}$, which measures
the difference of absorption of circularly polarized light
between the $K$ and $K'$ valleys,
\begin{equation}
 \tan\Theta_{\text{V},\pm}(\omega;\varepsilon_F)= 
 \frac{\text{Re}\,\sigma_{\pm}^{(+)}-\text{Re}\,\sigma_{\pm}^{(-)}}{\text{Re}\,\sigma_{\pm}^{(+)}+\text{Re}\,\sigma_{\pm}^{(-)}} \ .
\end{equation}
As expected, for $\Delta^+-\Delta^-=0$, $\Theta_{\text{H}}=0$ due to the
TR invariance, and $\Theta_{\text{V},\pm}\neq 0$ because of the breaking of inversion symmetry. On the contrary, for $\Delta^++\Delta^-=0$, $\Theta_{\text{H}}\neq 0$ and $\Theta_{\text{V},\pm}=0$
because TRS is broken in the system while retaining inversion symmetry. In the Haldane model, valley polarization and perfect
circular dichroism have been reported to occur exclusively,\cite{saito-2018} while 
the possibility of simultaneous phenomena has been explored recently within a modified Haldane model.\cite{Saito-Haldane}. In our model, for $|\Delta^+|\neq |\Delta^-|$ the valley polarization and circular dichroism
are achieved simultaneously, suggesting an alternative tunable way to realize them.


\subsection{Anomalous and valley Hall conductivities}

\begin{figure}[hb]
    \centering
    \includegraphics[scale=0.3]{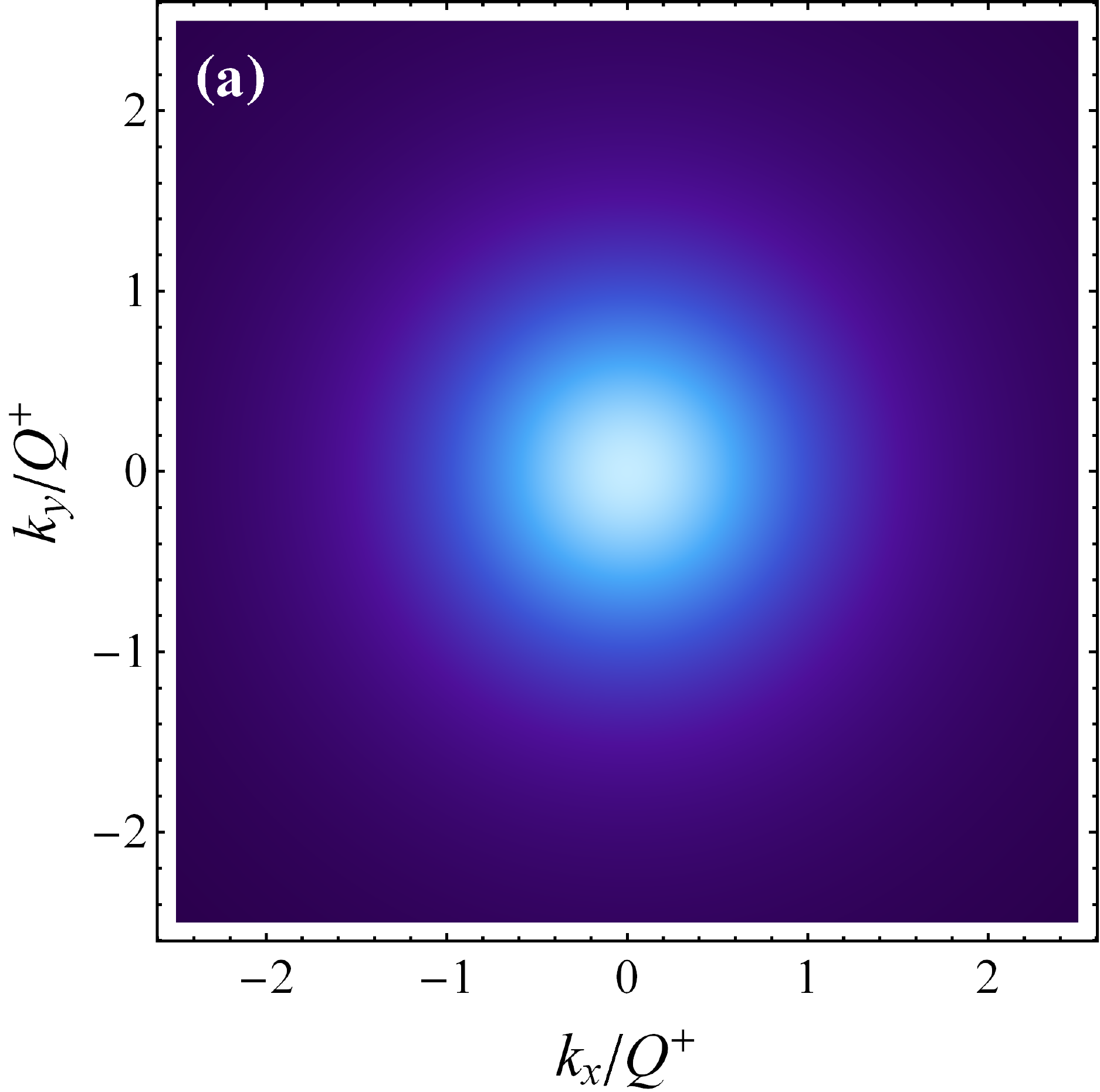}
    \includegraphics[scale=0.3]{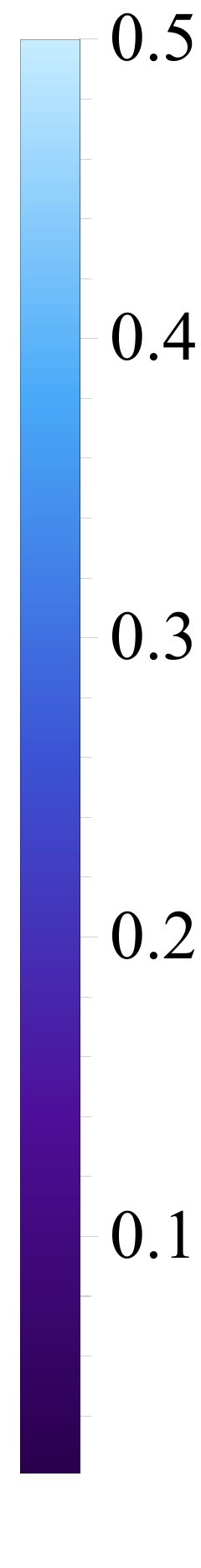}
    \includegraphics[scale=0.3]{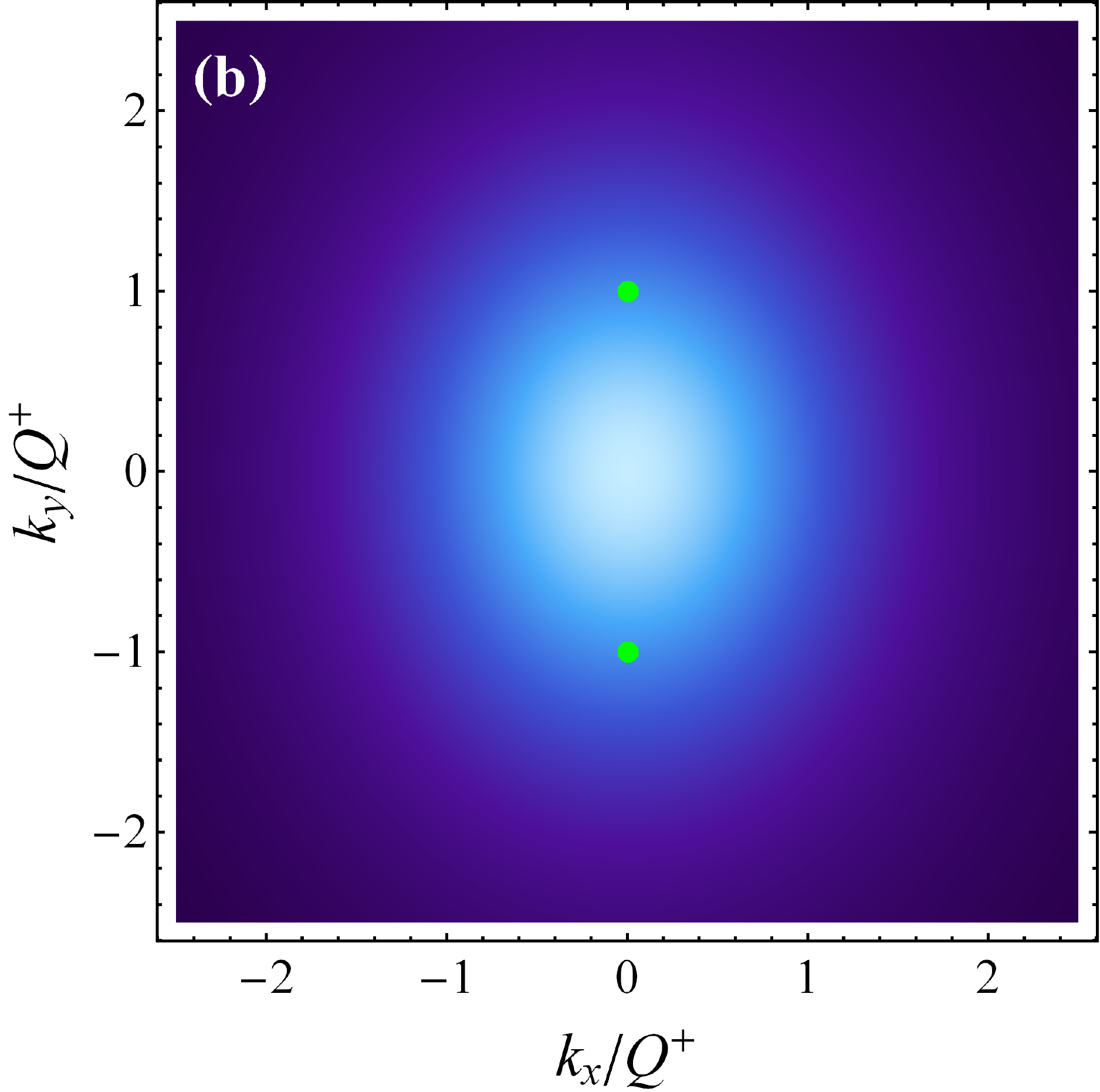}
    \includegraphics[scale=0.3]{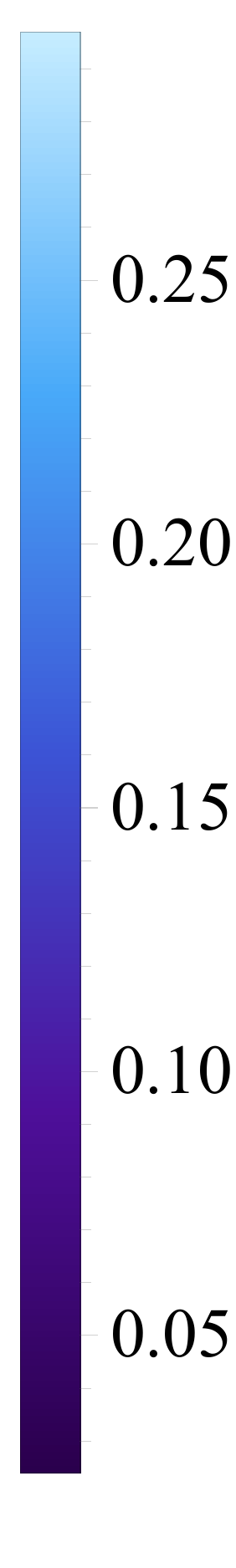}
    \caption{Berry curvature $-(\Delta^+/\alpha_F)^2\Omega^{+}_{+,z}({\bf k})$ for the states in the band $\varepsilon_+^+({\bf k})$ when (a) $v_t=0, v_x=v_y$,
    and (b) $v_t\neq 0, v_x\neq v_y$; the points $(0,\pm Q^+)$ are indicated. We take $\Delta^+=25\,$meV. The spread along $k_y$ directions reflects the anisotropy of the band.}
    \label{fig:my_labelBerry}
\end{figure}

The anomalous Hall conductivity (AHC) is defined by $\sigma^{AHE}=\sigma^{(+)}_{xy}(0)+\sigma^{(-)}_{xy}(0)$, which we obtain through
the well known formula
\begin{equation} \label{AHConduc1}
\sigma_{xy}^{(\xi)}(0)=-g_s\frac{e^2}{\hslash}\sum_{\lambda}\int\!\!\frac{d^2k}{(2\pi)^2}\,f(\varepsilon^{\xi}_{\lambda}({\bf k}))\,
\Omega^{\xi}_{\lambda,z}({\bf k}) \ ,
\end{equation}
where $\Omega^{\xi}_{\lambda,z}({\bf k})= -\xi\lambda\,\alpha_x\alpha_y\Delta^{\xi}/[2(d^{\xi}({\bf k}))^3]$ is the Berry curvature of a state in the band $\xi,\lambda$. 
Note that $\Omega^{\xi}_{-,z} = -\,\Omega^{\xi}_{+,z}$. In contrast to gapped graphene (Fig.\,\ref{fig:my_labelBerry}(a)), the curvature becomes smaller in magnitude, is no longer isotropic, and spreads over the $k_y$-axis between the points $(0,\pm Q^{\xi})$ (Fig.\,\ref{fig:my_labelBerry}(b)). 

At zero temperature, we have 
\begin{eqnarray}
\sigma^{AHE}(\varepsilon_F)  &=&  g_s\frac{e^2}{\hslash}\left[\int_{{\cal S}_+}\!\!\frac{d^2k}{(2\pi)^2}\,\Omega^{+}_{+,z}({\bf k}) 
+\int_{{\cal S}_-}\!\!\frac{d^2k}{(2\pi)^2}\,\Omega^{-}_{+,z}({\bf k})\right] \nonumber \\
  &=&  - g_s\frac{e^2}{2\hslash}\alpha_x\alpha_y\left[\Delta^+\int_{{\cal S}_+}\!\!\frac{d^2k}{(2\pi)^2}\,\frac{1}{(d^+({\bf k}))^3}
 -\Delta^-\int_{{\cal S}_-}\!\!\frac{d^2k}{(2\pi)^2}\,\frac{1}{(d^-({\bf k}))^3}\right] \ ,  \label{AnomalousSigma}
 \end{eqnarray}
where the integrals are taken over the sets ${\cal S}_{\xi}=\{{\bf k}\,|\, \varepsilon^{\xi}_-({\bf k})<\varepsilon_F<\varepsilon^{\xi}_+({\bf k})\}$.
The breaking of the TRS (through $\Delta^+\neq\Delta^-$) implies that $\Omega_{\lambda,z}^{+}({\bf k})\neq -\Omega_{\lambda,z}^{-}(-{\bf k})$ leading to
$\sigma^{AHE}\neq 0$.

In the following, we show results for the scenarios made possible by the unequal gaps, mentioned above:
\begin{enumerate}
\item $\Delta^+\neq 0$, $\Delta^-=0$,
$\ \sigma^{AHE}=\sigma_{xy}^{(+)}(0)$

$(i) \ |\varepsilon_F|<|\tilde{\Delta}^+|$
\begin{equation} \label{in_gap+}
\frac{\sigma^{AHE}}{-(e^2/h)}=\text{sgn}(\Delta^+) \ ,
\end{equation}
where $\text{sgn}(x)$ is the sign function. Thus a Hall plateau
can be observed \cite{hill2011valley}.

$(ii) \ |\Tilde{\Delta}^+|<|\varepsilon_F|<|\Delta^+|$
\begin{equation} \label{1iicase}
\frac{\sigma^{AHE}}{-(e^2/h)} = 
\text{sgn}(\Delta^+)\left(\frac{1}{2}-\frac{\beta_+}{\pi}\right) +\frac{\Delta^+}{\pi} \frac{\alpha_x}{\alpha_F}\frac{\alpha_y}{\alpha_F}\frac{\alpha_t}{\alpha_F}\!\! \int_{\theta_0-\theta_+^*}^{\theta_0+\theta_+^*}\!\!d\theta\,\frac{\sin\theta}{g^2(\theta)}f(\theta;\Delta^+) \ ,
 \end{equation}
 where
\begin{displaymath}
f(\theta;x)=\frac{\sqrt{\varepsilon_F^2g^2(\theta)-x^2[g^2(\theta)-h^2(\theta)]}}{\varepsilon_F^2g^2(\theta)+x^2h^2(\theta)} \ ,
\end{displaymath}
and $\beta_+$ is defined after Eq.(\ref{JDOS3}).

$(iii)\ |\varepsilon_F|>|\Delta^+|$
\begin{equation} \label{1iiicase}
\frac{\sigma^{AHE}}{-(e^2/h)}=\frac{\Delta^+}{\sqrt{\varepsilon_F^2+\gamma^2(\Delta^+)^2}} \ .
\end{equation}

In the next, without loss of generality we shall take $|\Delta^+|<|\Delta^-|$:
\item $|\varepsilon_F|<\text{min}\{|\tilde{\Delta}^+|,|\tilde{\Delta}^-|\}$. \\
From (\ref{AnomalousSigma}),
\begin{equation} \label{in_gap}
\frac{\sigma^{AHE}}{-(e^2/h)}=\text{sgn}(\Delta^+)-\text{sgn}(\Delta^-) \ .
\end{equation}
This expression was reported by Hill et al. \cite{hill2011valley} for gapped graphene with nonuniform gaps.
According to the result (\ref{in_gap}), the anomalous conductivity can be zero or
take the universal quantized value $\sigma^{AHE}=\pm 2e^2/h$.
\item $|\tilde{\Delta}^+|<|\Delta^+|<|\tilde{\Delta}^-|<|\Delta^-|$ 
\begin{itemize}
\item[$(i)$]  
$|\tilde{\Delta}^+|<|\varepsilon_F|<|\Delta^+|<|\tilde{\Delta}^-|<|\Delta^-|$. \\
\begin{equation} \label{3icase}
\frac{\sigma^{AHE}}{-(e^2/h)} = 
\text{sgn}(\Delta^+)\left(\frac{1}{2}-\frac{\beta_+}{\pi}\right) 
 +\frac{\Delta^+}{\pi} \frac{\alpha_x}{\alpha_F}\frac{\alpha_y}{\alpha_F}\frac{\alpha_t}{\alpha_F}2\!\! \int_{\theta_0-\theta_+^*}^{\theta_0+\theta_+^*}\!\!d\theta\,\frac{\sin\theta)}{g^2(\theta)}f(\theta;\Delta^+)
 - \ \text{sgn}(\Delta^-) \ .
\end{equation}
\item[$(ii)$]
 $|\tilde{\Delta}^+|<|\Delta^+|<|\varepsilon_F|<|\tilde{\Delta}^-|<|\Delta^-|$. \\
 \begin{equation} \label{3iicase}
\frac{\sigma^{AHE}}{-(e^2/h)}=\frac{\Delta^+}{\sqrt{\varepsilon_F^2+\gamma^2(\Delta^+)^2}}\ -\  \text{sgn}(\Delta^-)  \ .
\end{equation}
\item[$(iii)$] 
$|\tilde{\Delta}^+|<|\Delta^+|<|\tilde{\Delta}^-|<|\varepsilon_F|<|\Delta^-|$. \\
\begin{equation} \label{3iiicase}
\frac{\sigma^{AHE}}{-(e^2/h)}=   \frac{\Delta^+}{\sqrt{\varepsilon_F^2+\gamma^2(\Delta^+)^2}} -
\text{sgn}
(\Delta^-)\left(\frac{1}{2}-\frac{\beta_-}{\pi}\right) 
 - \frac{\Delta^-}{\pi}\frac{\alpha_x}{\alpha_F}\frac{\alpha_y}{\alpha_F}\frac{\alpha_t}{\alpha_F} \!\!
  \int_{\theta_0-\theta_-^*}^{\theta_0+\theta_-^*}\!\!d\theta\,\frac{\sin\theta}{g^2(\theta)}f(\theta;\Delta^-), 
\end{equation}
\item[$(iv)$] 
$|\tilde{\Delta}^+|<|\Delta^+|<|\tilde{\Delta}^-|<|\Delta^-|<|\varepsilon_F|$. \\
\begin{equation} \label{3ivcase}
\frac{\sigma^{AHE}}{-(e^2/h)}=\frac{\Delta^+}{\sqrt{\varepsilon_F^2+\gamma^2(\Delta^+)^2}}\ -\  
\frac{\Delta^-}{\sqrt{\varepsilon_F^2+\gamma^2(\Delta^-)^2}} \ .
\end{equation}
\end{itemize}
\item $|\tilde{\Delta}^+|<|\tilde{\Delta}^-|<|\varepsilon_F|<|\Delta^+|<|\Delta^-|$,
\begin{eqnarray} \label{4case}
\frac{\sigma^{AHE}}{-(e^2/h)} &=&  \left(\frac{1}{2}-\frac{\beta_+}{\pi}\right)\text{sgn}(\Delta^+) -
\left(\frac{1}{2}-\frac{\beta_-}{\pi}\right)\text{sgn}(\Delta^-) \\
&&  \hspace*{1cm} +\frac{1}{\pi}\frac{\alpha_x}{\alpha_F}\frac{\alpha_y}{\alpha_F}\frac{\alpha_t}{\alpha_F}\left[ 
\Delta^+\int_{\theta_0-\theta_+^*}^{\theta_0+\theta_+^*}\!\!d\theta\,\frac{\sin\theta}{g^2(\theta)}f(\theta;\Delta^+)
-\Delta^-\int_{\theta_0-\theta_-^*}^{\theta_0+\theta_-^*}\!\!d\theta\,\frac{\sin\theta}{g^2(\theta)}f(\theta;\Delta^-)\right]. \nonumber
\end{eqnarray}
\end{enumerate}

\begin{figure}[h!]
    \centering
    \includegraphics[scale=0.35]{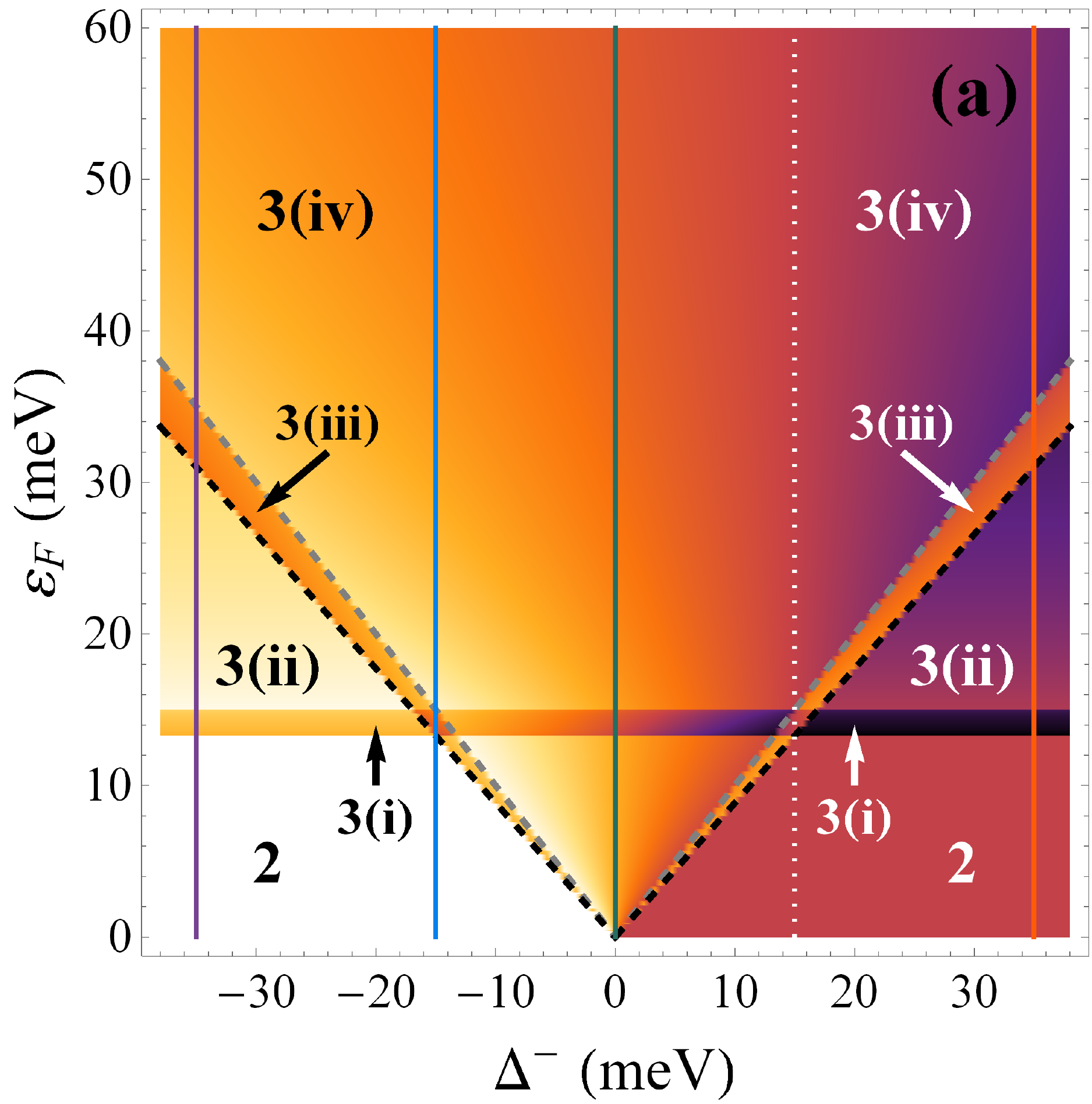}
    \includegraphics[scale=0.35]{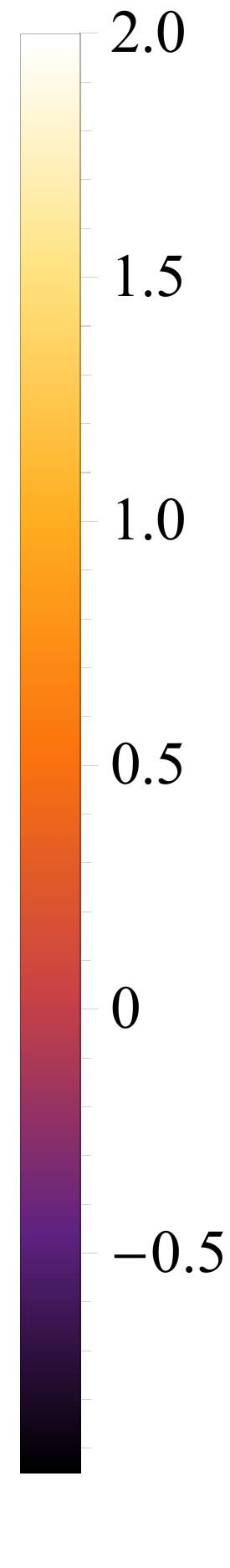}
    \includegraphics[scale=0.265]{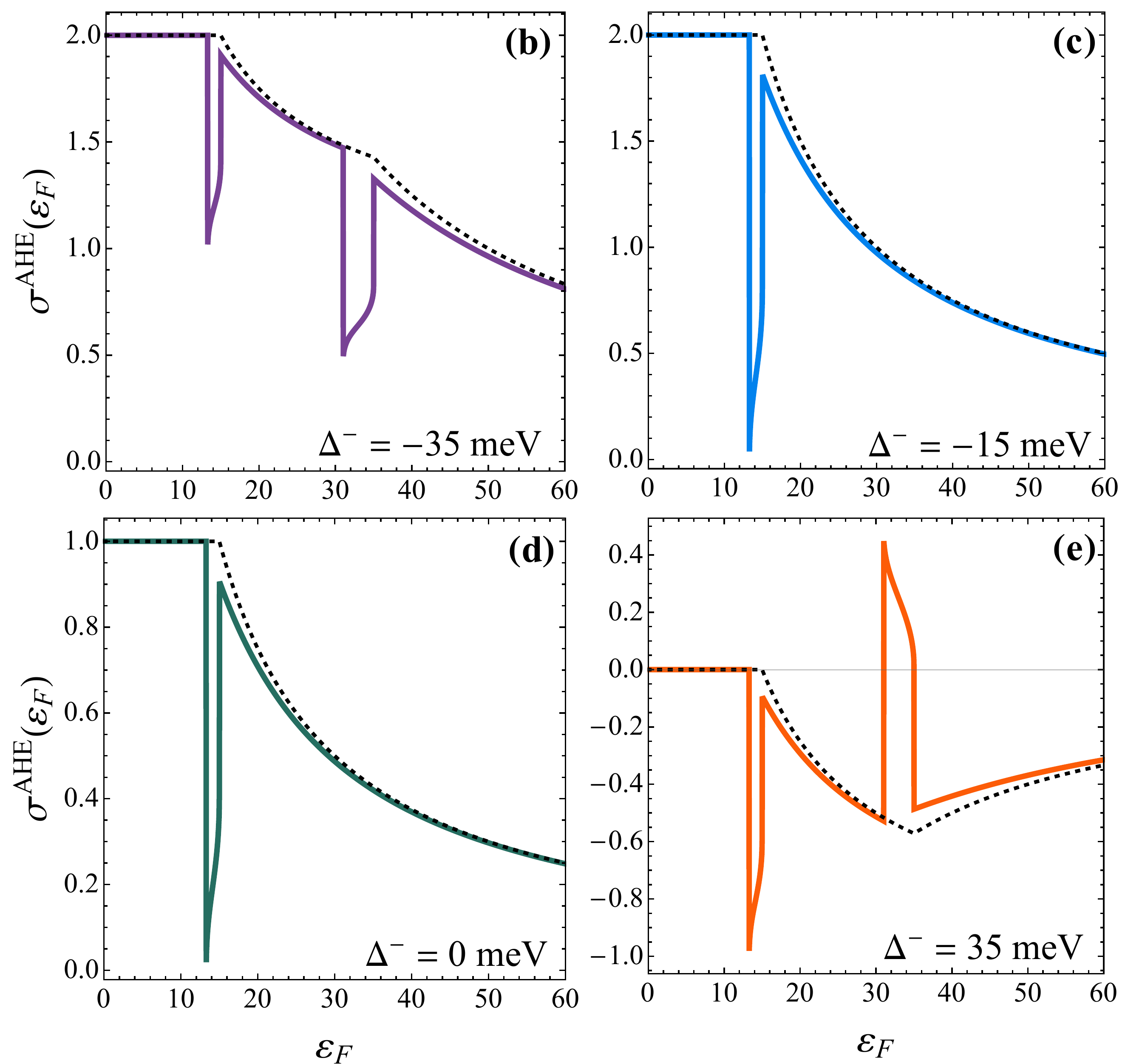}
    \caption{(a) Contour map of the anomalous Hall conductivity (AHC) $\sigma^{AHE}(\varepsilon_F;\Delta^-)$ (in units of $-e^2/h$), for $\Delta^+=15$ meV ($\tilde{\Delta}^+=13.3$\,meV). Black dashed lines correspond to $\varepsilon_F=\pm\tilde{\Delta}^-$ and gray dashed lines to $\varepsilon_F=\pm\Delta^-$. The horizontal strip indicates the indirect zone $\tilde{\Delta}^+<\varepsilon_F<\Delta^+$. The AHC vanishes along the vertical dotted line, defined by $\Delta^-=\Delta^+$. (b)-(e) Function $\sigma^{AHE}(\varepsilon_F)$ for the values of the gap parameter $\Delta^-$ indicated by the vertical lines in (a). The dotted lines show the AHC of gapped graphene with $\Delta^-\neq\Delta^+$.}
    \label{fig:my_labelAHE}
\end{figure}

In Fig.\,\ref{fig:my_labelAHE} we show the anomalous Hall conductivity as a function of Fermi energy for a continuous variation of $\Delta^-$, at a given value of the gap at the $K$
valley. Following Ref.\,\onlinecite{hill2011valley}, we consider positive and negative values of the quantities $\Delta^{\pm}$. Because 
$\sigma^{AHE}$ (\ref{AnomalousSigma}) is an even function of $\varepsilon_F$ we show only results for $\varepsilon_F\geqslant 0$. The labels on the figure \ref{fig:my_labelAHE}(a) indicate the region of each scenario, governed by the corresponding
equation (for $|\Delta^-|>|\Delta^+|$). For instance, the regions marked by number 2 correspond to the equation 
(\ref{in_gap}), with $\text{sgn}(\Delta^-)=-\text{sgn}(\Delta^+)$
at the left, giving the universal value $-2e^2/h$ for the AHC, and $\text{sgn}(\Delta^-)=\text{sgn}(\Delta^+)$ at the right, giving a null
value. The narrow strips labeled as $3(i)$ correspond to the AHC obtained from (\ref{3icase}), while its magnitude in
the triangular regions labeled as $3(ii)$ are given by (\ref{3iicase}), and so on. The AHC in the remaining regions which are not labeled, 
corresponding to the situation $|\Delta^-|<|\Delta^+|$, can be obtained from the same equations
(\ref{in_gap})-(\ref{3ivcase}) after the exchange $\Delta^- \leftrightarrow \Delta^+$. The small intersections between the narrow horizontal strip ($3(i)$) and the sectors marked as $3(iii)$ correspond to the case 4, Eq.(\ref{4case}), of overlapped indirect zones. We remark the need of the breaking of valley symmetry in order to have a finite Hall response
(see Eq.(\ref{AnomalousSigma})). Indeed,
along the line $\Delta^-=\Delta^+$
(dotted line in (a)) the time-reversal symmetry is recovered and the AHC vanishes.
Figures \ref{fig:my_labelAHE}(b)-(e) present the function $\sigma^{AHE}(\varepsilon_F)$
for several values of the gap parameter $\Delta^-$, obtained by the vertical cuts indicated in the contour map. The magnitudes on the vertical cut
at $\Delta^-=0$ (green line in (a), and (d))
start in the universal value $-e^2/h$ (Eq.(\ref{in_gap+})), then takes a reduced value in the narrow strip (Eq.(\ref{1iicase})), and decreases afterward according to Eq.(\ref{1iiicase}).
For gapped graphene with unequal gaps $\sigma^{AHE}/(-e^2/h)=\text{sgn}(\Delta^+)-\text{sgn}(\Delta^-)$ if $|\varepsilon_F|<\text{min}\{|\Delta^+|,|\Delta^-|\}$, 
$(\Delta^+/|\varepsilon_F|)-\text{sgn}(\Delta^-)$ if
$|\Delta^+|<|\varepsilon_F|<|\Delta^-|$, and $(\Delta^+-\Delta^-)/|\varepsilon_F|$ if $|\varepsilon_F|>\text{max}\{|\Delta^+|,|\Delta^-|\}$.
 For the sake of comparison, we have included in Fig.\,\ref{fig:my_labelAHE}(b)-(e) this result. We note
that the main deviation from graphene behavior occurs due to the indirect zones, which are caused by the tilting and the mass in each valley.


Similar expressions to (\ref{in_gap+})-(\ref{4case})
are derived for the conductivity response function 
$\sigma^{VHE}=\sigma^{(+)}_{xy}(0)-\sigma^{(-)}_{xy}(0)$,
which characterizes the valley Hall effect. A
contour map like that in Fig.\,\ref{fig:my_labelAHE}(a)
is obtained for the valley Hall conductivity (VHC) after
a reflection in the line $\Delta^-=0$. 
In particular, when
$\Delta^++\Delta^-=0$ the system presents inversion symmetry
and a corresponding null valley Hall response.
Moreover, the 
indirect zones introduce again the main modifications with respect to the valley response in the model of gapped graphene proposed by Hill et al. \cite{hill2011valley} ($\gamma=0, v_x=v_y$).

\subsection{Reflection and transmission} \label{RT}

From the electromagnetic scattering problem of optical reflection and refraction at a flat interface made of a 2D system, with
electrical conductivity  $\sigma_{ij}$, separating two homogeneous media with dielectric constants $\epsilon_1$ and $\epsilon_2$,
it is found that the optical reflectivity $R$ and transmissivity $T$ are given by 
\begin{eqnarray}
R(\omega,\theta_i,\phi;\varepsilon_F)&=&(|r_{pp}|^2+|r_{sp}|^2)\cos^2\phi+(|r_{ss}|^2+|r_{ps}|^2)\sin^2\phi  
+2\,\mbox{Re}(r_{pp}r_{ps}^*+r_{ss}r_{sp}^*)\sin\phi\cos\phi \\
\nonumber \\
T(\omega,\theta_i,\phi;\varepsilon_F)&=& F(\theta_i)
\left[(|t_{pp}|^2+|t_{sp}|^2)\cos^2\phi 
 +(|t_{ss}|^2+|t_{ps}|^2)\sin^2\phi +
 2\,\mbox{Re}(t_{pp}t_{ps}^*+t_{ss}t_{sp}^*)
\sin\phi\cos\phi\right] ,
\end{eqnarray}
where $F(\theta_i)=\sqrt{(\epsilon_2/\epsilon_1)-\sin^2\theta_i}/\cos\theta_i$,
$\theta_i$ is the angle of incidence and 
 the angle of polarization $\phi$ is measured from the plane of incidence; $p\, (s)$-polarization corresponds to $\phi=0\,(\pi/2)$.
The coefficients $r_{\mu\nu}(\omega,\theta_i)$ ($t_{\mu\nu}(\omega,\theta_i)$) 
are the Fresnel reflection (transmission) amplitudes corresponding to a $\nu$-polarized ($p$ or $s$) incident wave generating a $\mu$-polarized ($p$ or $s$)
reflected (transmitted) wave. In our problem, for the amplitudes involving polarization conversion we find $t_{sp}=r_{sp}=r_{ps} \propto \sigma_{xy}(\omega)$, and $t_{ps}=-\sqrt{\epsilon_2/\epsilon_1}\,r_{ps}/F(\theta_i)\propto\sigma_{xy}(\omega)$; for the conserved polarization cases,
$t_{ss}=1+r_{ss}, \,t_{pp}=\sqrt{\epsilon_2/\epsilon_1}(1-r_{pp})/F(\theta_i)$.
Explicit expressions for the amplitudes $r_{\mu\nu}(\omega,\theta_i)$ are given in the Appendix \ref{refraction}.
We can expect that for $\mu=p$ or $s$ incident polarization
$R(\omega)\approx |r_{\mu\mu}(\omega)|^2$ and $T(\omega)\approx F(\theta_i)|t_{\mu\mu}(\omega)|^2$, and that
their frequency dependence be mainly determined
by $\sigma_{xx}(\omega)=\sum_{\xi}\sigma_{xx}^{(\xi)}(\omega)$ for $\mu=p$ and by $\sigma_{yy}(\omega)=\sum_{\xi}\sigma_{yy}^{(\xi)}(\omega)$ for $\mu=s$, given that $|\sigma_{xy}/c|^2\ll 1$
(see (\ref{rpp})-(\ref{rss})).

\begin{figure}[h]
    \centering
    \includegraphics[scale=0.32]{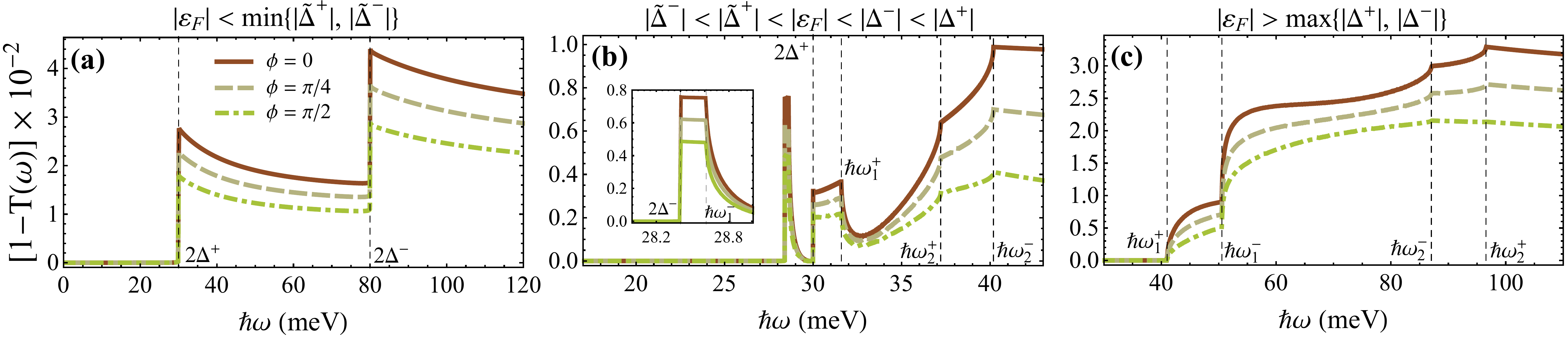}
    \caption{Optical opacity $1-T(\omega)$ for several values of the angle of polarization $\phi$, at normal incidence ($\theta_i=0$), 
    and for the free-standing sample ($\epsilon_1=\epsilon_2=1$). (a) Fermi level within the absolute gap (scenario 2), with $\Delta^+=15$ meV, $\Delta^-=40$ meV, $\varepsilon_F=0$ meV, (b) Fermi energy lying at overlapped indirect zones (scenario 4), with $\Delta^+=15$ meV, $\Delta^-=14.2$ meV,  $\varepsilon_F=13.5$ meV, and (c) Fermi level above the direct zones (scenario $3(iv)$) with $\Delta^+=15$ meV,  $\Delta^-=25$ meV,  $\varepsilon_F=27$ meV.}
    \label{fig:my_labelOpacity}
\end{figure}

This is illustrated in Fig.\,\ref{fig:my_labelOpacity} where the frequency dependence of the optical opacity $1-T(\omega;\phi)$ at
normal incidence of a free-standing sample ($F(\theta_i)=1$) is shown for
several scenarios according to the position of the Fermi level and relative magnitudes of the gaps. Given that
$R\sim 10^{-4}$, the absorbance $1-T-R$ is determined to a large extent by that quantity. Approximately $1-T(\omega)\approx (4\pi/c)\text{Re}[\sigma_{ii}(\omega)]$, with
$i=x\, (y)$ for $p\,(s)$-polarization. Indeed, in Fig.\,\ref{fig:my_labelOpacity} the form of the function
$\sigma^{(\xi)}_{xx}(\omega)$ for $\phi=0$ or of
$\sigma^{(\xi)}_{yy}(\omega)$ for $\phi=\pi/2$
can be easily identified after Fig.\,(\ref{fig:my_label_opt1})
(a)-(c). When $\varepsilon_F$ lies in the gaps,
\begin{equation}
1-T(\omega)\approx \frac{\pi\alpha}{2}\frac{v_i^2}{v_xv_y}
\left\{\left[1+\left(\frac{2\Delta^+}{\hslash\omega}\right)^2\right]\Theta(\hslash\omega-2|\Delta^+|)+
\left[1+\left(\frac{2\Delta^-}{\hslash\omega}\right)^2\right]\Theta(\hslash\omega-2|\Delta^-|)\right\} ,
\end{equation}
where $\alpha=e^2/\hslash c$ is the fine structure constant.
For $\varepsilon_F>\text{max}\{|\Delta^+|,|\Delta^-|\}$,
with $\omega>\text{max}\{\omega_2^+,\omega_2^-\}$,
\begin{equation}
1-T(\omega)\approx \frac{\pi\alpha}{2}\frac{v_i^2}{v_xv_y}
\left[2+\left(\frac{2\Delta^+}{\hslash\omega}\right)^2+
\left(\frac{2\Delta^-}{\hslash\omega}\right)^2\right] \ .
\end{equation}
For high enough frequencies, these results tends both to
$(2.3\%)(v_i^2/v_xv_y)$, where the value $\pi\alpha=2.3\%$
is the well known visual transparency of pristine graphene, defined only by fundamental constants\cite{nair2008fine}. That limit values are close to
$2.9\%$ for $i=x$ and $1.8\%$ for $i=y$. In Fig.\,\ref{fig:my_labelOpacity}(a), for $p$-polarization at $\hslash\omega\gtrsim 2\Delta^+$ and $\hslash\omega\gtrsim 2\Delta^-$, $1-T\approx 2.9\%$ and $4.5\%$ for $i=x$, and
$1.8\%$ and $2.9\%$ for $i=y$, respectively. Comparable
values are obtained in Fig.\,\ref{fig:my_labelOpacity}(c)
at $\omega_2^{\pm}$. On the other hand, for
$\varepsilon_F$ within overlapping indirect zones the transmission increases, as expected, with $1-T(\omega)< 1\%$
(Fig.\,\ref{fig:my_labelOpacity}(b)). For $\epsilon_2>1$ the
opacity increases. As an example, we obtain $1-T(\omega)\lesssim 3.8\%$
for the scenario of Fig.\,\ref{fig:my_labelOpacity}(b) when $\epsilon_2=2$.


\subsection{Rotation of polarization} \label{FK}

The breaking of the time reversal symmetry and the concomitant finite value of a transverse response lead to the 
well known phenomenon of polarization rotation of reflected and transmitted optical waves. The Kerr ($K$) and Faraday ($F$) angles
giving the azimuth of the ellipse of polarization can be obtained from the expression 
\begin{equation}\label{KF1}
\tan 2\theta^{\alpha}_{K/F}=\frac{2\text{Re}\{\chi^{\alpha}_{K/F}\}}{1-|\chi^{\alpha}_{K/F}|^2}
\end{equation}
where $\alpha=p\,(s)$ indicates the incident linearly $p\,(s)$-polarized light,  $\chi^p_K=-r_{sp}/r_{pp},\, \chi^s_K=r_{ps}/r_{ss}$
for the reflected light, and $\chi^p_F=t_{sp}/t_{pp},\, \chi^s_F=-t_{ps}/t_{ss}$ for the transmitted light. Typically $|\chi^{\alpha}_{K/F}|\ll 1$
and $\theta^{\alpha}_{K/F}\approx \text{Re}\{\chi^{\alpha}_{K/F}\}$.

\begin{figure}[h!]
    \centering
    \includegraphics[scale=0.46]{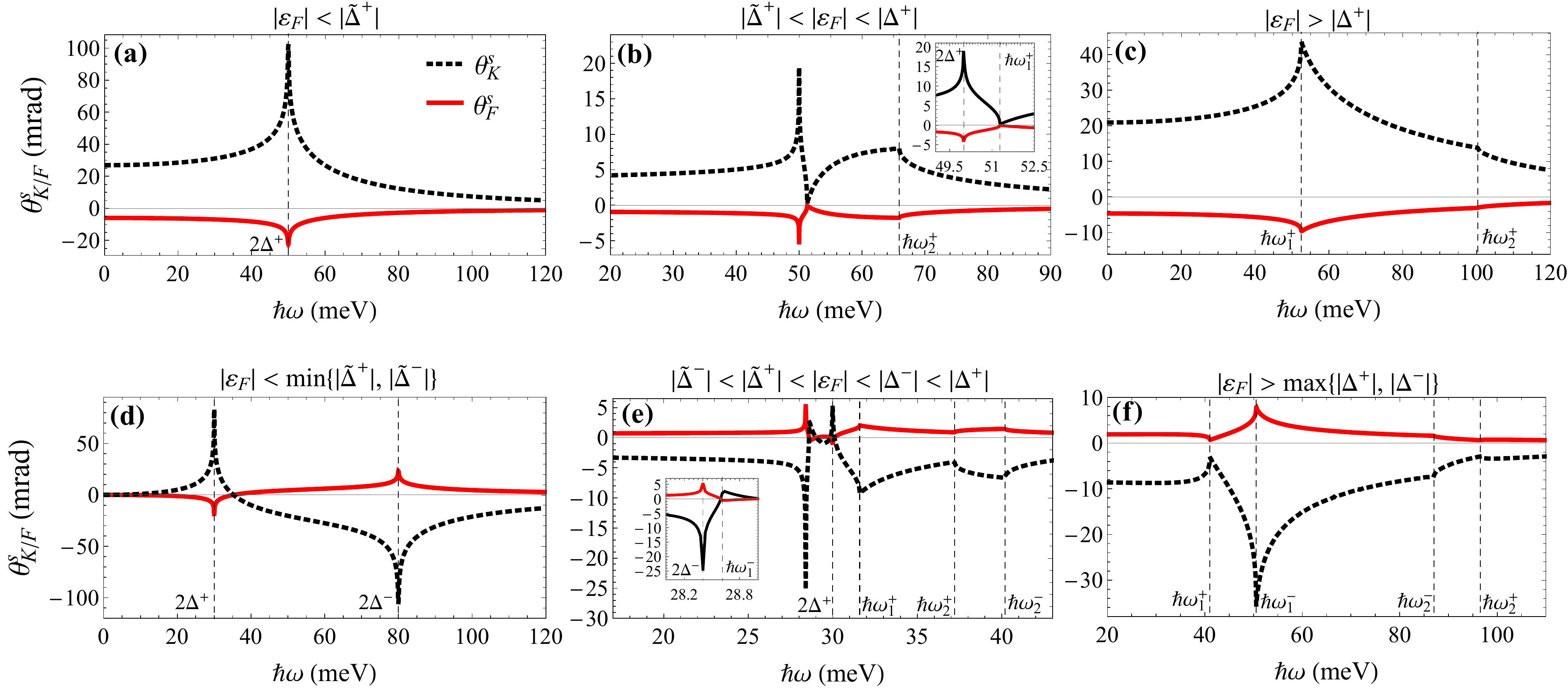}
    \caption{Kerr and Faraday rotations $\theta_{K/F}^s$ at normal incidence ($\theta_i=0$),
    and considering $\epsilon_1=1$, $\epsilon_2=2$ (SiO$_2$). In (a)-(c) it is considered $\Delta^+=25$ meV and $\Delta^-=0$, for different positions of the Fermi level as shown. (d) Fermi level within the absolute gap (scenario 2), with $\Delta^+=15$ meV, $\Delta^-=40$ meV, $\varepsilon_F=0$ meV, (e) Fermi energy lying at overlapped indirect zones (scenario 4), with $\Delta^+=15$ meV, $\Delta^-=14.2$ meV,  $\varepsilon_F=13.5$ meV, and (f) Fermi level above the direct zones (scenario $3(iv)$) with $\Delta^+=15$ meV,  $\Delta^-=25$ meV,  $\varepsilon_F=27$ meV.}
    \label{fig:my_labelKF}
\end{figure}

Figure \ref{fig:my_labelKF} shows results for $\theta_{K/F}^s(\omega)$ for three different positions of the Fermi energy when the $K'$ valley is gapless (top panels), 
and for the scenarios 2, 4, and $3(iv)$ (bottom panels). We consider normal incidence, and $\epsilon_1=1, \epsilon_2=2$. 
We note that the frequency dependence of these angles follows mainly that of the function $-\text{Re}[\sigma_{xy}(\omega;\varepsilon_F)]$
(see Fig.\,\ref{fig:my_labelH}). 
Indeed, given the smallness of $|\sigma_{ii}|/c$ and  $(|\sigma_{xy}|/c)^2$, we find $\theta_{K/F}^p(\omega)\approx \theta_{K/F}^s(\omega)$, 
and as a good approximation
\begin{eqnarray*}
\theta^s_{K}(\omega) &\approx&-\frac{8\pi\text{Re}[\sigma_{xy}(\omega)/c]}{\epsilon_2-1+2\pi\alpha\sqrt{\epsilon_2}} \\
\theta^s_{F}(\omega) &\approx& \frac{4\pi\text{Re}[\sigma_{xy}(\omega)/c]}{\sqrt{\epsilon_2}+1+\pi\alpha} \ ,
\end{eqnarray*}
after taking $\sigma_{ii}/c \approx \pi\sigma_0/4c=\alpha/4$.
As expected, near the resonances the Kerr and Faraday rotations increases, reaching magnitudes $\sim 10^{-2}-10^{-1}\,$rad and
$\sim 10^{-3}-10^{-2}\,$rad, respectively. Note also that at low frequency $\theta^s_{K/F}(0)$ is determined approximately by $\sigma^{AHE}$.
In the valley symmetry breaking mechanism suggested for graphene by Hill et al. \cite{hill2011valley}, a periodic magnetic flux opens gaps at both valleys  which can be tuned independently. In our case, that mechanism would allow to change the sign of $\Delta^-$ leading to 
$\theta^s_{K/F}(0) \approx \sigma^{VHE}$.

\begin{figure}
    \centering
    \includegraphics[scale=0.45]{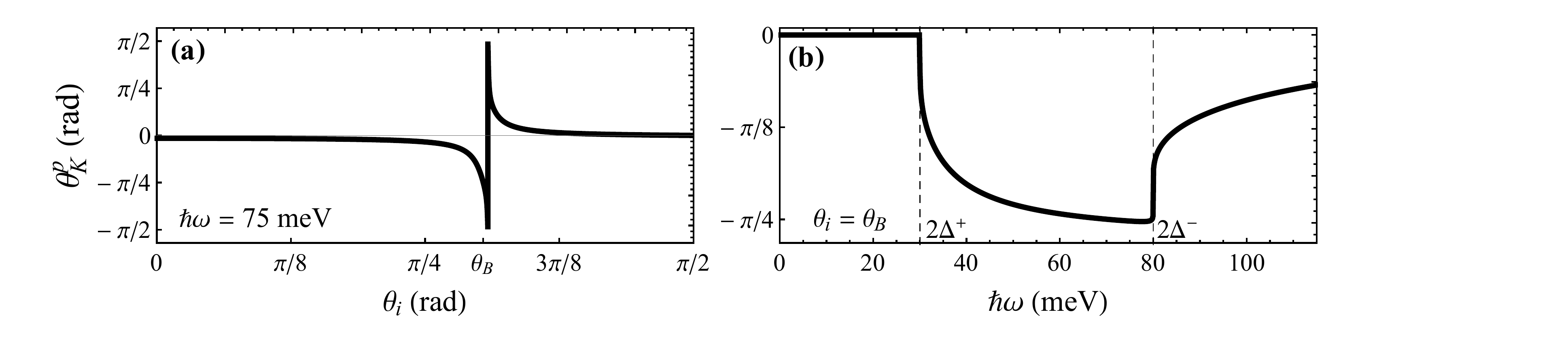}
    \caption{Kerr rotation $\theta_K^p(\omega,\theta_i)$ for Fermi level within the gap of the system, $\varepsilon_F<\text{min}\{|\tilde{\Delta}^+|,|\tilde{\Delta}^-|\}$ (scenario 2), with $\Delta^+=15\,$meV, $\Delta^-=40\,$meV,  $\varepsilon_F=0\,$meV, and $\epsilon_1=1$, $\epsilon_2=2$ (SiO$_2$).
    (a) Kerr angle as a function of the angle of incidence,
    at $\hslash\omega=75\,$meV, close to the onset $2\Delta^-$.
    For $\theta_i\approx\theta_B$, $\theta_K^p\approx\pi/2$. (b) Kerr angel spectrum at $\theta_i=\theta_B$.}
    \label{fig:brewster}
\end{figure}

Figure \ref{fig:brewster}(a) shows the Kerr rotation as a function of the incident angle, when the Fermi level lies within the gaps, at a frequency close to the onset for interband transitions at the $K'$ valley, $\hslash\omega\approx 2\Delta^-$. There is a strong enhancement of the rotation,
about $\pi/2$, for an incident angle defined approximately by
$(\sqrt{\epsilon_2}/\cos\theta_t)-(\sqrt{\epsilon_1}/
\cos\theta_i)=-(4\pi/c)\sigma_{xx}(\omega)$,
where $\theta_t$ is determined by the Snell's law (Appendix\,\ref{refraction}). 
That angle is very close to the Brewster value given by $\tan\theta_B=\sqrt{\epsilon_2/\epsilon_1}$. At $\theta_i=\theta_B$ we find
\begin{equation}
\chi_K^p(\omega;\theta_B)\approx -
\frac{2\sqrt{\epsilon_1}\sigma_{xy}}{\sqrt{\epsilon_1+\epsilon_2}\,\sigma_{xx}+\frac{4\pi}{c}(\sigma_{xx}\sigma_{yy}+\sigma^2_{xy})} \ ,
\end{equation}
which can be further approximated by $\chi_K^p\approx -2\cos\theta_B(\sigma_{xy}/\sigma_{xx})$; the Kerr angle is
found from (\ref{KF1}). An enhancement close to the Brewster angle has also been reported for bilayer graphene in the quantum anomalous Hall state\cite{szechenyi2016transfer}. 

In Fig.\,\ref{fig:brewster}(b) the spectrum for the Kerr
rotation at $\theta_i=\theta_B$ is displayed. 
For low frequencies, below the onsets of interband transitions, the conductivity response is determined by the
dispersive components. Indeed, $\sigma_{xy}$ is a real quantity while $\sigma_{xx}$ becomes imaginary,
yielding $\text{Re}(\sigma_{xy}/\sigma_{xx})\approx 0$ and a small value for the Kerr rotation.
Above $2\Delta^+$, the optical transitions in the $K$ valley increase the dissipative components $\text{Re}(\sigma_{xx}^{(+)})$ and 
$\text{Im}(\sigma_{xy}^{(+)})$, implying that the relative phase of $\sigma_{xy}$ and $\sigma_{xx}$ becomes increasingly small.
Accordingly, the angle $\theta_K^p$ starts to increase in magnitude and takes its largest (negative) value for $\hslash\omega\lesssim 2\Delta^-$, 
when the complex quantities $\sigma_{xy}$ and $\sigma_{xx}$ are closely in phase. However, for $\hslash\omega>2\Delta^-$ the Kerr angle
decreases in magnitude again for increasing frequency. 
This is due to the fact that the optical transitions at both valleys start to contribute, and for high enough frequency
$\sigma_{xy}\approx i\text{Im}(\sigma_{xy})$ and $\sigma_{xx}\approx\text{Re}(\sigma_{xx})$ become out of phase.


The characteristic spectral features displayed by these
results provide fingerprints of the anomalous transverse response, and suggest optical polarization rotation measurements as a contact-free probe of the valley symmetry breaking in the presence of tilting.

\section{Summary and Conclusions}\label{sec:Conc}

Following the study of Hill et al. \cite{hill2011valley} about a mechanism to break the valley symmetry in graphene and tune the gaps independently, 
we explored the  effect  of  valley-contrasting  gaps on the optical properties of two-dimensional anisotropic tilted Dirac systems like 
8-$Pmmn$ borophene  and some organic conductors. We employed a low-energy effective Hamiltonian with broken particle-hole and valley symmetries.
The energy spectrum is characterized by a valley index and the helicity of the states. Notably, 
the energy bands develop an indirect gap $2\tilde{\Delta}^\xi=2\Delta^{\xi}\sqrt{1-\gamma^2}$ in each valley,  which is lower
than the nominal gap $2\Delta^{\xi}$ of the untilted system and depends on the tilting and anisotropy through the parameter $\gamma=v_t/v_y$. 
As a consequence, when the Fermi level is outside a gap but close to the band edge of each valley, the corresponding Fermi contours can be displaced enough to provoke a dramatic change of the momentum space available for optical transitions. To explore this, 
we first calculated the joint density of states to probe the spectrum of interband transitions. When the Fermi energy lies within a gap, the JDOS
displays the characteristic linear frequency dependence above the only critical energy defined by the onset $2\Delta^{\xi}$. 
On the other hand, if the Fermi level is above $2\Delta^{\xi}$, the behavior of JDOS is no longer like that of graphene with the usual
threshold at $\hbar\omega=2\varepsilon_F$, but looks qualitatively similar to that of borophene case, presenting
two critical energies made possible by the tilt of the bands. However, given the new possibility of an indirect range
 $\tilde{\Delta}^{\xi}<\varepsilon_F<\Delta^{\xi}$, we found that the JDOS presents now three van Hove singularities in that range,
 and a strong reduction
 of the number of vertical transitions in a subregion located between photon energy intervals with graphene-like and a borophene-like behaviors.
 Accordingly, the intraband and interband parts of the optical conductivity tensor were also obtained from the Kubo formula. 
 We found an anisotropic response $\sigma^{(\xi)}_{xx}(\omega)\neq\sigma^{(\xi)}_{yy}(\omega)$ and a finite Hall component 
 $\sigma^{(\xi)}_{xy}(\omega)$ with spectral features determined by the set of interband critical points revealed by the JDOS. When the
 contributions of each valley are combined to obtain the total response, a number of spectra are obtained because of the
 multiple possibilities for the position of the Fermi level in the complete set of bands, opened by the presence of non uniform gaps.
 Similarly, the Drude weight is anisotropic and shows a sensitive dependence on the Fermi energy. To further characterize this scenario,
 we also calculated  the anomalous and valley Hall conductivities through the Berry curvature of the bands, and spectra of circular dichroism, optical opacity, and Kerr and Faraday angles of polarization rotation. 
 The plots of the AHC and VHC versus Fermi energy show behaviors that differ appreciably from the case without tilting due to the presence of indirect zones. Moreover, when the Fermi level is inside the absolute gap they can take universal values.
 Interestingly, we found that the existence of indirect zones makes possible to have almost perfect circular dichroism
 for right or left handed polarized light. Valley polarization
 can appear simultaneously, as recently reported within a modified Haldane model.
  With respect to the rotation of polarization, the Kerr and Faraday
 spectra display a strong dependence on the position of the Fermi level, signaled by the van Hove singularities of the JDOS,
 and reaching magnitudes about $10^{-3} - 10^{-2}\,$rad and $10^{-2} - 10^{-1}\,$rad, respectively. We observed an enhancement of Kerr rotation close to the Brewster angle of incidence, similar to that reported for bilayer graphene in the anomalous Hall state. We also found that by choosing appropriately the Fermi level position and tuning the exciting frequency close to the interband critical points, the optical transparency can deviate from the well known universal result  $\pi e^2/\hslash c=2.3\,\%$ of graphene by a factor that
 depends on the velocities 
 $v_x, v_y$ and gap parameters.
 
 In summary, the results show clear optical signatures of valley and electron-hole symmetry breaking in the optical
 properties, suggesting diverse optical ways to explore the simultaneous presence of anisotropy, tilt, and non uniform gaps in Dirac systems.

\section{Acknowledgements}
M.A.M. and R.C.-B. acknowledge V\'ictor Ibarra and Priscilla Iglesias for useful discussions on this work. M.A.M. and R.C.-B. thank to the 20va Convocatoria Interna (UABC).

\appendix
\section{Fermi lines} \label{FermiLines}
\begin{enumerate}
\item For $|\varepsilon_F|>|\Delta^{\xi}|$, the equation $\varepsilon^{\xi}_{\lambda}(k,\theta)=\lambda|\varepsilon_F|$ yields
the parametric curve
\begin{equation}
k^{\xi}_{\lambda,F}(\theta)=\frac{1}{\alpha_F}\frac{\sqrt{E(\theta)}-\xi\lambda|\varepsilon_F|h(\theta)}
{g^2(\theta)-h^2(\theta)}
\end{equation}
for $\theta\in [0,2\pi]$,
where $E(\theta)=\varepsilon_F^2g^2(\theta)-(\Delta^{\xi})^2[g^2(\theta)-h^2(\theta)]$. 
The energy difference $\varepsilon^{\xi}_+(k,\theta) -\varepsilon^{\xi}_-(k,\theta)=2d^{\xi}(k,\theta)$ at the Fermi
 curve is denoted by $\hslash\omega^{\xi}_{\lambda}(\theta)=2d^{\xi}(k^{\xi}_{\lambda,F}(\theta),\theta)$, and given by
\begin{equation}
\hslash\omega^{\xi}_{\lambda}(\theta) = 2\,\frac{|\varepsilon_F|g^2(\theta)-\xi\lambda h(\theta)\sqrt{E(\theta)}}{g^2(\theta)-h^2(\theta)}.
\end{equation}
\item For $|\tilde{\Delta}^{\xi}|\leqslant |\varepsilon_F|\leqslant |\Delta^{\xi}|$, the Fermi
line is given by the parametric curve
\begin{equation}
q_{\lambda,F}^{\xi,\pm}(\theta)=\frac{1}{\alpha_F}
\frac{\xi\lambda|\varepsilon_F|(-h(\theta))\pm\sqrt{E(\theta)}}{g^2(\theta)-h^2(\theta)}
\end{equation}
defined in the angular regions $|\theta-\theta_0|\leqslant\theta^*_{\xi}$, where $\theta_0=3\pi/2$ if $\xi\lambda=+$ and $\pi/2$ when $\xi\lambda=-$.
The angle $\theta^*_{\xi}$ is defined by the condition 
$E(\theta)\geqslant 0$.
Correspondingly, the energy difference between the conduction and valence band at the Fermi curve is 
$\hslash\nu^{\xi,\pm}_{\lambda}(\theta)=2d^{\xi}(q_{\lambda,F,}^{\xi,\pm}(\theta),\theta)$.
 Explicitly
\begin{equation} 
\hslash\nu^{\xi,\pm}_{\lambda}(\theta) = 2\,\frac{|\varepsilon_F|g^2(\theta)\pm\xi\lambda(-h(\theta))\sqrt{E(\theta)}}{g^2(\theta)-h^2(\theta)},
\end{equation}
with $\hslash\nu^{\xi,+}_{\lambda}(\theta)>\hslash\nu^{\xi,-}_{\lambda}(\theta)$.
\end{enumerate}



\section{Fresnel amplitudes}\label{refraction}

Here we sketch the solution of the electromagnetic problem defined in Section \ref{RT}. 
We consider harmonic plane waves ${\bf F}({\bf r},t)={\bf F}({\bf r})e^{-i\omega t}$ with ${\bf F}({\bf r})={\bf F}e^{i{\bf k}\cdot{\bf r}}$ satisfying
the Helmholtz equation $(\nabla^2+n^2k_0^2){\bf F}({\bf r})=0$, where the wave vector ${\bf k}=nk_0{\bf\hat{k}}$
lies in the $ZX$-plane, ${\bf k}\cdot{\bf F}=0$, $k_0=\omega/c$, and $\,n=\sqrt{\epsilon}$ is the index of refraction. It is convenient to introduce two vectors perpendicular 
to ${\bf\hat{k}}$ to span the vectorial amplitude ${\bf F}$. One of these vector is ${\bf\hat{s}}={\bf\hat{z}}\times{\bf\hat{x}}={\bf\hat{y}}$; for the other
we can take ${\bf\hat{p}}={\bf\hat{s}}\times{\bf\hat{k}}$. Thus we have defined the (real) orthogonal triad $({\bf\hat{p}},{\bf\hat{s}},{\bf\hat{k}})$ as
a basis to describe transverse plane waves. The incident electric field ${\bf F}={\bf E}^i$ from the medium with dielectric constant $\epsilon_1$ is
then written in terms of its $p$ and $s$ amplitudes as ${\bf E}^i = E_p^i{\bf\hat{p}}^i+E_s^i{\bf\hat{s}}$, where
${\bf\hat{p}}^i= {\bf\hat{s}}\times{\bf\hat{k}}^i=\frac{1}{k^i}(k_z^i{\bf\hat{x}}-k_x{\bf\hat{z}})=\cos\theta_i\,{\bf\hat{x}}-\sin\theta_i\,{\bf\hat{z}}$ and
 ${\bf k}^i= k_0\sqrt{\epsilon_1}\,\,{\bf\hat{k}}^i=k_x{\bf\hat{x}}+k_z^i{\bf\hat{z}}=k_0\sqrt{\epsilon_1}(\sin\theta_i\,{\bf\hat{x}}+\cos\theta_i\,{\bf\hat{z}})$
 are the polarization vector and wave vector of the incident wave, with $k^i=k_0\sqrt{\epsilon_1}$ and $\theta_i$ being the angle of incidence.
 Similarly, the reflected field is ${\bf E}^r= E_p^r{\bf\hat{p}}^r+E_s^r{\bf\hat{s}}$, where
 ${\bf\hat{p}}^r ={\bf\hat{s}}\times{\bf\hat{k}}^r=\frac{1}{k^i}(-k_z^r{\bf\hat{x}}-k_x{\bf\hat{z}})=-\cos\theta_i\,{\bf\hat{x}}-\sin\theta_i\,{\bf\hat{z}}$ and
 ${\bf k}^r =k_0\sqrt{\epsilon_1}\,\,{\bf\hat{k}}^r=k_x{\bf\hat{x}}-k_z^r{\bf\hat{z}}=k_0\sqrt{\epsilon_1}(\sin\theta_i\,{\bf\hat{x}}-\cos\theta_i\,{\bf\hat{z}})$;
 note that $k_z^r=k_z^i=k_0\sqrt{\epsilon_1}\,\cos\theta_i$. In the medium $\epsilon_2$, the transmitted field reads as
 ${\bf E}^t=E_p^t{\bf\hat{p}}^t+E_s^t{\bf\hat{s}}$, with
 ${\bf\hat{p}}^t={\bf\hat{s}}\times{\bf\hat{k}}^t=\frac{1}{k^t}(k_z^t{\bf\hat{x}}-k_x{\bf\hat{z}})=\cos\theta_t\,{\bf\hat{x}}-\sin\theta_t\,{\bf\hat{z}}$,
 ${\bf k}^t =k_0\sqrt{\epsilon_2}\,\,{\bf\hat{k}}^t=k_x{\bf\hat{x}}+k_z^t{\bf\hat{z}}=k_0\sqrt{\epsilon_2}(\sin\theta_t\,{\bf\hat{x}}+\cos\theta_t\,{\bf\hat{z}})$, $\,k^t=k_0\sqrt{\epsilon_2}$;
 the angle of refraction $\theta_t$ is determined by the Snell's law $k_x^i=k_x^t:\ \sqrt{\epsilon_1}\sin\theta_i=\sqrt{\epsilon_2}\sin\theta_t$.
 The corresponding magnetic fields are ${\bf B}^a=\frac{1}{k_0}{\bf k}^a\times {\bf E}^a$, $a=i,r,t$.
 
 The reflected and transmission amplitudes $E_{p,s}^r$, $E_{p,s}^t$ can be written 
  in terms of Fresnel amplitudes $r_{\mu\nu}, t_{\mu\nu}$,
 \begin{equation} \label{matrixR}
\left(\begin{array}{c}
  E_p^r \\
  \\
  E_s^r 
 \end{array}\right)=\left(\begin{array}{cc}
 r_{pp} & r_{ps} \\
 \\
  r_{sp} & r_{ss}
  \end{array}\right)\left(\begin{array}{c}
  E_p^i \\
  \\
  E_s^i
 \end{array}\right) \ \ , \ \ \ \ \ 
 \left(\begin{array}{c}
  E_p^t \\
  \\
  E_s^t 
 \end{array}\right)=\left(\begin{array}{cc}
 t_{pp} & t_{ps} \\
 \\
  t_{sp} & t_{ss}
  \end{array}\right)\left(\begin{array}{c}
  E_p^i \\
  \\
  E_s^i
 \end{array}\right) \ ,
\end{equation}
(in the basis $\{\bf\hat{p},\bf\hat{s}\}$). They  should satisfy the boundary conditions
(1) ${\bf\hat{z}}\cdot[\epsilon_2{\bf E}^t-\epsilon_1({\bf E}^i+{\bf E}^r)]=4\pi\sigma_s(\omega)$, (2) ${\bf\hat{z}}\times [{\bf E}^t-({\bf E}^i+{\bf E}^r)]={\bf 0}$, (3) ${\bf\hat{z}}\cdot [{\bf B}^t-({\bf B}^i+{\bf B}^r)]=0$, and (4) ${\bf\hat{z}}\times[{\bf B}^t-({\bf B}^i+{\bf B}^r)]=\frac{4\pi}{c}{\bf J}_s(\omega)$,
where the induced surface charge density $\sigma_s$ is related to the surface current $J_{s,i}(\omega)=\sigma_{ij}(\omega)E_j(z=0)$,
$\,i,j=x,y$, through the continuity equation, which implies $k_xJ_{s,x}(\omega)=\omega\sigma_s(\omega)$. Given that
 $E_x(z=0)=E_x^i(0)+E_x^r(0)=E_x^t(0)=(k_z^t/k^t)E_p^t=(k_z^i/k^i)(E_p^i-E_p^r)$ and
 $E_y(z=0)=E_y^i(0)+E_y^r(0)=E_y^t(0)=E_s^t=E_s^i+E_s^r$, 
\begin{eqnarray*}
J_{s,x}(\omega) &=&  \sigma_{xx}(\omega)\frac{k_z^t}{k^t}E_p^t+\sigma_{xy}(\omega)E_s^t=\sigma_{xx}(\omega)\frac{k_z^i}{k^i}(E_p^i-E_p^r)+
\sigma_{xy}(\omega)(E_s^i+E_s^r)  \label{Jx} \\
J_{s,y}(\omega) &=& \sigma_{yx}(\omega)\frac{k_z^t}{k^t}E_p^t+\sigma_{yy}(\omega)E_s^t=\sigma_{yx}(\omega)\frac{k_z^i}{k^i}(E_p^i-E_p^r)+
\sigma_{yy}(\omega)(E_s^i+E_s^r) \ .  \label{Jy}
\end{eqnarray*}
The algebraic system of equations defined by the boundary conditions can be rearranged, leading to
\begin{equation*}
\left(\begin{array}{cc}
M_{pp} & M_{ps} \\
M_{sp} & M_{ss}
\end{array}\right)\left(\begin{array}{c}
E_p^r \\ E_s^r
\end{array}\right)=\left(\begin{array}{cc}
F_{pp} & F_{ps} \\
F_{sp} & F_{ss}
\end{array}\right)\left(\begin{array}{c}
E_p^i \\ E_s^i
\end{array}\right) \ ,
\end{equation*}
for the reflexion amplitudes, and
\begin{equation*}
\left(\begin{array}{cc}
M_{pp} & -\eta M_{ps} \\ \\
-\eta^{-1}M_{sp} & M_{ss}
\end{array}\right)\left(\begin{array}{c}
E_p^t \\  \\ E_s^t
\end{array}\right)=2k_z^i\left(\begin{array}{c}
\sqrt{\epsilon_1\epsilon_2}\,E_p^i \\  \\ E_s^i
\end{array}\right) \ ,
\end{equation*}
for the transmission field amplitudes,
where $\eta=\sqrt{\epsilon_2/\epsilon_1}/F(\theta_i)$, 
$F(\theta_i)=k_z^t/k_z^i=\sqrt{(\epsilon_2/\epsilon_1)-\sin^2\theta_i}/\cos\theta_i$.
We shall not write the matrix $M_{\mu\nu}$ nor the source matrix $F_{\mu\nu}$ for brevity.
Comparison with (\ref{matrixR}) allows the Fresnel amplitudes to be identified. We display the result for $r_{\mu\nu}$ only,
\begin{eqnarray}
r_{pp}  &=& \frac{(\epsilon_2k_z^i-\epsilon_1k_z^t+4\pi\frac{\sigma_{xx}}{\omega}k_z^ik_z^t)(k_z^i+k_z^t+\frac{4\pi}{c}k_0\sigma_{yy})-\left(\frac{4\pi}{c}\right)^2\sigma_{xy}\sigma_{yx}k_z^ik_z^t}
{(\epsilon_2k_z^i+\epsilon_1k_z^t+4\pi\frac{\sigma_{xx}}{\omega}k_z^ik_z^t)(k_z^i+k_z^t+\frac{4\pi}{c}k_0\sigma_{yy})-\left(\frac{4\pi}{c}\right)^2
\sigma_{xy}\sigma_{yx}k_z^ik_z^t}  \label{rpp} \\ \nonumber \\
r_{sp}  &=& -\frac{\frac{8\pi}{c}\sigma_{yx}\sqrt{\epsilon_1}\,k_z^ik_z^t}
{(\epsilon_2k_z^i+\epsilon_1k_z^t+4\pi\frac{\sigma_{xx}}{\omega}k_z^ik_z^t)(k_z^i+k_z^t+\frac{4\pi}{c}k_0\sigma_{yy})-\left(\frac{4\pi}{c}\right)^2
\sigma_{xy}\sigma_{yx}k_z^ik_z^t} \label{rsp} \\  \nonumber \\
r_{ps} &=&  \frac{\frac{8\pi}{c}\sigma_{xy}\sqrt{\epsilon_1}\,k_z^ik_z^t}
{(\epsilon_2k_z^i+\epsilon_1k_z^t+4\pi\frac{\sigma_{xx}}{\omega}k_z^ik_z^t)(k_z^i+k_z^t+\frac{4\pi}{c}k_0\sigma_{yy})-\left(\frac{4\pi}{c}\right)^2
\sigma_{xy}\sigma_{yx}k_z^ik_z^t} \label{rps} \\ \nonumber \\ 
r_{ss}  &=&  \frac{(\epsilon_2k_z^i+\epsilon_1k_z^t+4\pi\frac{\sigma_{xx}}{\omega}k_z^ik_z^t)(k_z^i-k_z^t-\frac{4\pi}{c}k_0\sigma_{yy})+\left(\frac{4\pi}{c}\right)^2\sigma_{xy}\sigma_{yx}k_z^ik_z^t} 
{(\epsilon_2k_z^i+\epsilon_1k_z^t+4\pi\frac{\sigma_{xx}}{\omega}k_z^ik_z^t)(k_z^i+k_z^t+\frac{4\pi}{c}k_0\sigma_{yy})-\left(\frac{4\pi}{c}\right)^2
\sigma_{xy}\sigma_{yx}k_z^ik_z^t} \ . \label{rss}
\end{eqnarray}
We can see that if $\sigma_{yx}=-\sigma_{xy}$ then $r_{sp}=r_{ps}$.
Note also that if $|\sigma_{xy}\sigma_{yx}|/c^2\ll 1$ then $r_{pp}$ ($r_{ss}$) involves only the component $\sigma_{xx}$ ($\sigma_{yy}$). For example, at normal incidence and for a free
standing sample $T(\omega)\approx |t_{\mu\mu}|^2\approx 1-(4\pi/c)\text{Re}[\sigma_{ii}(\omega)]$, 
 with $i=x$ for $\mu=p$ and $i=y$ for $\mu=s$, where
$|\sigma_{ii}(\omega)|/c\ll 1$ is assumed for high enough frequencies in the range of interband transitions.

\bibliographystyle{unsrt}
\bibliography{biblio.bib}
\end{document}